\documentclass[aps,preprint,amssymb,showpacs,tightenlines]{revtex4}
\usepackage{pstricks,pst-node,pst-plot}
\newpsobject{showgrid}{psgrid}{subgriddiv=1, griddots=5, gridlabels=6pt}

\usepackage{amsmath}
\usepackage{amssymb}
\usepackage{amsthm}

\newcommand{\R}{\mathbb{R}}
\newcommand{\C}{\mathbb{C}}

\newcommand{\<}{\langle}
\renewcommand{\>}{\rangle}
\newcommand{\half}{\tfrac{1}{2}}

\newcommand{\third}{\tfrac{1}{3}}

\newcommand{\sixth}{\tfrac{1}{6}}
\newcommand{\quarter}{\tfrac{1}{4}}
\newcommand{\thquarters}{\tfrac{3}{4}}
\newcommand{\twthirds}{\tfrac{2}{3}}

\theoremstyle{plain} %% This is the default 
\newtheorem{theorem}{Theorem}

\newtheorem{lemma}{Lemma}
\newtheorem{proposition}{Proposition}
\newtheorem{corollary}{Corollary}
\begin{document} 

\title{On the one-particle reduced density matrices of
a pure three-qutrit quantum state}
\author{Atsushi Higuchi}
\affiliation{Department of Mathematics, University of York, 
Heslington, York YO10 5DD, U.K.\\ email:ah28@york.ac.uk}

\date{25 September, 2003}

\begin{abstract} 
We present
a necessary and sufficient condition for three qutrit
density matrices to be the one-particle reduced density matrices
of a pure three-qutrit quantum state.  The condition consists of
seven classes of inequalities satisfied by the eigenvalues of these
matrices.  One of them 
is a generalization of a known inequality
for the qubit case.  Some of these inequalities are proved 
algebraically whereas the proof of the others uses
the fact that a continuous function of the state must have a
minimum.  Construction of states satisfying these inequalities
relies on a representation of the convex set of the allowed set of 
eigenvalues in terms of corner points.  We also present a result
for a more general quantum system concerning the nature of the
boundary surface of the set of the allowed set of eigenvalues.
\end{abstract}


\maketitle

\section{Introduction} \label{S:Intro}

Recently there has been much activity in clarifying
the nature of entanglement in quantum systems of finite
degrees of freedom in connection with quantum computing,
and there have been some results
about properties of reduced density matrices of a 
pure state.  In particular,
a necessary and sufficient condition has been found 
for $n$ one-qubit density matrices to be the reduced
density matrices of
a pure $n$-qubit quantum state~\cite{hss,bravyi} for all $n$. 
[The phrase ``reduced density matrix"
is used to mean ``one-particle reduced density matrix". It will be
abbreviated as RDM below.]
In this paper, as a
continuation of this line of investigation we find a
necessary and sufficient condition for three one-qutrit
density matrices to be the RDMs 
of a pure three-qutrit quantum state. (Throughout
this paper ``a state" means ``a pure state" unless otherwise stated.)

We consider the space of states spanned by the basis vectors
$|ijk\> \equiv |i\>\otimes |j\>\otimes |k\>$, where $i$, $j$ and
$k$ are $1$, $2$ or $3$. The one-particle states
$|1\>$, $|2\>$ and $|3\>$ are assumed to be orthonormal.
A general state can be given by
$$
|\Psi\>= \sum_{i=1}^3\sum_{j=1}^3\sum_{k=1}^3 c_{ijk}|ijk\>\,.
$$
We require that $\<\Psi\,|\,\Psi\>=\sum_{ijk}|c_{ijk}|^2 = 1$.
Thus, the state $|\Psi\>$ is a normalized 
three-qutrit quantum state.  The three 
RDMs are given by
$\rho^{(1)}_{ii'} = \sum_{J,K} c_{iJK}\overline{c_{i'JK}}$,
$\rho^{(2)}_{jj'} = \sum_{I,K} c_{IjK}\overline{c_{Ij'K}}$ and
$\rho^{(3)}_{kk'} = \sum_{I,J} c_{IJk}\overline{c_{IJk'}}$.
We denote the eigenvalues of the
$a$th RDM, $\rho_{ii'}^{(a)}$, 
by $\lambda_1^{(a)}$, $\lambda_2^{(a)}$ and $\lambda_3^{(a)}$ with
the ordering $\lambda_1^{(a)}\leq \lambda_2^{(a)}\leq\lambda_3^{(a)}$.
We now state the main theorem of this paper.
\begin{theorem}
A necessary and sufficient condition for nine nonnegative
numbers $\lambda_i^{(a)}$, $i=1,2,3$, $a=1,2,3$ satisfying
$\lambda_1^{(a)}\leq \lambda_2^{(a)} \leq \lambda_3^{(a)}$ and
$\lambda_1^{(a)}+\lambda_2^{(a)}+\lambda_3^{(a)} = 1$ to be the
eigenvalues of the three RDMs of
a three-qutrit quantum state is given by the following inequalities:
\begin{eqnarray}
\lambda_2^{(a)} + \lambda_1^{(a)} & \leq &
\lambda_2^{(b)} + \lambda_1^{(b)} + \lambda_2^{(c)} + \lambda_1^{(c)}, 
\label{ineq1} \\
\lambda_3^{(a)}+\lambda_1^{(a)} & \leq & 
\lambda_2^{(b)} + \lambda_1^{(b)} + \lambda_3^{(c)} + \lambda_1^{(c)}, 
\label{ineq2} \\
\lambda_2^{(a)}+\lambda_3^{(a)} & \leq & 
\lambda_2^{(b)} + \lambda_1^{(b)} + \lambda_2^{(c)} + \lambda_3^{(c)},
\label{ineq3} \\
2\lambda_2^{(a)}+\lambda_1^{(a)} & \leq &
2\lambda_2^{(b)} + \lambda_1^{(b)} + 2\lambda_2^{(c)} 
+ \lambda_1^{(c)}, \label{ineq4} \\
2\lambda_1^{(a)} + \lambda_2^{(a)} & \leq &
2\lambda_2^{(b)} + \lambda_1^{(b)} + 2 \lambda_1^{(c)} 
+ \lambda_2^{(c)}, \label{ineq5} \\
2\lambda_2^{(a)} + \lambda_3^{(a)} & \leq &
2 \lambda_2^{(b)} + \lambda_1^{(b)} + 2 \lambda_2^{(c)} 
+ \lambda_3^{(c)}\,,\label{ineq6} \\
2\lambda_2^{(a)} + \lambda_3^{(a)} & \leq & 
2\lambda_1^{(b)} + \lambda_2^{(b)} + 2\lambda_3^{(c)} + \lambda_2^{(c)}
\,, \label{ineq7}
\end{eqnarray}
where $(abc)$ are any permutations of $(123)$. 
\label{main}
\end{theorem}

\noindent
If $\lambda_1^{(a)}=0$ for all $a$, then by writing
$\lambda_3^{(a)} = 1-\lambda_2^{(a)}$, these inequalities can be 
expressed in terms of $\lambda_2^{(a)}$.  We find that
the only nontrivial
inequalities resulting from Theorem~\ref{main} are
$\lambda_2^{(a)} \leq \lambda_2^{(b)} + \lambda_2^{(c)}$
for all permutations $(abc)$ of $(123)$.
This conclusion
agrees with Refs.~\cite{hss,bravyi}.  In Ref.~\cite{bravyi} Bravyi
solved the corresponding 
problem for the system $\C^2\otimes \C^2\otimes \C^4$.
His result can be used to find a necessary and sufficient condition
for three density matrices to be RDMs
of the quantum system $\C^2\otimes \C^2\otimes \C^3$.
This condition can be compared with that obtained from 
Theorem~\ref{main} by letting $\lambda_1^{(1)} = \lambda_1^{(2)}=0$.
These conditions can be shown to be identical.

The rest of the paper is organized as follows.  In 
Sec.~\ref{corner} we represent the set of the allowed set of
eigenvalues as a convex set determined uniquely by the corner points.
In Sec.~\ref{ineqone} we prove a generalization of inequality
(\ref{ineq1}).  Inequality (\ref{ineq1}) itself follows from this
as a corollary.  In Sec.~\ref{ineqtwothree} we provide a proof of
inequalities (\ref{ineq2}) and (\ref{ineq3}).  Our proof of
inequalities (\ref{ineq1})--(\ref{ineq3}) 
is algebraic.  The proof of the
other inequalities uses the fact that a continuous function 
of the (ordered) set of coefficients $(c_{ijk})$ with
the constraint $\sum_{i,j,k}|c_{ijk}|^2 = 1$ must
have a minimum because the domain can be identified with 
the $53$-dimensional sphere of radius one, 
which is compact.  We explain
our method in Sec.~\ref{perturb} and use it
to prove inequalities (\ref{ineq4}) and
(\ref{ineq5}) in Sec.~\ref{ineqfourfive}. 
In Sec.~\ref{ineqsixseven} we prove inequalities
(\ref{ineq6}) and (\ref{ineq7}).  [Our proof of inequality
(\ref{ineq6}) is rather lengthy.] We then construct states for each
set of eigenvalues satisfying (\ref{ineq1})--(\ref{ineq7})
in Secs.~\ref{constr1} and ~\ref{constr2}.  The corner-point
representation given 
in Sec.~\ref{corner} is useful in this construction.
We conclude this paper in Sec.~\ref{conclude} by showing that
the set of the allowed set of eigenvalues for a more general
quantum system is likely to be bounded by hyperplanes.

\section{The corner-point representation of the set of 
the allowed set of eigenvalues}
\label{corner}

Let us write a set of possible eigenvalues as
\begin{equation}
[\lambda_1^{(1)}\lambda_2^{(1)}\lambda_3^{(1)},\lambda_1^{(2)}
\lambda_2^{(2)}\lambda_3^{(2)},\lambda_1^{(3)}\lambda_2^{(3)}
\lambda_3^{(3)}]\,.  \label{eigenpoint}
\end{equation}
Since the combinations $001$ and $0\half\half$ and 
$\third\third\third$ will appear often, we abbreviate them
as $O\equiv 001$, $A\equiv 0\half\half$ and 
$B\equiv \third\third\third$.
A set of eigenvalues can be regarded as a point in 
a six-dimensional space with coordinates $\lambda_1^{(a)}$ and
$\lambda_2^{(a)}$, $a=1,2,3$.  For this reason we call 
a set given by (\ref{eigenpoint}) an eigenvalue point, or an
E-point in short.  In this representation the set of 
E-points defined in Theorem~\ref{main} is a convex set
bounded by hyperplanes.  Therefore this set, which will be denoted
by $S$, can also be specified by
giving all the corner points, i.e. the points on the boundary
of the set where there is no tangent line to the boundary.

If we impose only the conditions 
$0\leq \lambda_1^{(a)}\leq \lambda_2^{(a)} \leq \lambda_3^{(a)}$
and $\lambda_1^{(a)}+\lambda_2^{(a)} + \lambda_3^{(a)}=1$, then
the set of all allowed E-points are bounded by the hyperplanes
$\lambda_1^{(a)}= 0$,
$\lambda_1^{(a)}= \lambda_2^{(a)}$ and
$\lambda_2^{(a)}=\lambda_3^{(a)} = 1-\lambda_1^{(a)}
-\lambda_2^{(a)}$, $a=1,2,3$.
The set of E-points satisfying these conditions is of the form
$L \times L \times L$, where $L$ is given by the triangle shown
in Fig.~1.
\begin{center}
\begin{pspicture}(-2,-2)(4,3)
%\showgrid
\psline{->}(-.5,0)(4,0)
\psline{->}(0,-1)(0,2)
\psline{-}(1.5,1.5)(1,1)
\psline[linewidth=0.5mm](0,0)(1.5,1.5)
\psline[linewidth=0.5mm](0,0)(2.25,0)
\psline[linewidth=0.5mm](1.5,1.5)(2.25,0)
\uput[-90](0,-1.5){Fig.~1.\ The set $L$.}
\uput[-135](0,0){$O$}
\uput[45](1.5,1.5){$B$}
\uput[-90](4,0){$\lambda_2^{(a)}$}
\uput[180](0,2){$\lambda_1^{(a)}$}
\uput[-90](2.2,0){$A$}
\psccurve[linewidth=0.0mm,fillstyle=vlines]%
(0,0)(.1,.1)(.3,.3)(.6,.6)(1.4,1.4)(1.5,1.5)(1.65,1.2)(1.8,.9)%
(1.95,.6)(2.1,.3)(2.25,0)(2.15,0)(2.0,0)(.4,0)(.3,0)(.2,0)(0,0)
\end{pspicture}
\end{center}
The corner points of the set $L\times L\times L$ are
\begin{eqnarray}
&& \left[O,O,O\right],\ 
\left[O,O,A\right],\ 
\left[O,O,B\right],\ 
\left[O,A,A\right],\
\left[O,A,B\right],\ 
\left[O,B,B\right],\nonumber \\ 
&& \left[A,A,A\right],\ 
\left[A,A,B\right],
\left[A,B,B\right],\
\left[B,B,B\right]
\label{corner1}
\end{eqnarray}
and the E-points obtained from these by qutrit permutations.
(For example, the E-points $[A,B,A]$ and $[B,A,A]$ are obtained
from $[A,A,B]$ by qutrit permutations.)

If $\lambda_3^{(a)}=1$ and $\lambda_2^{(a)} = \lambda_1^{(a)}=0$
for some $a$, the state reduces to a two-qutrit system.
It is well known that the two RDMs must have
the same eigenvalues for any two-particle state.  It can
readily be verified 
that this fact is compatible with Theorem~\ref{main}.
In fact inequalities (\ref{ineq1}) and (\ref{ineq2}) imply this fact
and the rest are satisfied if $\lambda_3^{(a)}=1$,
$\lambda_i^{(b)}=\lambda_i^{(c)}$, $i=1,2,3$ for some $(abc)$.
This implies that the E-points $[O,O,A]$, $[O,O,B]$ and
$[O,A,B]$ are not in the set $S$.
All the other E-points in the list (\ref{corner1}) satisfy the
inequalities of this theorem.  Therefore they will remain corner
points of $S$.

Let us give the corner points of $S$.  This will be useful when we
construct quantum 
states satisfying the inequalities of Theorem~\ref{main} in
Secs.~\ref{constr1} and \ref{constr2}.
\begin{theorem}
The convex set $S$ of 
the E-points characterized by Theorem~\ref{main} has the
following corner points and those obtained by qutrit permutations
from them:
\begin{eqnarray}
&& \left[O,O,O\right],\ 
\left[O,A,A\right],\ 
\left[O,B,B\right],\ 
\left[A,A,A\right],\ 
\left[A,A,B\right],\ 
\left[A,B,B\right],\
\left[B,B,B\right],\nonumber \\
&& \left[B,0\third\twthirds,0\third\twthirds\right],\  
\left[A,0\quarter
\thquarters,\quarter\quarter\half\right],\
\left[A,B, \sixth\sixth\twthirds\right],\ 
\left[A,\sixth\sixth\twthirds,\sixth\sixth\twthirds\right].
\label{corner2}
\end{eqnarray}
\label{cnr}
\end{theorem}

\noindent
Since the E-points in this list that are corner points of
$L \times L\times L$, i.e. the first seven E-points, 
remain corner points
of $S$, we only need to show that the corner points of $S$ that are
not corner points of $L\times L\times L$ are given by the last four
E-points in (\ref{corner2}) and those obtained by qutrit permutations
from them. 
Since each of these new corner points 
must satisfy one of {\em equalities} 
(\ref{ineq1})--(\ref{ineq7}), i.e. equalities obtained by replacing
inequality signs by equal signs, the following proposition,
which lists all the corner points on the hyperplane given by each
of equalities (\ref{ineq1})--(\ref{ineq7}), will imply Theorem
\ref{cnr}.
\begin{proposition}
The intersection of the set $S$ and the hyperplane bounding each of 
the inequalities listed in Theorem~\ref{main} is a
5-simplex.  These 5-simplices
are characterized by the corner points given as follows [for
$(abc)=(123)$]:
\begin{eqnarray}
{\rm (\ref{ineq1})} & : &
\left[O,O,O\right],\ 
\left[B,O,B\right],\ 
\left[B,B,O\right],\ 
\left[A,O,A\right],\ 
\left[A,A,O\right],\ 
\left[B,0\third\twthirds,0\third\twthirds\right]; \nonumber \\
{\rm (\ref{ineq2})} & : &
\left[O,O,O\right],\ 
\left[B,O,B\right],\ 
\left[A,O,A\right],\
\left[O,A,A\right],\ 
\left[\sixth\sixth\twthirds,\sixth\sixth\twthirds,A\right],\
\left[\quarter\quarter\half,0\quarter\thquarters,A\right]; \nonumber
\\
{\rm (\ref{ineq3})} & : &
\left[O,O,O\right]\,\ 
\left[B,O,B\right],\ 
\left[A,O,A\right],\ 
\left[0\third\twthirds,0\third\twthirds,B\right],\
\left[A,0\quarter\thquarters,\quarter\quarter\half\right],\
\left[A,\sixth\sixth\twthirds,B\right]; \nonumber \\
{\rm (\ref{ineq4})} & : &
\left[O,O,O\right],\ 
\left[B,O,B\right],\ 
\left[B,B,O\right],\
\left[A,O,A\right],\ 
\left[A,A,O\right],\ 
\left[A,\sixth\sixth\twthirds,\sixth\sixth\twthirds\right]; \nonumber
\\
{\rm (\ref{ineq5})} & : &
\left[O,O,O\right],\
\left[B,O,B\right],\ 
\left[B,B,O\right],\
\left[A,O,A\right],\ 
\left[B,0\third\twthirds,0\third\twthirds\right],\
\left[B,\sixth\sixth\twthirds,A\right]; \nonumber \\
{\rm (\ref{ineq6})} & : &
\left[O,O,O\right],\
\left[B,O,B\right],\ 
\left[A,O,A\right],\
\left[A,0\quarter\thquarters,\quarter\quarter\half\right],\ 
\left[A,\sixth\sixth\twthirds,\sixth\sixth\twthirds\right],\
\left[A,\sixth\sixth\twthirds,B\right]; \nonumber \\
{\rm (\ref{ineq7})} & : &
\left[A,A,B\right],\ 
\left[B,O,B\right],\ 
\left[A,O,A\right],\
\left[0\third\twthirds,0\third\twthirds,B\right],\ 
\left[A,0\quarter\thquarters,\quarter\quarter\half\right],\
\left[A,\sixth\sixth\twthirds,B\right]. \nonumber
\end{eqnarray}
\label{prop111}
\end{proposition}

\begin{proof}
Each set of E-points in this list clearly forms a 5-simplex
since there are six E-points each.  It is straightforward to check that
each simplex is nondegenerate, i.e. five-dimensional.
To verify that these simplices form
the boundaries, i.e. to show that there are no other E-points contained
in the set $S$ on each
hyperplane, it is sufficient to show that the six sides
of each 5-simplex are on boundary hyperplanes of $S$.  
This will be true
if any five E-points chosen from the set of corner points of each
5-simplex are on a single boundary hyperplane of $S$.  
This fact can readily 
be verified.  We list the hyperplanes bounding each 5-simplex together
with the E-point to be removed.  For example,
for equality (\ref{ineq1}), 
``$\left[O,O,O\right]\,:\, \lambda_2^{(1)}=\lambda_3^{(1)}$"
below means that the E-points 
$\left[B,O,B\right]$,
$\left[B,B,O\right]$,
$\left[A,O,A\right]$,
$\left[A,A,O\right]$ and
$\left[B,0\third\twthirds,0\third\twthirds\right]$
are on the hyperplane $\lambda_2^{(1)}=\lambda_3^{(1)}$.
\begin{eqnarray}
&& {\rm Equality\ (\ref{ineq1})},(abc)=(123) \nonumber\\
&&\left[O,O,O\right]\,:\,\lambda_2^{(1)}=\lambda_3^{(1)};\ \
\left[B,O,B\right]\,:\,\lambda_1^{(3)}=0;\nonumber \\
&& \left[B,B,O\right]\,:\, \lambda_1^{(2)}=0;\ \
\left[A,O,A\right]\,:\,{\rm (\ref{ineq5})},(abc)=(132);\nonumber \\
&& \left[A,A,O\right]\,:\, {\rm (\ref{ineq5})},(abc)=(123);\ \
\left[B,0\third\twthirds,0\third\twthirds\right]\,:\,
{\rm (\ref{ineq4})},(abc)=(123);
\nonumber
\end{eqnarray}
\begin{eqnarray}
&& {\rm Equality\ (\ref{ineq2})},(abc)=(123) \nonumber \\
&& \left[O,O,O\right]\,:\,\lambda_2^{(3)}=\lambda_3^{(3)};\ \ 
\left[B,O,B\right]\,:\,\lambda_1^{(3)}=0;\nonumber \\
&& \left[A,O,A\right]\,:\,\lambda_1^{(1)}=\lambda_2^{(1)};\ \
\left[O,A,A\right]\,:\,{\rm (\ref{ineq6})},(abc)=(321);\nonumber \\
&& \left[\sixth\sixth\twthirds,\sixth\sixth\twthirds,A\right]\,:\,
\lambda_1^{(2)}=0;\ \
\left[\quarter\quarter\half,0\quarter\thquarters,A\right]\,:\,
{\rm (\ref{ineq4})},(abc)=(321);
\nonumber
\end{eqnarray}
\begin{eqnarray}
&& {\rm Equality\ (\ref{ineq3})},(abc)=(123) \nonumber \\
&& \left[O,O,O\right]\,:\, {\rm (\ref{ineq7})},(abc)=(123);\ \
\left[B,O,B\right]\,:\, \lambda_1^{(1)}=0;\nonumber \\
&& \left[A,O,A\right]\,:\, \lambda_1^{(3)}=\lambda_2^{(3)};\ \
\left[0\third\twthirds,0\third\twthirds,B\right]\,:\,
{\rm (\ref{ineq6})},(abc)=(123);
\nonumber \\
&& \left[A,0\quarter\thquarters,\quarter\quarter\half\right]\,:\,
{\rm (\ref{ineq5})},(abc)=(321);\ \
\left[A,\sixth\sixth\twthirds,B\right]\,:\,
\lambda_1^{(2)}=0;
\nonumber
\end{eqnarray}
\begin{eqnarray}
&& {\rm Equality\ (\ref{ineq4})},(abc)=(123) \nonumber \\
&& \left[O,O,O\right]\,:\, 
\lambda_2^{(1)}=\lambda_3^{(1)};\ \
\left[B,O,B\right]\,:\, {\rm (\ref{ineq2})},(abc)=(231);\nonumber \\
&& \left[B,B,O\right]\,:\, {\rm (\ref{ineq2})},(abc)=(321);\ \
\left[A,O,A\right]\,:\, \lambda_1^{(3)}=\lambda_2^{(3)};\nonumber \\
&& \left[A,A,O\right]\,:\, \lambda_1^{(2)}=\lambda_2^{(2)};\ \
\left[A,\sixth\sixth\twthirds,\sixth\sixth\twthirds\right]\,:\,
{\rm (\ref{ineq1})},(abc)=(123);
\nonumber
\end{eqnarray}
\begin{eqnarray}
&& {\rm Equality\ (\ref{ineq5})},(abc)=(123) \nonumber \\
&& \left[O,O,O\right]\,:\,
\lambda_2^{(1)}=\lambda_3^{(1)};\ \
\left[B,O,B\right]\,:\, \lambda_1^{(3)}=0;\nonumber \\
&& \left[B,B,O\right]\,:\, {\rm (\ref{ineq3})},(abc)=(321);\ \
\left[A,O,A\right]\,:\,\lambda_1^{(1)}=\lambda_2^{(1)};\nonumber \\
&& \left[B,0\third\twthirds,0\third\twthirds\right]\,:\,
\lambda_1^{(2)}=\lambda_2^{(2)};\ \
\left[B,\sixth\sixth\twthirds,A\right]\,:\,
{\rm (\ref{ineq1})},(abc)=(123);
\nonumber 
\end{eqnarray}
\begin{eqnarray}
&& {\rm Equality\ (\ref{ineq6})},(abc)=(123) \nonumber \\
&& \left[O,O,O\right]\,:\, \lambda_2^{(1)}=\lambda_3^{(1)};\ \
\left[B,O,B\right]\,:\, \lambda_1^{(1)}=0;\nonumber \\
&& \left[A,O,A\right]\,:\, \lambda_1^{(3)}=\lambda_2^{(3)};\ \
\left[A,0\quarter\thquarters,\quarter\quarter\half\right]\,:\,
\lambda_1^{(2)}=\lambda_2^{(2)}; \nonumber \\
&& \left[A,\sixth\sixth\twthirds,\sixth\sixth\twthirds\right]\,:\,
{\rm (\ref{ineq3})},(abc)=(123);\ \
\left[A,\sixth\sixth\twthirds,B\right]\,:\,
{\rm (\ref{ineq2})},(abc)=(321);
\nonumber
\end{eqnarray}
\begin{eqnarray}
&& {\rm Equality\ (\ref{ineq7})},(abc)=(123) \nonumber \\
&& \left[A,A,B\right]\,:\, {\rm (\ref{ineq3})},(abc)=(123);\ \ 
\left[B,O,B\right]\,:\,\lambda_1^{(1)}=0;\nonumber \\
&& \left[A,O,A\right]\,:\,
\lambda_1^{(3)}=\lambda_2^{(3)};\ \
\left[0\third\twthirds,0\third\twthirds,B\right]\,:\,
\lambda_2^{(1)}=\lambda_3^{(1)};
\nonumber \\
&& \left[A,0\quarter\thquarters,\quarter\quarter\half\right]\,:\,
\lambda_2^{(3)}=\lambda_3^{(3)};\ \ 
\left[A,\sixth\sixth\twthirds,B\right]\,:\,
\lambda_1^{(2)}=0.
\nonumber
\end{eqnarray}
\end{proof}

\section{Proof of a generalization of inequality (\ref{ineq1})}
\label{ineqone}

Now we start proving the inequalities in Theorem~\ref{main}, thus
establishing that they are necessary conditions for the
$\lambda_i^{(a)}$ to be the eigenvalues of the RDMs
of a three-qutrit quantum state.
Inequality (\ref{ineq1}) is a corollary of a more 
general result stated in Ref.~\cite{hss}. We present a proof of
the latter.
\begin{theorem}
Consider an $n$-qudit state given by
$$
|\Psi\rangle = \sum_{1\leq i_1,\ldots,i_n \leq d}c_{i_1 i_2\cdots i_n}
|i_1\>\otimes |i_2\>\otimes
\cdots \otimes |i_n\>\,,
$$
where the set $\{ |1\>,|2\>,\ldots,|d\> \}$ is an orthonormal
basis for a $d$-dimensional vector space.
Let $\lambda_i^{(a)}$, $i=1,2,\ldots,d$,
be the eigenvalues of the $a$th RDM
with $\lambda_i^{(a)} \leq \lambda_j^{(a)}$ if $i < j$.  Then
\begin{equation}
\sum_{a=1}^{n-1}\sum_{i=1}^{d-1} \lambda_i^{(a)}
\geq \lambda_1^{(n)}+\lambda_2^{(n)}+\cdots + \lambda_{d-1}^{(n)}\,.
\end{equation}
\label{general}
\end{theorem}

\begin{proof}
The basis states can be chosen in such a way
that the RDMs are all 
diagonal~\cite{tony,brun1,brun2}.  We work in this basis.
Define
$$
G \equiv \sum_{a=1}^{n-1}\sum_{i=1}^{d-1}
\lambda_i^{(a)}\,.
$$
Note that
$$
\lambda_i^{(a)} = \sum_{i_1,\ldots,i_{a-1},i_{a+1},\ldots i_n}
|c_{i_1 \cdots i_{a-1}i\, i_{a+1}\cdots i_n}|^2\,.
$$
Let us write 
$$
c_{i_1 i_2\cdots i_{n-1} i} = 
\sqrt{\lambda_i^{(n)}}a_{i_1 i_2\cdots i_{n-1}i}\,.
$$
If $\lambda_i^{(n)} > 0$ for all $i$, we have
\begin{eqnarray}
&& \sum_{i_1 i_2\cdots i_{n-1}}|a_{i_1 i_2\cdots i_{n-1}i}|^2 = 1\,,
\label{normal}\\
&& \sum_{i_1 i_2\cdots i_{n-1}}a_{i_1 i_2 \cdots i_{n-1}i}\,
\overline{a_{i_1 i_2 \cdots i_{n-1} i'}} = 0\ \ {\rm if}\ i\neq i'\,.
\label{ortho}
\end{eqnarray}
The latter follows from the fact that the $n$th RDM is diagonal.
If $\lambda_j^{(n)} = 0$ for some $j$, then the coefficients
$a_{i_1 i_2\cdots i_n{-1}j}$ are undefined.  However, we can choose
them so that Eqs.~(\ref{normal}) and (\ref{ortho}) are satisfied for
all $i$ and $i'$ even in this case.  For $a\neq n$ we have
$$
\lambda_i^{(a)} = \sum_{i_n=1}^{d}\lambda_{i_n}^{(n)}
\sum_{i_1, \cdots, i_{a-1}, i_{a+1}, \cdots,i_{n-1}}
|a_{i_1 \cdots i_{a-1} i\,i_{a+1}\cdots i_{n}}|^2\,.
$$
Then we find
$$
G = \sum_{a=1}^{n-1}\sum_{i=1}^{d-1}
\lambda_i^{(a)} =
\sum_{i_1,i_2\cdots i_{n-1},i_n}
g(i_1,i_2,\ldots,i_n)\lambda_{i_n}^{(n)}
|a_{i_1 i_2\cdots i_n}|^2\,, 
$$
where $g(i_1,i_2,\ldots,i_n) \geq 1$ unless $i_1=i_2=\cdots=i_{n-1}=d$,
and where $g(d,d,\ldots,d,i_n)=0$.  Thus
\begin{eqnarray}
G & \geq & \sum_{i_n=1}^{d}
\lambda_{i_n}^{(n)} \left\{ \left(\sum_{i_1,\cdots i_{n-1}}
|a_{i_1 i_2\cdots i_n}|^2\right) - |a_{dd\cdots d\,i_n}|^2\right\} 
\nonumber \\
& = & \sum_{i_n=1}^{d} \lambda_{i_n}^{(n)}
(1-|a_{dd\cdots d\,i_n}|^2)\nonumber \\
& = & 1 - \sum_{i_n=1}^d \lambda_{i_n}^{(n)}|a_{dd\cdots d\,i_n}|^2\,,
\label{FFF}
\end{eqnarray}
where we have used (\ref{normal}).  Eqs.~(\ref{normal})
and (\ref{ortho}) allow us to consider $a_{i_1 i_2\cdots i_{n-1} i}$
as components of orthonormal vectors labelled by $i$
in $d^{n-1}$ dimensions.  This implies, in particular, that
$(a_{dd\cdots d1},a_{dd\cdots d2},\ldots,a_{dd\cdots dd})$ is part
of a $d^{n-1}\times d^{n-1}$ unitary matrix.
Hence we have
\begin{equation}
\sum_{i_n=1}^d |a_{dd\cdots d\,i_n}|^2 \leq 1\,. \label{straint}
\end{equation}
The quantity $\sum_{i_n=1}^d\lambda_{i_n}^{(n)}|a_{dd\cdots d\,i_n}|^2$
takes the maximum value under the constraint (\ref{straint}) if
$|a_{dd\cdots d\,d}|=1$ and $a_{dd\cdots d\,i_n}=0$ for $i_n \neq d$
since $\lambda_d^{(n)}$ is the largest among $\lambda_{i_n}^{(n)}$ by
definition. Then from (\ref{FFF}) we find
$$
G \geq 1-\lambda_{d}^{(n)} = \lambda_1^{(n)}+\lambda_2^{(n)}
+ \cdots \lambda_{d-1}^{(n)}
$$
as required.
\end{proof}

\noindent
Inequality (\ref{ineq1}) follows from this theorem by letting
$d=3$ and $n=3$.

\section{Proof of inequalities (\ref{ineq2}) and (\ref{ineq3})}
\label{ineqtwothree}

We work in the basis for which the RDMs
are all diagonal unless otherwise stated.
We let $(abc)=(231)$ in inequality (\ref{ineq3}) and
$(abc)=(132)$ in inequality 
(\ref{ineq2}) to find
\begin{eqnarray}
\lambda_2^{(2)}+\lambda_3^{(2)} & \leq &
\lambda_1^{(3)}+\lambda_2^{(3)} + \lambda_2^{(1)}+\lambda_3^{(1)}\,, 
\nonumber \\
\lambda_3^{(1)}+\lambda_1^{(1)} & \leq &
\lambda_1^{(3)}+\lambda_2^{(3)}+\lambda_3^{(2)} + \lambda_1^{(2)}\,.
\nonumber 
\end{eqnarray}
These are equivalent to the following proposition.
\begin{proposition}
$\lambda_3^{(3)}-\lambda_1^{(2)} \leq \lambda_2^{(1)}+\lambda_3^{(1)}$
and $\lambda_3^{(3)}-\lambda_1^{(2)} \leq
\lambda_2^{(1)}+\lambda_3^{(2)}$.
\end{proposition}

\begin{proof}
First we note that
$$
\lambda_3^{(3)}-\lambda_1^{(2)}
\leq |c_{123}|^2+|c_{133}|^2+|c_{223}|^2+|c_{233}|^2+
|c_{323}|^2+|c_{333}|^2\,.
$$
Define
$$
x \equiv  |c_{123}|^2+|c_{133}|^2+|c_{223}|^2 
+ |c_{233}|^2+|c_{323}|^2+|c_{333}|^2\,. 
$$
Then it is sufficient to show that
\begin{eqnarray}
x & \leq & \lambda_2^{(1)}+\lambda_3^{(1)}\,, \label{first}\\
x & \leq & \lambda_2^{(1)}+\lambda_3^{(2)}\,. \label{second}
\end{eqnarray}
To prove (\ref{first}) we define $b_{ijk}$ by
$c_{ijk} \equiv \sqrt{\lambda_i^{(1)}}\,b_{ijk}$
if $\lambda_i^{(1)}\neq 0$ for all $i$.
Then we have
$$
\sum_{j=1}^3 \sum_{k=1}^3b_{ijk}\overline{b_{i'jk}} = \delta_{ii'}\,.
$$
If $\lambda_i^{(1)}=0$ for some $i$, 
then $b_{ijk}$ are arbitrary, but we can still 
choose them to satisfy these equations. Hence the following matrix
is part of a $9\times 9$ unitary matrix:
$$
M_1 \equiv \left(\begin{array}{ccc}
b_{133} & b_{233} & b_{333} \\
b_{123} & b_{223} & b_{323} \end{array}\right)\,.
$$
Thus, we have the following inequalities:
\begin{eqnarray}
|b_{133}|^2 + |b_{233}|^2 + |b_{333}|^2 & \leq & 1\,,\label{A}\\
|b_{123}|^2 + |b_{223}|^2 + |b_{323}|^2 & \leq & 1\label{B}
\end{eqnarray}
and
\begin{equation}
|b_{i33}|^2 + |b_{i23}|^2  \leq  1\ \ {\rm for\ all}\ i\,.\label{A1}
\end{equation}
Note that
\begin{equation}
x =  \lambda_1^{(1)}(|b_{133}|^2 + |b_{123}|^2)
+ \lambda_2^{(1)}(|b_{233}|^2 + |b_{223}|^2)
+ \lambda_3^{(1)}(|b_{333}|^2 + |b_{323}|^2)\,. \label{X1}
\end{equation}
{}From (\ref{A}) and (\ref{B}) we find
$$
|b_{133}|^2+|b_{123}|^2 \leq 2 - 
(|b_{233}|^2+|b_{223}|^2 + |b_{333}|^2 + |b_{323}|^2)\,.
$$
By using this in (\ref{X1}) we obtain
\begin{eqnarray}
x & \leq & 2\lambda_1^{(1)}
+ (\lambda_2^{(1)}-\lambda_1^{(1)})
(|b_{233}|^2 + |b_{223}|^2)+ (\lambda_3^{(1)}-\lambda_1^{(1)})
(|b_{333}|^2 + |b_{323}|^2) \nonumber \\
& \leq & \lambda_2^{(1)}+\lambda_3^{(1)}\,,
\end{eqnarray}
where we have used (\ref{A1}) with $i=2$ and $3$.  
Thus we have established inequality (\ref{first}).

To establish (\ref{second}) we first make a unitary transformation
on the first and second rows of $M_1$ and obtain
$$
\tilde{M}_1 = \left( \begin{array}{ccc}
\overline{\alpha}b_{133} + \overline{\beta}b_{123} &
\overline{\alpha}b_{233} + \overline{\beta}b_{223} &
\overline{\alpha}b_{333} + \overline{\beta}b_{323} \\
-\beta b_{133}+\alpha b_{123} & 
-\beta b_{233} + \alpha b_{223} &
-\beta b_{333} + \alpha b_{323} \end{array} \right)\,,
$$
where $|\alpha|^2 + |\beta|^2 = 1$.  We choose $\alpha$ and $\beta$
so that $-\beta b_{333}+\alpha b_{323}=0$.  If 
$b_{333}= c_{333}/\sqrt{\lambda_3^{(1)}} \neq 0$ or
$b_{323} = c_{323}/\sqrt{\lambda_3^{(1)}} \neq 0$, 
then $\alpha$ and $\beta$ can be chosen as follows:
\begin{eqnarray}
\alpha & = & 
\frac{c_{333}}{\sqrt{|c_{333}|^2 + |c_{323}|^2}}\,,\label{al}\\
\beta & = & 
\frac{c_{323}}{\sqrt{|c_{333}|^2 + |c_{323}|^2}}\,. \label{be}
\end{eqnarray}
If $c_{333}=c_{323}=0$, then $(\alpha,\beta)$ can be chosen, for
example, to be $(1,0)$.  Then we have from the second row
of $\tilde{M}_1$
$$
|-\beta b_{133}+\alpha b_{123}|^2 + |-\beta b_{233} + \alpha b_{223}|^2
\leq 1\,.
$$
By multiplying this by $\lambda_2^{(1)}$ and remembering that
$\lambda_1^{(1)} \leq \lambda_2^{(1)}$ we obtain
\begin{equation}
|-\beta c_{133} + \alpha c_{123}|^2 + |-\beta c_{233} + \alpha
c_{223}|^2
\leq \lambda_2^{(1)}\,.  \label{a1}
\end{equation}
Next we define $d_{ijk}$ by 
$c_{ijk} \equiv \sqrt{\lambda_j^{(2)}}\,d_{ijk}$.
Then the following matrix is part of a unitary matrix
if $\lambda_2^{(2)}\neq 0$:
$$
M_2 \equiv \left(\begin{array}{ccc}
d_{133} & d_{233} & d_{333} \\
d_{123} & d_{223} & d_{323} \end{array}\right)\,.
$$
If $\lambda_2^{(2)}=0$, we can still choose $d_{ijk}$ 
so that $M_2$ is part of a unitary matrix.  Again we perform a unitary
transformation on the first and second rows as follows:
$$
\tilde{M}_2 = \left( \begin{array}{ccc}
\overline{\alpha'}d_{133} + \overline{\beta'}d_{123} &
\overline{\alpha'}d_{233} + \overline{\beta'}d_{223} &
\overline{\alpha'}d_{333} + \overline{\beta'}d_{323} \\
-\beta' d_{133}+\alpha' d_{123} & 
-\beta' d_{233} + \alpha' d_{223} &
-\beta' d_{333} + \alpha' d_{323} \end{array} \right)\,,
$$
where
\begin{eqnarray}
\alpha' & = & \frac{\sqrt{\lambda_3^{(2)}}\,\alpha}
{\sqrt{\lambda_3^{(2)}|\alpha|^2 + \lambda_2^{(3)}|\beta|^2}}\,,
\label{alphap}\\
\beta' & = & \frac{\sqrt{\lambda_2^{(2)}}\,\beta}
{\sqrt{\lambda_3^{(2)}|\alpha|^2 + \lambda_2^{(3)}|\beta|^2}}\,. 
\label{betap}
\end{eqnarray}
{}From the first row of $\tilde{M}_2$ we obtain
$$
|\overline{\alpha'}d_{133} + \overline{\beta'}d_{123}|^2 +
|\overline{\alpha'}d_{233} + \overline{\beta'}d_{223}|^2 +
|\overline{\alpha'}d_{333} + \overline{\beta'}d_{323}|^2 \leq 1\,.
$$
By substituting (\ref{alphap}) and (\ref{betap}), and by
recalling the definition of $d_{ijk}$, we find
$$
|\overline{\alpha}c_{133}+\overline{\beta}c_{123}|^2
+|\overline{\alpha}c_{233}+\overline{\beta}c_{223}|^2
+ |\overline{\alpha}c_{333}+\overline{\beta}c_{323}|^2
\leq \lambda_3^{(2)}|\alpha|^2 + \lambda_2^{(2)}|\beta|^2\,.
$$
Adding this and (\ref{a1}) and using 
(\ref{al}) and (\ref{be}) we find
$$
|c_{133}|^2 + |c_{123}|^2 + |c_{233}|^2 + |c_{223}|^2
+ |c_{333}|^2 + |c_{323}|^2 \leq \lambda_2^{(1)}
+ \lambda_3^{(2)}|\alpha|^2 + \lambda_2^{(2)}|\beta|^2\,.
$$
Since the left-hand side is equal to $x$,
we obtain $x\leq \lambda_2^{(1)}+\lambda_3^{(2)}$ by
using $|\alpha|^2 + |\beta|^2 = 1$.
This proves inequality (\ref{second}).
\end{proof}

\noindent 
The following result will be useful later.
\begin{corollary}
$\lambda_2^{(b)} + \lambda_3^{(c)} \geq \lambda_2^{(a)}$ for
all permutations $(abc)$ of $(123)$.
\label{coro}
\end{corollary}

\begin{proof}
Inequality (\ref{ineq3}) implies
\begin{eqnarray}
2(\lambda_2^{(b)} + \lambda_3^{(c)}) & \geq &
\lambda_2^{(b)} + \lambda_1^{(b)} + \lambda_3^{(c)} + \lambda_2^{(c)}
\nonumber \\
& \geq & \lambda_3^{(a)} + \lambda_2^{(a)} \nonumber \\
& \geq & 2\lambda_2^{(a)}\,. \nonumber
\end{eqnarray}
{}From this the corollary immediately follows.
\end{proof}

\section{The method for proving inequalities 
(\ref{ineq4})--(\ref{ineq7})}
\label{perturb}

All our inequalities have the form
\begin{equation}
F \equiv \sum_{i=1}^3 \left(p_i \lambda_i^{(1)}
+ q_i\lambda_i^{(2)} + r_i\lambda_i^{(3)}\right)\,, \label{func}
\end{equation}
where $p_i$, $q_i$ and $r_i$ are constants.
The function $F$ of the 27 complex variables $c_{ijk}$ with constraint
$\sum_{i,j,k}|c_{ijk}|^2 = 1$ is a continuous map from a
$53$-dimensional unit sphere, which is compact, to $\R$.
Therefore there must be a minimum value.  For all remaining
inequalities we will show that the minimum of the corresponding
function $F$ cannot be negative.

Consider a three-qutrit state $|\Psi\>$ given as in 
Sec.~\ref{S:Intro}
with three diagonal RDMs.  Then the coefficients $c_{ijk}$ 
satisfy the 
orthogonality relations, $c_{iJK}\overline{c_{i'JK}}=0$,
$c_{IjK}\overline{c_{Ij'K}} = 0$ and
$c_{IJk}\overline{c_{IJk'}} = 0$, where the repeated indices are
summed over.   The state gives the minimum value of
$F$ only if the variation of $F$ is nonnegative 
under any variation $\delta c_{ijk}$ (not necessarily maintaining
the orthogonality relations) of the coefficients $c_{ijk}$.
(``A variation" means ``a first-order variation" throughout
this paper.)

To examine the variation of the eigenvalues 
$\lambda_i^{(a)}$ under a variation of $c_{ijk}$
we use first-order perturbation theory in quantum mechanics. 
An RDM for the $a$th qutrit,
$\rho^{(a)}_{ii'}$, of a given quantum state corresponds to
the ``unperturbed Hamiltonian", and
the variation of an RDM
under a variation of $c_{ijk}$ to the ``perturbation".  
The eigenvalues $\lambda_i^{(a)}$ correspond to the
``energy eigenvalues".  If there are no degenerate
eigenvalues, then the
variation of $\lambda_i^{(a)}$ can be read off from
the diagonal elements of $\delta \rho^{(a)}_{ii'}$.
We obtain the following result by this argument.
\begin{proposition}
Suppose that $\lambda_1^{a)} < \lambda_2^{(a)} < \lambda_3^{(a)}$.
Then the variation of the eigenvalues, 
$\delta\lambda_i^{(a)}$, under a variation $\delta c_{ijk}$ of the
coefficients $c_{ijk}$ satisfying
$\sum_{I,J,K}{\rm Re}\,(\delta c_{IJK}\overline{c_{IJK}})=0$ 
are given by
\begin{eqnarray}
\delta\lambda_i^{(1)} 
& = & 2\sum_{J,K}{\rm Re}\,(\delta c_{iJK}\overline{c_{iJK}})\,,
\nonumber \\
\delta\lambda_j^{(2)} 
& = & 2\sum_{I,K}{\rm Re}\,(\delta c_{IjK}\overline{c_{IjK}})\,,
\nonumber \\
\delta\lambda_k^{(3)} 
& = & 2\sum_{I,J}{\rm Re}\,(\delta c_{IJk}\overline{c_{IJk}})\,.
\nonumber
\end{eqnarray}
\label{prop1}
\end{proposition}

\begin{proof}
It is enough to note that the right-hand sides give the
variation of
the diagonal elements of the RDMs.
\end{proof}

\noindent
For the function $F$ given by (\ref{func}) we define
the level $L(i,j,k)$ of the triple $[ijk]$ by
\begin{equation}
L(i,j,k) \equiv p_i + q_j + r_k\,. \label{level}
\end{equation}
\begin{lemma}
Suppose that the function $F$ of $c_{IJK}$ defined by (\ref{func})
has its minimum value at a state with nondegenerate eigenvalues 
$\lambda_I^{(a)}$ for all $a$.  If $c_{ijk}$ and
$c_{i'j'k'}$ are nonzero for this state, then
$L(i,j,k) = L(i',j',k')$.
\label{lemma1}
\end{lemma}

\begin{proof}
Assume first that $i\neq i'$, $j \neq j'$ and $k \neq k'$.
Consider the variation of coefficients $c_{IJK}$ given by
$$
\delta c_{i j k} =  \alpha c_{i'j' k'}\,,\ \ \
\delta c_{i' j' k'} =  -\overline{\alpha} c_{i j k}
$$
and $\delta c_{IJK}=0$ if 
$\left[IJK\right]\neq \left[ijk\right],\ \left[ i'j'k'\right]$.
Then by Proposition~\ref{prop1} we have the following
variation of $\lambda_I^{(a)}$:
\begin{eqnarray}
\delta\lambda_{i}^{(1)} = \delta\lambda_{j}^{(2)}
= \delta\lambda_{k}^{(3)} & = & \Delta\,, \nonumber \\
\delta\lambda_{i'}^{(1)} = \delta\lambda_{j'}^{(2)}
= \delta\lambda_{k'}^{(3)} & = & - \Delta\,, \nonumber
\end{eqnarray}
where $\Delta = 2{\rm Re}\,
(\alpha c_{i'j' k'}\overline{c_{i j k}})$, with all other
$\delta \lambda_I^{(a)}$ vanishing.
Hence the variation of $F$ is
\begin{eqnarray}
\delta F & = & (p_{i} + q_{j}+r_{k} - p_{i'} - q_{j'} 
- r_{k'})\Delta\nonumber \\
& = & \left[ L(i,j,k)-L(i',j',k')\right]\Delta\,. \nonumber
\end{eqnarray}
It can readily be verified that this formula is valid even if
$i=i'$, $j=j'$ or $k=k'$.  Since $c_{i j k}\neq 0$
and $c_{i' j' k'}\neq 0$, we can make $\Delta$ positive or negative
by adjusting the phase of $\alpha$, thus making $\delta F$ negative
if $L(i,j,k)\neq L(i',j',k')$.  Therefore, if the
function $F$ takes
its minimum value 
at a state with nondegenerate eigenvalues,
and if $c_{ij k}$ and $c_{i'j'k'}$ are both
nonzero, then we must have
$L(i,j,k)=L(i',j',k')$.
\end{proof}

Let us define
\begin{eqnarray}
P_4 & = & 2\lambda_2^{(2)} + \lambda_1^{(2)}
+ 2\lambda_2^{(3)} + \lambda_1^{(3)}
- 2\lambda_2^{(1)} -\lambda_1^{(1)}\,, \label{P4}\\
P_5 & = & 2\lambda_2^{(2)} + \lambda_1^{(2)}
+ 2\lambda_1^{(3)} + \lambda_2^{(3)}
- 2\lambda_1^{(1)} -\lambda_2^{(1)}\,,\label{P5}\\
P_6 & = & 2\lambda_2^{(2)} + \lambda_1^{(2)}
+ 2\lambda_2^{(3)} + \lambda_3^{(3)}
- 2\lambda_2^{(1)} -\lambda_3^{(1)}\,,\label{P6}\\
P_7 & = & 2\lambda_1^{(2)} + \lambda_2^{(2)}
+ 2\lambda_3^{(3)} + \lambda_2^{(3)}
- 2\lambda_2^{(1)} -\lambda_3^{(1)}\,. \label{P7}
\end{eqnarray}
Our task is to show that the function 
$P_m$ cannot have a negative minimum
for any $m$.
It is useful to note that if the quantum state 
reduces to a two-qutrit system, then $P_m \geq 0$ as we have
mentioned before.  We state this fact as a lemma.
\begin{lemma}
If $\lambda_3^{(a)} =1$ and $\lambda_2^{(a)}=\lambda_1^{(a)}=0$ 
for some $a$, then $P_m \geq 0$ for all $m$.
\label{lemma2}
\end{lemma}

\noindent
Let $L(i,j,k)$ be the level of $[ijk]$ defined by (\ref{level})
for the function $P_m$.
It can readily be seen that
\begin{equation}
P_m = \sum_{i,j,k}L(i,j,k)|c_{ijk}|^2\,. \label{Pm}
\end{equation}
If $P_m$ takes its minimum value at a state with nondegenerate
eigenvalues, the level $L(i,j,k)$ must be the same for all $[ijk]$
with $c_{ijk}\neq 0$ by Lemma~\ref{lemma1}.  
This fact can be used to establish that
$P_m$ cannot have a negative minimum at a state with nondegenerate
eigenvalues.
\begin{proposition}
If $P_m$ defined by (\ref{P4})--(\ref{P7}) for some $m$ has a
negative minimum, then the state where this minimum occurs must
have at least one pair of degenerate eigenvalues, i.e.
$\lambda_i^{(a)}=\lambda_{i+1}^{(a)}$ for some $a$ and $i$.
\label{propprop}
\end{proposition}

\begin{proof}
All $P_m$ have the following general form:
\begin{equation}
P_m = 
2\lambda_{j_3}^{(2)} + \lambda_{j_2}^{(2)}
+ 2\lambda_{k_3}^{(3)}+\lambda_{k_2}^{(3)}
- 2\lambda_{i_1}^{(1)} -\lambda_{i_2}^{(1)}\,.  \label{Pfun}
\end{equation}
Let $i_3$ be the number in the set $\{ 1,2,3\}$
which differs from $i_1$ and $i_2$, and similarly for $j_1$ and
$k_1$.  There are only four triples $[ijk]$ with negative levels
$L(i,j,k)$, which are
\begin{eqnarray}
&& \left[i_1j_1 k_1\right]\ {\rm with}\ L = -2\,,\nonumber \\
&& \left[i_2j_1 k_1\right]\,,\ 
\left[i_1 j_2 k_1\right]\,,\ \left[i_1 j_1 k_2\right]\ {\rm with}\ 
L = -1\,. \nonumber
\end{eqnarray}
Suppose that all eigenvalues are nondegenerate.  Then
for the state where the negative minimum occurs, 
the level $L$ of the triples $[ijk]$
such that $c_{ijk}\neq 0$ must all be $-2$ or $-1$ by
Lemma~\ref{lemma1}. We must have
$\lambda_3^{(a)} > \lambda_2^{(a)} > 0$ for all $a$. This cannot
be the case if there is only one 
nonzero coefficient $c_{ijk}$.  Hence we
cannot have $L=-2$.
Suppose $L=-1$. Then, the coefficients $c_{ijk}$ other than
$c_{i_2j_1k_1}$, $c_{i_1j_2k_1}$ and
$c_{i_1j_1k_2}$ would be zero.  Since $\lambda_3^{(a)}$ and
$\lambda_2^{(a)}$ must be nonzero, we must have
$i_3=j_3=k_3=1$.  This is not satisfied by any $P_m$.  Hence
if $P_m$ has a negative minimum, then the state where the minimum
occurs must
have at least one pair of degenerate eigenvalues.
\end{proof}

Next we establish some general results for degenerate cases.
Note that if there are degenerate eigenvalues, 
we need to make sure that
the variation of the
off-diagonal elements in the degenerate sector vanishes in order to
use the perturbation theory argument by taking the variation of
only the
diagonal elements of the RDMs into account.  

\

\noindent
{\bf Remark}. We say
that the variation of the RDMs 
is {\em effectively diagonal} 
if it has no off-diagonal elements in the degenerate sector.
\begin{lemma}
Suppose that 
$\lambda_{i}^{(1)} = \lambda_{i+1}^{(1)} \neq \lambda_{i'}^{(1)}$ 
with no other degeneracy at a state where the function $F$
defined by (\ref{func}) has its minimum.
Suppose further that 
$(c_{ijk},c_{i+1,jk})$ and $c_{i' j'k'}$ are nonzero, i.e.
$c_{ijk}\neq 0$ or $c_{i+1,jk}\neq 0$, and $c_{i'j'k'}\neq 0$,
for this state.  Then
$$
L(i,j,k) \leq L(i',j',k') \leq L(i+1,j,k)\,. 
$$
\label{lemma3}
\end{lemma}

\begin{proof}
Since $\lambda_{i}^{(1)} = \lambda_{i+1}^{(1)}$, a unitary
transformation in the degenerate sector, i.e. 
between $|i\>$ and $|i+1\>$ in the first qutrit,
leaves the RDMs diagonal.  Hence we can let
$c_{i+1,jk} = 0$ and $c_{ijk} \neq 0$ using such a unitary 
transformation.  Then consider the variation given in
the proof of Lemma~\ref{lemma1}.  Then 
$$
\delta \rho_{i,i+1}^{(1)} = \delta c_{ijk}\cdot \overline{c_{i+1,jk}}
+ c_{ijk}\overline{\delta c_{i+1,jk}} = 0\,,
$$
i.e. the variation of the RDM is effectively diagonal. Define
$\Delta \equiv 2{\rm Re}\,\left(
\alpha c_{i'j'k'}\overline{c_{ijk}}\right)$.
We have $\delta\lambda_j^{(2)} = \delta\lambda_k^{(3)} = \Delta$ and
$\delta\lambda_{i'}^{(1)} = 
\delta\lambda_{j'}^{(2)} = \delta\lambda_{k'}^{(3)} = - \Delta$ as before.
However, since $\lambda_i^{(1)}= \lambda_{i+1}^{(1)}$, we must have
$\delta\lambda_{i+1}^{(1)} \geq \delta\lambda_i^{(1)}$ by definition.
Hence we have
$\delta\lambda_{i+1}^{(1)} = \Delta$, $\delta\lambda_i^{(1)} = 0$ if
$\Delta > 0$ and
$\delta\lambda_{i+1}^{(1)} = 0$, $\delta\lambda_i^{(1)} = \Delta$ if
$\Delta < 0$.
Hence
\begin{eqnarray}
\delta F & = & (p_{i+1} + q_j + r_k - p_{i'} - q_{j'} - r_{k'})\Delta\ \
{\rm if}\ \ \Delta > 0\,, \nonumber \\
\delta F & = & (p_{i} + q_j + r_k - p_{i'} - q_{j'} - r_{k'})\Delta\ \
{\rm if}\ \ \Delta < 0\,. \nonumber
\end{eqnarray}
By requiring that 
$\delta F \geq 0$ for both cases we find the desired inequalities.
\end{proof}

\noindent
{\bf Remark}.  Analogous statements for the cases with 
$\lambda_j^{(2)} = \lambda_{j+1}^{(2)} \neq \lambda_{j'}^{(2)}$ and 
with
$\lambda_k^{(3)} = \lambda_{k+1}^{(3)} \neq \lambda_{k'}^{(3)}$ 
can readily be
established.  

\

\noindent
{\bf Remark}.
We often use a unitary transformation in the degenerate sector, {\em 
which keeps the RDM unchanged}.
A unitary transformation on a qutrit is assumed to be in the
degenerate sector unless otherwise stated in this and the next two
sections.

\begin{lemma}
Suppose that 
$\lambda_{i}^{(1)} = \lambda_{i+1}^{(1)} \neq \lambda_{i'}^{(1)}$ and
$\lambda_{j}^{(2)} = \lambda_{j+1}^{(2)} \neq \lambda_{j'}^{(2)}$
with no other degeneracy at a state where the function $F$
has its minimum.
Suppose further that 
$(c_{ijk},c_{i+1,jk},c_{i,j+1,k},c_{i+1,j+1,k})$ and $c_{i' j'k'}$ 
are nonzero for this state. 
Then
$$
L(i,j,k) \leq L(i',j',k') \leq L(i+1,j+1,k)\,.
$$
\label{lemma4}
\end{lemma}

\begin{proof}
Since $\lambda_i^{(1)} = \lambda_{i+1}^{(1)}$ and
$\lambda_j^{(2)} = \lambda_{j+1}^{(2)}$, 
we can use a unitary transformation
to have $c_{i+1,jk} = c_{i,j+1,k} = 0$.  Then consider the
variation given in the proof of Lemma~\ref{lemma1}
and define $\Delta$ as in that lemma.
Then, as in the previous lemma, the variation of 
the RDMs is effectively diagonal, and we have
$\delta \lambda_{i}^{(1)} = \delta\lambda_j^{(2)} = 0$, 
$\delta \lambda_{i+1}^{(1)} = \delta\lambda_{j+1}^{(2)} = \Delta$ if
$\Delta > 0$ and 
$\delta \lambda_{i}^{(1)}=\delta\lambda_j^{(2)} = \Delta$, 
$\delta \lambda_{i+1}^{(1)} = \delta\lambda_{j+1}^{(2)} =0$ 
if $\Delta < 0$.  Thus,
\begin{eqnarray}
\delta F & = & (p_{i+1}+q_{j+1} + r_k - p_{i'} - q_{j'} - r_{k'})
\Delta\ \
{\rm if}\ \ \Delta > 0\,, \nonumber \\
\delta F & = & (p_{i} + q_j + r_k - p_{i'} - q_{j'} - r_{k'})\Delta\ \
{\rm if}\ \ \Delta < 0\,. \nonumber
\end{eqnarray}
By requiring that $\delta F \geq 0$ 
for both cases we find the desired inequalities.
\end{proof}

\begin{lemma}
For the function $F$ defined by (\ref{func}) suppose that
$\lambda_i^{(1)} = \lambda_{i+1}^{(1)} \neq \lambda_{i'}^{(1)}$ and
$\lambda_{j'}^{(2)} = \lambda_{j'+1}^{(2)} \neq \lambda_j^{(2)}$ and
that there is no other degeneracy.  If
$(c_{ijk},c_{i+1,jk})$ and
$(c_{i'j'k'},c_{i',j'+1,k'})$ are nonzero at a state where the
function $F$ is minimized, then
$$
L(i,j,k)  \leq L(i',j'+1,k')\ \ and\ \ 
L(i',j',k') \leq L(i+1,j,k)\,.
$$
\label{lemma5}
\end{lemma}

\begin{proof}
We can use unitary transformations 
to have $c_{i,j+1,k} = c_{i',j'+1,k'} = 0$ and 
$c_{ijk}\neq 0$, $c_{i'j'k'}\neq 0$ without changing the RDMs.
Then by using the variation given in the proof of
Lemma~\ref{lemma1} we find
\begin{eqnarray}
\delta F = (p_{i+1} + q_j + r_k - p_{i'} - q_{j'} - r_{k'})\Delta\ \ 
{\rm if}\ \ \Delta > 0\,,\nonumber \\
\delta F = (p_{i} + q_j + r_k - p_{i'} - q_{j'+1} - r_{k'})\Delta\ \ 
{\rm if}\ \ \Delta < 0\,. \nonumber
\end{eqnarray}
By requiring $\delta F \geq 0$ for both cases we find the desired
inequalities.
\end{proof}

\begin{lemma}
Suppose that 
$\lambda_{i}^{(1)} = \lambda_{i+1}^{(1)} \neq \lambda_{i'}^{(1)}$,
$\lambda_{j}^{(2)} = \lambda_{j+1}^{(2)} \neq \lambda_{j'}^{(2)}$ and
$\lambda_{k}^{(3)} = \lambda_{k+1}^{(3)} \neq \lambda_{k'}^{(3)}$
at a state where the function $F$ defined by (\ref{func}) has its
minimum.
Suppose further that at least one of the eight 
coefficients $c_{IJK}$ with $I=i,i+1$, $J=j,j+1$ and $K=k,k+1$
is nonzero and that $c_{i' j'k'}$ is also
nonzero for this state. 
Then 
$$
L(i,j,k) \leq L(i',j',k') \leq L(i+1,j+1,k+1)\,.
$$
\label{lem4}
\end{lemma}

\begin{proof}
Since $\lambda_i^{(1)} = \lambda_{i+1}^{(1)}$,
$\lambda_j^{(2)} = \lambda_{j+1}^{(2)}$ and 
$\lambda_k^{(3)} = \lambda_{k+1}^{(3)}$, 
a unitary transformation can be used to maximize 
$|c_{ijk}|^2$.  This makes $c_{i+1,jk}$, $c_{i,j+1,k}$ and
$c_{ij,k+1}$ vanish. (This argument was used to construct a
generalized Schmidt decomposition in Refs.~\cite{tny,hlry}.)
Then consider the variation given in the proof of
Lemma~\ref{lemma1}.  The variation of the
RDMs is effectively diagonal, and we have
$\delta \lambda_{i}^{(1)} = \delta\lambda_j^{(2)}
=\delta\lambda_k^{(3)} = 0$, 
$\delta \lambda_{i+1}^{(1)} = \delta \lambda_{j+1}^{(2)} = 
\delta\lambda_{k+1}^{(3)} = \Delta$ if
$\Delta > 0$ and $\delta \lambda_{i}^{(1)} = \lambda_j^{(2)} 
= \delta\lambda_k^{(3)}= \Delta$, 
$\delta \lambda_{i+1}^{(1)} = \delta\lambda_{j+1}^{(2)} = 
\delta\lambda_{k+1}^{(3)} = 0$ 
if $\Delta < 0$.  Thus,
\begin{eqnarray}
\delta F & = & (p_{i+1}+q_{j+1}+r_{k+1}-p_{i'}-q_{j'}-r_{k'})\Delta\ \
{\rm if}\ \ \Delta > 0\,, \nonumber \\
\delta F & = & (p_{i} + q_j + r_k - p_{i'} - q_{j'} - r_{k'})\Delta\ \
{\rm if}\ \ \Delta < 0\,.\nonumber
\end{eqnarray}
By requiring that $\delta F \geq 0$ for both cases we find 
the desired inequalities.
\end{proof}

\noindent
The following proposition will also be useful. 
\begin{proposition}
Suppose that $\lambda_{i}^{(1)}=\lambda_{i+1}^{(1)}$ with 
no other degeneracy at a state where the function $F$ defined
by (\ref{func}) has its minimum value.  
Define the reduced level $R^{(1)}(j,k)$ by
\begin{equation}
R^{(1)}(j,k) \equiv q_j + r_k\,.
\end{equation}
If $R^{(1)}(j,k) \neq R^{(1)}(j',k')$, then
\begin{equation}
c_{ijk}\overline{c_{ij'k'}} + c_{i+1,jk}\overline{c_{i+1,j'k'}} = 0\,,
\label{ortho2}
\end{equation}
i.e. the vectors $(c_{ijk},c_{i+1,jk})$ and
$(c_{ij'k'},c_{i+1,j'k'})$ are orthogonal to each other,
for this state.
\label{prop5}
\end{proposition}

\begin{proof}
We consider the following variation:
\begin{eqnarray}
&& \delta c_{ijk} = \alpha c_{ij'k'}\,,\ \
\delta c_{i+1,jk} = \alpha c_{i+1,j'k'}\,,\nonumber \\
&& \delta c_{ij'k'} = -\overline{\alpha} c_{ijk}\,,\ \ 
\delta c_{i+1,j'k'} = -\overline{\alpha} c_{i+1,jk} \nonumber
\end{eqnarray}
with all other $\delta c_{IJK}$ vanishing.
Then we find that the variation of the
RDM for the first qutrit vanishes.
We also find that if we define
$$
\Delta \equiv {\rm Re}\,[\alpha(c_{ijk}\overline{c_{ij'k'}}
+ c_{i+1,jk} \overline{c_{i+1,j'k'}})]\,,
$$
then
$\delta\lambda_{j}^{(2)}= - \delta \lambda_{j'}^{(2)} = \Delta$
and
$\delta\lambda_{k}^{(3)}= -\delta \lambda_{k'}^{(3)} = \Delta$.
Hence
$$
\delta F = (q_j+r_k - q_{j'}-r_{k'})\Delta = 
[R^{(1)}(j,k)-R^{(1)}(j',k')]\Delta\,.
$$
If Eq.~(\ref{ortho2}) does not hold and if
$R^{(1)}(j,k)\neq R^{(1)}(j',k')$, then the parameter
$\alpha$ can be adjusted 
so that $\delta F < 0$.  Hence, at a state where $F$ has
its minimum value we must have either (\ref{ortho2}) or
$R^{(1)}(j,k)=R^{(1)}(j',k')$. 
\end{proof}

\noindent
{\bf Remark}. The statements obtained from 
Lemmas~\ref{lemma4} and \ref{lemma5} and Proposition~\ref{prop5}
by qutrit permutations are also valid.

\

It is useful to treat 
some cases with only one pair of degenerate eigenvalues in general.
\begin{lemma}
Suppose that the function $P_m$ given by (\ref{Pfun}) 
in the proof of
Proposition~\ref{propprop} has a negative minimum at a state with
$\lambda_{i_2}^{(1)} = \lambda_{i_3}^{(1)}$ and without any other 
degeneracy.  Then $\lambda_{j_3}^{(2)} = \lambda_{k_3}^{(3)} = 0$.
\label{lemma7}
\end{lemma}

\begin{proof}
The only triples with negative levels are
$[i_1j_1k_1]$, $[i_2j_1 k_1]$, $[i_1j_2k_1]$ and $[i_1j_1k_2]$.  If
$(c_{i_1j_1k_1},c_{i_1j_2k_1},c_{i_1j_1k_2})$ and
$c_{ij_3k}$ are nonzero, then the preceding lemmas imply
that there must be some $i'$ (which is not necessarily equal to $i$)
such that $L(i',j_3,k)$ is less than $L(i_1,j_1,k_1)$,
$L(i_1,j_2,k_1)$ or $L(i_1,j_1,k_2)$.  This is impossible because
$L(i',j_3,k) \geq 0$ for all $i'$ and $k$.  
Hence, if $c_{ij_3k}\neq 0$
for some $i$ and $k$, then
$c_{i_1j_1k_1}=c_{i_1j_2k_1}=c_{i_1j_1k_2}=0$.  Then a
unitary transformation can be used
to have $c_{i_2j_1k_1}=0$ without changing the RDMs.
Thus, we can make all
coefficients $c_{ijk}$ with $L(i,j,k)< 0$ vanish.  Hence from 
(\ref{Pm}) we conclude that $P_m \geq 0$ 
if $\lambda_{j_3}^{(2)} > 0$.  Thus,
if the function $P_m$ were to have a negative minimum, then 
we must have $\lambda_{j_3}^{(2)} = 0$.
In a similar manner it can be shown that
$\lambda_{k_3}^{(3)} = 0$.
\end{proof}

\begin{corollary}
Suppose that the function $P_m$ given by (\ref{Pfun}) 
in the proof of Proposition~\ref{propprop}
has a negative minimum at a state with
$\lambda_{j_2}^{(2)} = \lambda_{j_3}^{(2)}$ 
($\lambda_{k_2}^{(3)} = \lambda_{k_3}^{(3)})$ 
and without any other degeneracy.
Then $\lambda_{i_3}^{(1)}= \lambda_{k_3}^{(3)}= 0$ 
($\lambda_{i_3}^{(1)} = \lambda_{j_3}^{(2)} = 0$).
\label{prevcoro}
\end{corollary}

\begin{proof}
The function $P_m$ can be rewritten as
$$
P_m = 1+\lambda_{i_3}^{(1)}-\lambda_{i_1}^{(1)}
+ \lambda_{j_3}^{(2)}-\lambda_{j_1}^{(2)}
+ \lambda_{k_3}^{(3)}-\lambda_{k_1}^{(3)}\,. 
$$
Then it is clear that the proof of Lemma~\ref{lemma7} applies here
after performing a qutrit permutations.
\end{proof}

\begin{lemma}
If the function $P_m$ given by (\ref{Pfun}) has a negative
minimum at a state with $\lambda_{i_1}^{(1)} = \lambda_{i_2}^{(1)}$
and with no other degeneracy, then $\lambda_{i_3}^{(1)} = 0$ or
$\lambda_{j_3}^{(2)} = \lambda_{k_3}^{(3)} = 0$.
\label{lemma8}
\end{lemma}

\begin{proof}
To have $P_m < 0$ we must have $c_{i_1j_1k_1}\neq 0$,
$c_{i_2j_1k_1}\neq 0$, $c_{i_1j_2k_1}\neq 0$ or $c_{i_1j_1k_2}\neq 0$.
Then from the preceding 
lemmas we find that if $c_{i_3jk}\neq 0$, then
$L(i_3,j,k)\leq 0$.  This is true only for $j=j_1$, $k=k_1$.  Thus,
if we assume that
$\lambda_{i_3}^{(1)}\neq 0$, then $c_{i_3j_1k_1}\neq 0$, and 
orthogonality relations give $c_{i_2j_1k_1}=c_{i_1j_1k_1}=0$.
If the vectors $(c_{i_1j_2k_1},c_{i_2j_2k_1})$ and
$(c_{i_1j_1k_2},c_{i_2j_1k_2})$ are linearly independent, then there
cannot be any other nonzero coefficients of the form $c_{i_1jk}$ or
$c_{i_2jk}$ by Proposition~\ref{prop5} because 
$R^{(1)}(j,k)< 0$ for 
$\left[jk\right]=[j_1k_1]$, $[j_1k_2]$ and $[j_2k_1]$, and
$R^{(1)}(j,k)\geq 0$ for all other $\left[jk\right]$.  
This implies that
$\lambda_{j_3}^{(2)} = \lambda_{k_3}^{(3)} = 0$.  If these vectors
are linearly dependent, a unitary transformation can be used 
to have
$c_{i_1j_1k_2}=c_{i_1j_2k_1} = 0$.  Then all coefficients $c_{ijk}$
with $L(i,j,k) < 0$ vanish.  Hence $P_m \geq 0$.
\end{proof}

\begin{corollary}
If the function $P_m$ given by (\ref{Pfun}) has a negative
minimum at a state with $\lambda_{j_1}^{(2)} = \lambda_{j_2}^{(2)}$
($\lambda_{k_1}^{(3)}=\lambda_{k_2}^{(3)}$)
and with no other degeneracy, then $\lambda_{j_3}^{(2)} = 0$ 
($\lambda_{k_3}^{(3)} = 0$) or
$\lambda_{i_3}^{(1)} = \lambda_{k_3}^{(3)} = 0$
($\lambda_{i_3}^{(1)} = \lambda_{j_3}^{(2)}=0$).
\label{coro7}
\end{corollary}

\section{Proof of inequalities (\ref{ineq4}) and (\ref{ineq5})}
\label{ineqfourfive}

Now we are in a position to prove inequalities (\ref{ineq4}) and
(\ref{ineq5}).  The following Proposition, which was used in
Ref.~\cite{hss}, will be useful.
\begin{proposition}
If $i_1 < i_2$, then 
$|c_{i_1 jk}|^2 + |c_{i_2 jk}|^2 \leq \lambda_{i_2}^{(1)}$.
\label{prop6}
\end{proposition}

\begin{proof} 
If $\lambda_{i_1}^{(1)} = 0$, then 
$c_{i_1 jk} = 0$ for all $[jk]$.  Since we have 
$|c_{i_2 jk}|^2 \leq \lambda_{i_2}^{(1)}$, the desired inequality is
satisfied.  So we assume $\lambda_{i_1}^{(1)} \neq 0$.
By applying the Cauchy-Schwarz inequality to the 
orthogonality relation 
$c_{i_1JK}\overline{c_{i_2 JK}} = 0$, we obtain
$$
|c_{i_1 jk}|^2|c_{i_2 jk}|^2 \leq 
(\lambda_{i_1}^{(1)} - |c_{i_1 jk}|^2)
(\lambda_{i_2}^{(1)} - |c_{i_2 jk}|^2)\,.
$$
Thus,
$$
\lambda_{i_2}^{(1)}|c_{i_1 jk}|^2 + \lambda_{i_1}^{(1)}|c_{i_2jk}|^2
\leq \lambda_{i_2}^{(1)}\lambda_{i_1}^{(1)}\,.
$$
Hence
$$
|c_{i_1jk}|^2 + |c_{i_2jk}|^2
\leq \frac{\lambda_{i_2}^{(1)}}{\lambda_{i_1}^{(1)}}|c_{i_1jk}|^2
+ |c_{i_2jk}|^2 \leq \lambda_{i_2}^{(1)}\,.
$$
\end{proof}

\noindent
{\bf Remark}.  We can prove similarly that 
$|c_{ij_1k}|^2 + |c_{ij_2k}|^2 \leq \lambda_{j_2}^{(2)}$ if
$j_1 < j_2$ and that
$|c_{ijk_1}|^2 + |c_{ijk_2}|^2 \leq \lambda_{k_2}^{(3)}$ if
$k_1 < k_2$.

\

\noindent
We need the following lemma to prove $P_4\geq 0$, i.e.
inequality (\ref{ineq4}).
\begin{lemma}
The function $P_4$ has its minimum value at a state with $c_{233} = 0$.
\label{lemma9}
\end{lemma}

\begin{proof}
If $\lambda_2^{(1)}$, $\lambda_3^{(2)}$ or $\lambda_3^{(3)}$ is 
degenerate with another eigenvalue, then we can use a 
unitary transformation to have $c_{233}=0$.  Hence, if
$c_{233}\neq 0$ at all states where the minimum of $P_4$ occurs, 
then the
only possible degeneracies are $\lambda_1^{(2)}=\lambda_2^{(2)}$ and
$\lambda_1^{(3)}=\lambda_2^{(3)}$.  Suppose that $c_{233}$ and
another coefficient $c_{ijk}$ are nonzero.  Then we must have
$L(i,j',k') \leq L(2,3,3)$ for some $[ij'k']$ with
$\lambda_{j'}^{(2)}=\lambda_j^{(2)}$ and
$\lambda_{k'}^{(3)}=\lambda_{k}^{(3)}$.  This is impossible because
$[233]$ is the only triple with $L(2,3,3)\leq -2$.  Hence there must
be a state with $c_{233} = 0$ where $P_4$ has its minimum value.
\end{proof}

\begin{proposition}
The function $P_4$ is nonnegative.
\end{proposition}

\begin{proof}
We write $P_4$ in the following form
\begin{equation}
P_4 = \lambda_1^{(1)} +
2\lambda_3^{(1)}+\lambda_2^{(2)}-\lambda_3^{(2)}
+ \lambda_2^{(3)}-\lambda_3^{(3)}\,. \label{P4mod}
\end{equation}
By Proposition~\ref{prop6} and the remark following it we obtain
\begin{eqnarray}
|c_{133}|^2 + |c_{333}|^2 & \leq & \lambda_3^{(1)}\,, \nonumber \\
|c_{213}|^2 + |c_{223}|^2 & \leq & \lambda_2^{(2)}\,, \nonumber \\
|c_{231}|^2 + |c_{232}|^2 & \leq & \lambda_2^{(3)}\,. \nonumber
\end{eqnarray}
By using these inequalities in (\ref{P4mod}) we find
\begin{eqnarray}
P_4 & \geq & |c_{133}|^2 + |c_{333}|^2 + |c_{213}|^2 
+ |c_{223}|^2 + |c_{231}|^2 + |c_{232}|^2 \nonumber \\
&& + \lambda_1^{(1)}+\lambda_3^{(1)}-\lambda_3^{(2)}
-\lambda_3^{(3)}\,.
\nonumber
\end{eqnarray}
By letting $\lambda_1^{(1)}=\sum_{j,k} |c_{1jk}|^2$, and similarly
for $\lambda_3^{(1)}$, $\lambda_3^{(2)}$ and $\lambda_3^{(3)}$, we
find
\begin{eqnarray}
P_4 & \geq & |c_{111}|^2 +|c_{112}|^2 + |c_{121}|^2 + |c_{122}|^2
-2|c_{233}|^2 \nonumber \\
&& + |c_{311}|^2 + |c_{312}|^2 + |c_{321}|^2 + |c_{322}|^2\,.
\label{nice}
\end{eqnarray}
By Lemma~\ref{lemma9} we can choose a state with 
$c_{233}=0$ to achieve the minimum of
$P_4$.  Then, Eq.~(\ref{nice}) shows that this minimum is nonnegative.
\end{proof}

Our next task is to prove $P_5 \geq 0$.
\begin{lemma}
If $c_{133}=c_{132}=0$, then $P_5 \geq 0$.
\label{lemma10}
\end{lemma}

\begin{proof}
The function $P_5$ can be written as
\begin{equation}
P_5 = \lambda_2^{(1)} +
2\lambda_3^{(1)}+\lambda_2^{(2)}-\lambda_3^{(2)}
+ \lambda_1^{(3)}-\lambda_3^{(3)}\,. \label{P5mod}
\end{equation}
By Proposition~\ref{prop6} and the remark following it we have
\begin{eqnarray}
|c_{233}|^2 + |c_{333}|^2 & \leq & \lambda_3^{(1)}\,, \nonumber \\
|c_{113}|^2 + |c_{123}|^2 & \leq & \lambda_2^{(2)}\,. \nonumber
\end{eqnarray}
By using these inequalities in (\ref{P5mod}) we find
\begin{eqnarray}
P_5 & \geq & |c_{233}|^2 + |c_{333}|^2 + |c_{113}|^2 + |c_{123}|^2
\nonumber \\
&& + \lambda_2^{(1)}+\lambda_3^{(1)}-\lambda_3^{(2)}+\lambda_1^{(3)}
-\lambda_3^{(3)}\,. \nonumber 
\end{eqnarray}
By using the expressions of
$\lambda_2^{(1)}$, $\lambda_3^{(1)}$, $\lambda_3^{(2)}$ and
$\lambda_3^{(3)}$ as sums of $|c_{ijk}|^2$ we find
\begin{eqnarray}
P_5 & \geq & \lambda_1^{(3)} - |c_{131}|^2 - |c_{132}|^2 - 2|c_{133}|^2
\nonumber \\
&& +|c_{211}|^2 + |c_{212}|^2 + |c_{221}|^2 + |c_{222}|^2 \nonumber \\
&& + |c_{311}|^2 + |c_{312}|^2 + |c_{321}|^2 + |c_{322}|^2\,. \nonumber
\end{eqnarray}
Since $\lambda_1^{(3)}\geq |c_{131}|^2$, we have $P_5 \geq 0$ if
$c_{133}=c_{132}=0$.
\end{proof}

\begin{lemma}
The function $P_5$ is nonnegative if $\lambda_2^{(2)}=\lambda_3^{(2)}$,
$\lambda_1^{(3)}=\lambda_2^{(3)}$ or $\lambda_2^{(3)}=\lambda_3^{(3)}$.
\end{lemma}

\begin{proof}
If $\lambda_2^{(2)} =\lambda_3^{(2)}$, then the inequality
$P_5\geq 0$ is equivalent to
\begin{equation}
2\lambda_3^{(2)} + \lambda_1^{(2)} + 2\lambda_1^{(3)} + \lambda_2^{(3)}
- 2\lambda_1^{(1)} - \lambda_2^{(1)}\geq 0. \label{temp}
\end{equation}
Since $\lambda_3^{(2)} \geq 1/3 \geq \lambda_1^{(1)}$, we have
$\lambda_3^{(2)}+\lambda_1^{(3)} \geq \lambda_1^{(1)}$.  Hence 
inequality (\ref{temp}) will follow from
$$
\lambda_3^{(2)} + \lambda_1^{(2)} + \lambda_1^{(3)} + \lambda_2^{(3)}
- \lambda_1^{(1)} - \lambda_2^{(1)}\geq 0\,. 
$$
This follows from inequality (\ref{ineq1}) by letting 
$(abc)=(123)$ and
noting that $\lambda_3^{(2)} \geq \lambda_2^{(2)}$.
If $\lambda_1^{(3)} = \lambda_2^{(3)}$, then
\begin{eqnarray}
P_5 & = & 2\lambda_2^{(2)} + \lambda_1^{(2)}
+ 2\lambda_2^{(3)} + \lambda_1^{(3)} 
- 2\lambda_1^{(1)} - \lambda_2^{(1)}
\nonumber \\
& = & P_4 + \lambda_2^{(1)} - \lambda_1^{(1)}\,.\nonumber
\end{eqnarray}
This is nonnegative because $P_4 \geq 0$.  Finally,
let $\lambda_2^{(3)} = \lambda_3^{(3)}$. 
{}From Corollary~\ref{coro} we have
$$
\lambda_2^{(2)} + \lambda_3^{(1)} \geq \lambda_2^{(3)}\,.
$$
By adding this to inequality (\ref{ineq3}) with $(abc)=(321)$
we obtain
$$
2\lambda_2^{(2)} + \lambda_1^{(2)}
+ 2\lambda_3^{(1)} + \lambda_2^{(1)}\geq  
2\lambda_2^{(3)} + \lambda_3^{(3)} 
= 2\lambda_3^{(3)} + \lambda_2^{(3)}\,.
$$
This is equivalent to the inequality $P_5 \geq 0$.
\end{proof}

\noindent
Thus we only need to deal with the cases with
$\lambda_1^{(1)}=\lambda_2^{(1)}$,
$\lambda_2^{(1)}=\lambda_3^{(1)}$
or $\lambda_1^{(2)} = \lambda_2^{(2)}$.

\begin{lemma}
Suppose that the minimum of $P_5$ is negative.  Then
it does not occur in any of the following
three cases:
\begin{enumerate}
\item $\lambda_1^{(1)}=\lambda_2^{(1)}$ with no other
degeneracy;
\item $\lambda_2^{(1)}=\lambda_3^{(1)}$ with no other
degeneracy;
\item $\lambda_1^{(2)}=\lambda_2^{(2)}$ with
no other degeneracy.
\end{enumerate}
\end{lemma}

\begin{proof}
Case~1.  By Lemma~\ref{lemma8} we have either $\lambda_3^{(1)}=0$ or
$\lambda_2^{(2)}=\lambda_1^{(3)}=0$.  The first case is impossible.
The second case leads to $\lambda_3^{(2)}=1$.  Then, by Lemma
\ref{lemma2} we have $P_5 \geq 0$.

Case~2. By Lemma~\ref{lemma7} we must have 
$\lambda_2^{(2)}=\lambda_1^{(3)}=0$.  This implies that
$\lambda_3^{(2)}=1$.  By Lemma~\ref{lemma2} we have
$P_5 \geq 0$.

Case~3.  By Corollary~\ref{prevcoro} we must have 
$\lambda_3^{(1)}=\lambda_1^{(3)}=0$.  This is impossible.
\end{proof}

\noindent
The following lemma completes the proof of
inequality (\ref{ineq5}).
\begin{lemma}
Suppose that the minimum of $P_5$ is negative.  Then
it does not occur in any of the following
three cases:
\begin{enumerate}
\item $\lambda_1^{(1)}=\lambda_2^{(1)}=\lambda_3^{(1)}$;
\item $\lambda_1^{(1)}=\lambda_2^{(1)}$ and
$\lambda_1^{(2)}=\lambda_2^{(2)}$ with no
other degeneracy;
\item $\lambda_2^{(1)}=\lambda_3^{(1)}$ and
$\lambda_1^{(2)}=\lambda_2^{(2)}$ with no
other degeneracy.
\end{enumerate}
\label{lemma13}
\end{lemma}

\begin{proof}
We will show that there is a state with $c_{133}=c_{132}=0$
where the function $P_5$ has its minimum if it occurs 
in any of the three cases.  Then by Lemma~\ref{lemma10}
we conclude that the minimum must be nonnegative.

Case 1.  Any unitary transformation on the first qutrit
leaves the RDMs diagonal.
Consider the vectors $(c_{133},c_{233},c_{333})$ and
$(c_{132},c_{232},c_{332})$.  By a unitary transformation
on the first qutrit we can set $c_{133}=c_{233}=c_{132}=0$.

Case 2. Suppose that $(c_{132},c_{232})= (0,0)$ or
$(c_{133},c_{233})= (0,0)$.  Then 
both $c_{132}$ and $c_{133}$ can be made
zero by a unitary transformation
on the first qutrit.  Hence, all we need to show is that if
$(c_{132},c_{232})\neq (0,0)$, then $(c_{133},c_{233})=0$
at a state where the function $P_5$ is minimized.
Suppose $(c_{132},c_{232})\neq (0,0)$ and $c_{33k}\neq 0$
for some $k$.  Then by Lemma~\ref{lemma3} we have
$L(3,3,k) \leq L(2,3,2)=0$, which implies that
$r_k=0$, i.e. $k=3$.  Next, suppose that
$(c_{31k},c_{32k})\neq (0,0)$.  Then, by Lemma~\ref{lemma5} 
we have $L(3,1,k) \leq L(2,3,2)$, i.e.
$r_k \leq -1$.  This is impossible.
Hence $(c_{31k},c_{32k})=(0,0)$ for all $k$.  Thus, $c_{333}$ is
the only nonzero coefficient of the form $c_{3jk}$.  Then by
orthogonality relations we have $(c_{133},c_{233})=(0,0)$.

Case 3.  Note that the triple $[133]$ is not linked to
the degenerate sector of the RDMs in the
sense that $\lambda_1^{(1)}\neq \lambda_2^{(1)}$, 
$\lambda_3^{(2)}\neq \lambda_2^{(2)}$ and
$\lambda_3^{(3)}\neq \lambda_2^{(3)}$.  Since $L(1,3,3)=-2$ and 
$L(i,j,k) > -2$ for all other triples $[ijk]$, if $c_{133}\neq 0$,
then $c_{ijk} = 0$ for any other $[ijk]$.  This implies that
$\lambda_3^{(1)}=0$, which is a 
contradiction.  Thus, we can assume that $c_{133}=0$.
Suppose that $c_{132}\neq 0$.  If $(c_{33k},c_{23k})\neq (0,0)$
as well, then by Lemma~\ref{lemma3} we have
$L(2,3,k)\leq L(1,3,2)=-1$, hence
$r_k=0$, i.e. $k=3$.  Next, if 
$(c_{32k},c_{31k},c_{22k},c_{21k})\neq (0,0,0,0)$, then by Lemma
\ref{lemma4} we have
$L(2,1,k)\leq L(1,3,2)$. This
is impossible for any $k$.  Hence 
$c_{32k}=c_{31k}=c_{22k}=c_{21k}=0$.  
Thus, the only nonzero
coefficient of the form $c_{3jk}$ ($c_{2jk}$) is $c_{333}$ ($c_{233}$).
Then, by orthogonality we must have $c_{233}=0$ since 
$\lambda_3^{(1)}\neq 0$.  Then
$\lambda_2^{(1)}=0$ and hence $\lambda_3^{(1)}\neq \lambda_2^{(1)}$,
contradicting the assumption.
\end{proof}

\section{Proof of inequalities (\ref{ineq6}) and (\ref{ineq7})}
\label{ineqsixseven}

For the case where there are two pairs of degenerate eigenvalues, 
it is useful to
consider the variation of the eigenvalues for some variation of
the coefficients $c_{IJK}$ in a general setting.
\begin{proposition}
Suppose that $\lambda_i^{(1)}=\lambda_{i+1}^{(1)}$ and
$\lambda_j^{(2)}=\lambda_{j+1}^{(2)}$, and that there is no other
degeneracy.  Suppose further that
$$
\left(\begin{array}{cc} c_{ijk} & c_{i,j+1,k} \\
c_{i+1,jk} & c_{i+1,j+1,k} \end{array} \right)\ and\ 
(c_{ij'k'}, c_{i+1,j'k'})\ are\ nonzero.
$$
Let $c_{i,j+1,k} = c_{i+1,j,k}=0$ by using unitary transformations, 
and let
\begin{eqnarray}
&& \delta c_{ijk} = \alpha_1 c_{ij'k'} + \alpha_2 c_{i+1,j'k'},\ \
\delta c_{i+1,j+1,k} = \beta_1 c_{ij'k'} +
\beta_2 c_{i+1,j'k'}\,, \nonumber \\
&& \delta c_{ij'k'} = -\overline{\alpha_1} c_{ijk} -
\overline{\beta_1} c_{i+1,j+1,k}\,,\ \ 
\delta c_{i+1,j'k'} = - \overline{\alpha_2}c_{ijk} -
\overline{\beta_2} c_{i+1,j+1,k}\,, \nonumber
\end{eqnarray}
where $j' \neq j$ and $j'\neq j+1$, but $k'$ may or may not be equal
to $k$.
Define
$a_{I+1} \equiv 2{\rm Re}\,(\alpha_{I+1} 
\overline{c_{i+I,j'k'}}c_{ijk})$ and
$b_{I+1} \equiv 2{\rm Re}\,(\beta_{i+I}
\overline{c_{i+I,j'k'}}c_{i+1,j+1,k})$, $I=0,1$.
Then it is possible to arrange the variation of the coefficients
$c_{i,j+1,k}$ and $c_{i+1,jk}$ so that the variation
of the RDMs is effectively diagonal 
without making $a_1$, $a_2$, $b_1$ and
$b_2$ all vanish.  
If $k\neq k'$,
the variation of the eigenvalues of the
RDMs is given as follows:
\begin{eqnarray}
&& \delta \lambda_i^{(1)} = -|a_2-b_1|\,,\ \
\delta \lambda_{i+1}^{(1)} = |a_2-b_1|\,, \nonumber \\
&& \delta \lambda_j^{(2)} = {\rm min}\,(a_1+a_2,b_1+b_2),\ \ 
\delta \lambda_{j+1}^{(2)} = {\rm max}\,(a_1+a_2,b_1+b_2)\,, 
\nonumber \\
&& \delta\lambda_{j'}^{(2)} = -a_1 -a_2 - b_1 - b_2\,,\ \ 
\delta\lambda_k^{(3)} = a_1+a_2+b_1 +b_2\,, \nonumber \\
&& \delta \lambda_{k'}^{(3)} =  -a_1 -a_2 -b_1 -b_2\,.
\nonumber
\end{eqnarray}
If $k=k'$, then $\delta\lambda_k^{(3)}=\delta\lambda_{k'}^{(3)} = 0$,
and the variation of the other eigenvalues remains the same. 
\label{prop9}
\end{proposition}

\begin{proof}
By requiring the variation of 
the RDMs to be effectively diagonal, we find
\begin{eqnarray}
c_{ijk}\overline{\delta c_{i+1,jk}} + 
\delta c_{i,j+1,k}\overline{c_{i+1,j+1,k}}
& = & -\delta\left[c_{ij'k'}\overline{c_{i+1,j'k'}}\right]\,,
\nonumber \\
\overline{c_{ijk}}\delta c_{i,j+1,k}
+ \overline{\delta c_{i+1,jk}} c_{i+1,j+1,k} & = & 0\,. \nonumber
\end{eqnarray}
Thus, if $|c_{ijk}|^2 \neq |c_{i+1,j+1,k}|^2$,
then we can solve for $\delta c_{i+1,jk}$ and $\delta c_{i,j+1,k}$
to make the variation of the RDMs effectively diagonal.
If $|c_{ijk}|^2 = |c_{i+1,j+1,k}|^2$, we
first use phase transformations to have $c_{ijk}=c_{i+1,j+1,k}$.
Then any unitary transformation on the first qutrit 
can be compensated by a unitary transformation on
the second qutrit while maintaining the condition
 $c_{i+1,jk}=c_{i,j+1,k}=0$.  Thus we can make $c_{i+1,j'k'}$
vanish as well.  Then by letting 
$\delta c_{i+1,j'k'} = \delta c_{i+1,jk} =\delta c_{i,j+1,k}=0$,
the variation of the RDMs can be kept effectively
diagonal.  We have
$\alpha_2=\beta_2 = 0$ and $a_2=b_2=0$ as a consequence
but can have $a_1\neq 0$
or $b_1\neq 0$ by the assumption on the
coefficients. (Since we have made $c_{i+1,j'k'}$ vanish,
the equation $\alpha_2=\beta_2=0$ does not give rise to any further
constraint on the variation of the eigenvalues.) 
The calculation of the variation of the eigenvalues is
straightforward. 
\end{proof}

\noindent
{\bf Remark}. This proposition remains valid after a qutrit 
permutation.

\

Now we prove $P_6 \geq 0$, i.e. inequality (\ref{ineq6}). 
Our proof is rather lengthy.
\begin{lemma}
If $\lambda_1^{(1)}=\lambda_2^{(1)}$, 
$\lambda_2^{(2)}=\lambda_3^{(2)}$ or $\lambda_2^{(3)}=\lambda_3^{(3)}$,
then $P_6 \geq 0$.
\end{lemma}

\begin{proof}
Inequality (\ref{ineq3}) with $(abc)=(123)$ reads
\begin{equation}
\lambda_2^{(2)}+\lambda_1^{(2)}+\lambda_2^{(3)}+\lambda_3^{(3)}
\geq \lambda_2^{(1)}+\lambda_3^{(1)}\,. \label{ineq3pr}
\end{equation}
From inequality (\ref{ineq1}) we have
\begin{eqnarray}
\lambda_2^{(2)}+\lambda_2^{(3)} & \geq &
\frac{1}{2}\left(\lambda_2^{(2)}+\lambda_1^{(2)}
+\lambda_2^{(3)}+\lambda_1^{(3)}\right)\nonumber \\
& \geq & \frac{1}{2}\left(\lambda_2^{(1)} + \lambda_1^{(1)}\right)
\nonumber \\
& \geq & \lambda_1^{(1)}\,. \nonumber
\end{eqnarray}
By adding these two inequalities we obtain
$$
2\lambda_2^{(2)}+\lambda_1^{(2)}
+ 2\lambda_2^{(3)}+\lambda_3^{(3)} \geq \lambda_1^{(1)}
+ \lambda_2^{(1)}+\lambda_3^{(1)}\,.
$$
This is equivalent to inequality (\ref{ineq6}) if 
$\lambda_1^{(1)}=\lambda_2^{(1)}$.
Next we recall that 
$\lambda_3^{(2)}+\lambda_2^{(3)} \geq \lambda_2^{(1)}$ 
(Corollary~\ref{coro}).  By adding this to (\ref{ineq3pr}) we have
$$
\lambda_3^{(2)}+\lambda_2^{(2)}+\lambda_1^{(2)}
+ 2\lambda_2^{(3)} + \lambda_3^{(3)} \geq 2\lambda_2^{(1)}
+ \lambda_3^{(1)}\,.
$$
This is equivalent to inequality (\ref{ineq6}) if 
$\lambda_2^{(2)}=\lambda_3^{(2)}$.  
Next, if we add the
inequality $\lambda_2^{(2)}+\lambda_3^{(3)} \geq \lambda_2^{(1)}$
to inequality (\ref{ineq3pr}), we have
$$
2\lambda_2^{(2)} + \lambda_1^{(2)} + \lambda_2^{(3)}+2\lambda_3^{(3)}
\geq 2\lambda_2^{(1)}+\lambda_3^{(1)}\,.
$$
This is equivalent to inequality (\ref{ineq6}) if
$\lambda_2^{(3)}=\lambda_3^{(3)}$.
\end{proof}

\noindent
The following lemmas will be useful in proving $P_6\geq 0$.

\begin{lemma}
The function $P_6$ is nonnegative if $c_{231} = c_{233}=c_{331}=0$.
\label{lemma16}
\end{lemma}

\begin{proof}
By Proposition~\ref{prop6} we have 
$|c_{211}|^2 + |c_{221}|^2 \leq \lambda_2^{(2)}$.  Hence
\begin{eqnarray}
P_6 & = & \lambda_1^{(1)}-\lambda_2^{(1)}
+ 2\lambda_2^{(2)} + \lambda_1^{(2)}+\lambda_2^{(3)}-\lambda_1^{(3)}
\nonumber \\
& \geq & \lambda_1^{(1)}-\lambda_2^{(1)}
+ \lambda_1^{(2)}+\lambda_2^{(2)}+\lambda_2^{(3)}-\lambda_1^{(3)}
+ |c_{221}|^2 + |c_{211}|^2\,. \nonumber
\end{eqnarray}
Then, if we substitute 
$\lambda_i^{(1)} = \sum_{JK}|c_{iJK}|^2$ and
the similar formulae for $\lambda_j^{(2)}$ and $\lambda_k^{(3)}$,
the only coefficients
that contribute negatively to $P_6$ are
$c_{231}$, $c_{233}$, $c_{331}$.  Hence, if these coefficients vanish,
then $P_6 \geq 0$.
\end{proof}

\begin{lemma}
The function $P_6$ is nonnegative if $c_{232}=c_{233}=c_{332}=0$.
\label{lemmanice}
\end{lemma}

\begin{proof}
Note first that
$$
P_6 \geq \tilde{P}_6 \equiv
2\lambda_2^{(2)}+\lambda_1^{(2)}
+ 2\lambda_1^{(3)}+\lambda_3^{(3)}-2\lambda_2^{(1)}-\lambda_3^{(1)}\,.
$$
It can be shown that $\tilde{P}_6\geq 0$ under the assumption of this
lemma by the argument in the proof of the previous lemma with
$\lambda_1^{(3)}\leftrightarrow \lambda_2^{(3)}$.
\end{proof}

\begin{lemma}
If $c_{233}=c_{333}=0$, then $P_6 \geq 0$.
\label{lemma213}
\end{lemma}
\begin{proof}
Since $\lambda_2^{(2)}+\lambda_3^{(3)} \geq \lambda_2^{(1)}$ by
Corollary~\ref{coro}, we have
\begin{eqnarray}
P_6 & \geq & \lambda_2^{(2)}+\lambda_1^{(2)}
+ 2\lambda_2^{(3)} - \lambda_2^{(1)} - \lambda_3^{(1)} \nonumber \\
& \geq & \lambda_2^{(2)}+\lambda_1^{(2)}
+ \lambda_2^{(3)}+\lambda_1^{(3)} -\lambda_2^{(1)}-\lambda_3^{(1)}\,.
\nonumber
\end{eqnarray}
Using the formulae for the eigenvalues in terms of $|c_{ijk}|^2$
in the last expression, 
we find that only $|c_{233}|^2$ and $|c_{333}|^2$ have negative
coefficients.  Hence $P_6 \geq 0$ if $c_{233}=c_{333}=0$.
\end{proof}

First we treat the cases with only one pair of degenerate eigenvalues.
\begin{lemma}
If the function $P_6$ has a negative minimum, then
it does not occur at a state with any of the following
properties:
\begin{enumerate}
\item $\lambda_2^{(1)} = \lambda_3^{(1)}$ with no other 
degeneracy;
\item $\lambda_1^{(2)} =\lambda_2^{(2)}$ with no other
degeneracy;
\item $\lambda_1^{(3)} = \lambda_2^{(3)}$ with no other
degeneracy.
\end{enumerate}
\end{lemma}

\begin{proof}
Case 1. By Lemma~\ref{lemma8} we have $\lambda_1^{(1)}=0$ or
$\lambda_2^{(2)}=\lambda_2^{(3)}=0$.  The latter case leads to
$\lambda_3^{(1)}=\lambda_3^{(2)}=\lambda_3^{(3)}=1$, and the 
inequality is obviously satisfied.  Hence we can assume that
$\lambda_1^{(1)}=0$, i.e. $c_{1jk}=0$ for all $\left[jk\right]$.
Then $\lambda_2^{(1)}=\lambda_3^{(1)}=1/2$.
If $(c_{231},c_{331})=(0,0)$, then by letting $c_{233}=0$
by a unitary transformation on the
first qutrit and using Lemma~\ref{lemma16} we have
$P_6 \geq 0$.  
If $(c_{231},c_{331})\neq (0,0)$, then all other vectors
of the form $(c_{2jk},c_{3jk})$ must be orthogonal to
$(c_{231},c_{331})$ by Proposition~\ref{prop5}.  
Thus, if we let $c_{331}=0$ by using
a unitary transformation on the first
qutrit, then $c_{2jk}=0$ for all $\left[jk\right]$ except
$\left[31\right]$. Hence
$|c_{231}|^2=\lambda_2^{(1)}=1/2$, and
$\lambda_1^{(3)} \geq |c_{231}|^2=1/2$, which is a contradiction.

Case 2. Corollary~\ref{prevcoro} implies that
$\lambda_1^{(1)}=\lambda_2^{(3)}=0$, and by Lemma~\ref{lemma2}
we have $P_6 \geq 0$.

Case 3.  Assume that $c_{233}\neq 0$.  Then, since there is no 
$\left[ij\right]$
satisfying $L(i,j,3)=-1$ except $\left[ij\right]=\left[23\right]$
and since the triple $[233]$ 
is not connected to the degenerate sector,
the $c_{233}$ is the
only nonzero coefficient of the form $c_{ij3}$.  Then,
by orthogonality relations we have $c_{231}=c_{232}=0$.  If 
$(c_{ij1},c_{ij2})\neq (0,0)$, by Lemma~\ref{lemma4} we must have
$L(i,j,1)\leq L(2,3,3)$, i.e. $p_i + q_j\geq -1$.  Hence 
$\left[ij\right]=\left[23\right]$,
which has been rejected, $\left[21\right]$ or $\left[33\right]$.  Then 
$\lambda_2^{(3)}=0$, and we have $P_6 \geq 0$ by Lemma~\ref{lemma2}.
Thus, we can assume that $c_{233}=0$.
If $(c_{231},c_{232})=(0,0)$ [$(c_{331},c_{332})=(0,0)$],
then we can have $c_{331}=0$ [$c_{231}=0$] in addition
by a unitary transformation 
on the third qutrit.  Then $P_6 \geq 0$ by Lemma~\ref{lemma16}.
Assume that $(c_{231},c_{232})$ and
$(c_{331},c_{332})$ are both nonzero.
These vectors are orthogonal to each other by
Proposition~\ref{prop5} applied to the case with 
degeneracy in the third qutrit.
Then the only other possible
nonzero vector of the form $(c_{ij1},c_{ij2})$ is
$(c_{211},c_{212})$ by the same proposition.
For $c_{ij3}$ to be nonzero we must have 
$L(i,j,3)\leq {\rm min}\,[L(2,3,2),L(2,1,2),L(3,3,2)]=1$.  
All nonzero coefficients of the form $c_{ij3}$ must have
the same value of $L(i,j,3)$.  
In order not to have $\lambda_2^{(2)}=0$ (and hence
$P_6\geq 0$ by Lemma~\ref{lemma2})
we must have $L(i,j,3)=1$ and the possibly nonzero coefficients
of the form $c_{ij3}$ are $c_{223}$, $c_{313}$ and $c_{133}$.
(We mean, strictly speaking, that $c_{ij3}$ is nonzero only if
$[ij3]$ is in the set $\{[223],[313],[133]\}$.)
Then $c_{233}=c_{333}=0$, and
by Lemma~\ref{lemma213} we have $P_6 \geq 0$.
\end{proof}

\noindent
{\bf Remark}.  We abbreviate the phrase 
``possibly nonzero coefficient" as PNC below.

\

\noindent
The following lemma is used in the most degenerate case.
\begin{lemma}
If $c_{231}=c_{211}=c_{233}=c_{213}=0$, then $P_6 \geq 0$.
\label{lemmaOK}
\end{lemma}

\begin{proof}
First we note inequality (\ref{ineq3}) with $(abc)=(123)$:
$$
\lambda_2^{(2)}+\lambda_1^{(2)}+\lambda_2^{(3)}
+\lambda_3^{(3)}-\lambda_2^{(1)}-\lambda_3^{(1)} \geq 0\,.
$$
The inequality $P_6\geq 0$ follows from this if
$$
\lambda_2^{(2)}+\lambda_2^{(3)} - \lambda_2^{(1)} \geq 0\,.
$$
By using the expression for $\lambda_2^{(a)}$ in terms
of $|c_{ijk}|^2$ we find that this inequality is 
satisfied if
$c_{211}=c_{213}=c_{231}=c_{233}=0$, as required.
\end{proof}

\noindent
Now we treat the most degenerate case.
\begin{proposition}
If the minimum value of $P_6$ is negative, then
it does not occur at a state with
$\lambda_2^{(1)}=\lambda_3^{(1)}$, 
$\lambda_1^{(2)}=\lambda_2^{(2)}$, $\lambda_1^{(3)}=\lambda_2^{(3)}$
and with no other degeneracy.
\end{proposition}

\begin{proof}
Note that we can make $c_{233}$, $c_{231}$ and
$c_{211}$ vanish by successively using unitary transformations on
the first, third and second qutrits.  
Then the only PNC
with a negative level is $c_{331}$.
We will show that if $c_{331}\neq 0$, then $c_{213}=0$.
Then by Lemma~\ref{lemmaOK} we have $P_6\geq 0$.
Thus, we assume that
$$
\left( \begin{array}{cc} c_{331} & c_{332} \\
c_{231} & c_{232} \end{array} \right)\ \ {\rm and}\ \ 
\left( \begin{array}{cc} c_{323} & c_{313} \\
c_{223} & c_{213} \end{array} \right)  
$$
are nonzero and derive a contradiction.

Suppose that the first matrix has rank less than two.
Then after making
$c_{233}$ vanish by using a unitary transformation
on the first qutrit, we can make
$c_{231}$ and $c_{331}$ vanish simultaneously 
by a unitary transformation on
the third qutrit.  Then, by Lemma~\ref{lemma16}
we have $P_6\geq 0$.
Thus, we can assume that the first matrix
has rank two. We set $c_{231}=c_{332}=0$ by unitary transformations
on the first and third qutrits
and let $c_{313}=0$ by
using a unitary transformation 
on the second qutrit.  Then, the above matrices become
$$
\left( \begin{array}{cc} c_{331} & 0 \\
 0 & c_{232} \end{array} \right)\ \ {\rm and}\ \ 
\left( \begin{array}{cc} c_{323} & 0  \\
c_{223} & c_{213} \end{array} \right)
$$
with $c_{331}\neq 0$ and $c_{232}\neq 0$.
Let us consider the following variation:
\begin{eqnarray}
\delta c_{323} & = & -\alpha_1 c_{331} -\alpha_2 c_{232}\,,\nonumber \\
\delta c_{213} & = & -\beta_1 c_{331} - \beta_2 c_{232}\,,\nonumber \\
\delta c_{223} & = & -\gamma_1 c_{331} -\gamma_2 c_{232}\,,\nonumber \\
\delta c_{331} & = & 
\overline{\alpha_1} c_{323} + \overline{\beta_1} c_{213}
+ \overline{\gamma_1} c_{223}\,, \nonumber \\
\delta c_{232} & = & \overline{\alpha_2}c_{323} + 
\overline{\beta_2} c_{213} + \overline{\gamma_2} c_{223}.
\nonumber
\end{eqnarray}
By requiring that the variation of the
RDMs be effectively diagonal we have
\begin{eqnarray}
c_{331}\overline{\delta c_{231}} + \delta c_{332}\overline{c_{232}}
+ \delta c_{313}\overline{c_{213}} & = & - \delta\left[
c_{323}\overline{c_{223}}\right]\,, \nonumber \\
\delta c_{313} \overline{c_{323}} & = & -
\delta\left[c_{213}\overline{c_{223}}\right]\,, \nonumber \\
c_{232}\overline{\delta c_{231}} 
+ \delta c_{332}\overline{c_{331}}& = & 0\,. \nonumber 
\end{eqnarray}
These equations can be consistently solved for $\delta c_{231}$,
$\delta c_{332}$ and $\delta c_{313}$ provided that
$c_{323}\neq 0$ and $|c_{331}|^2 \neq |c_{232}|^2$. (Notice that
the variations $\delta c_{231}$, $\delta c_{332}$ and $\delta c_{313}$
do not contribute to the variation of the diagonal
elements.)  If 
$c_{323} = 0$, then we can use a unitary transformation on the
second qutrit to eliminate $c_{223}$ as well.  Then 
$\delta[c_{323}\overline{c_{223}}]=\delta[c_{213}\overline{c_{223}}]
=0$ if we set $\gamma_1=\gamma_2=0$. Then
we can simply let $\delta c_{332}=\delta c_{313}=\delta c_{231}=0$.
If $|c_{331}|^2=|c_{232}|^2$, then we can use a phase transformation
to have $c_{331}=c_{232}$.  Then any unitary transformation in the
first qutrit can be compensated by a unitary transformation in the
third qutrit to keep $c_{331}$ and $c_{232}$ unchanged and keep
the condition $c_{231}=c_{332}=0$.  This means that we can diagonalize
the second matrix as well, making
$c_{223}$ vanish in addition.  Then we again have
$\delta[c_{323}\overline{c_{223}}]=\delta[c_{213}\overline{c_{223}}]
= 0$ if we let $\gamma_1=\gamma_2=0$.  Then we can let
$\delta c_{332}=\delta c_{313}=\delta c_{231}=0$.  

Let us introduce the following definitions:
$a_i \equiv 2{\rm Re}\,(\alpha_i C^{(i)}\overline{c_{323}})$,
$b_i \equiv 2{\rm Re}\,(\beta_i C^{(i)}\overline{c_{213}})$ and
$c_i \equiv 2{\rm Re}\,(\gamma_i C^{(i)}\overline{c_{223}})$,
where $C^{(1)}\equiv c_{331}$ and $C^{(2)} \equiv c_{232}$.
The variation of the function $P_6$ can be given as follows:
$$
\delta P_6 = \delta\lambda_1^{(1)}-\delta\lambda_2^{(1)}
+ \delta \lambda_2^{(2)}-\delta\lambda_3^{(2)}
+\delta \lambda_2^{(3)} -\delta\lambda_1^{(3)}\,.
$$
Assume first that $c_{323}\neq 0$ and
$|c_{331}|^2 \neq |c_{232}|^2$ and that the variation of
the RDMs is made effectively diagonal.  Then, we have
\begin{eqnarray}
\delta P_6 & = & 
|a_2 - b_1 -c_1| - {\rm min}\,(a_1+a_2+c_1+c_2,b_1+b_2)\nonumber \\
&& -(a_1+b_1+c_1+a_2+b_2+c_2) \nonumber \\
&& + |a_2+b_2+c_2-a_1-b_1-c_1|\,. \nonumber
\end{eqnarray}
If $a_1$ can be made nonzero,
then 
so can $a_2$, and vice versa, and similarly for $b_1$, $b_2$, $c_1$
and $c_2$ because $c_{331}$ and $c_{232}$ are both nonzero.  
Hence we can always have $a_2+b_2+c_2=a_1+b_1+c_1$ by adjusting
$\alpha_i$, $\beta_i$ and $\gamma_i$, $i=1,2$.
We can also choose all coefficients to be nonnegative.  Then,
$$
\delta P_6 \leq |a_2 -b_1-c_1| - (a_1+b_1+c_1+a_2+b_2+c_2)\,.
$$
If $a_1$ and $a_2$ can be made nonzero, then by choosing 
the others to be zero (by letting $\beta_i=\gamma_i=0$, $i=1,2$)
we find $\delta P_6 \leq - a_1 < 0$.  If $a_1=a_2=0$, then
$\delta P_6 \leq -b_2-c_2$.  
Since $b_2$ or $c_2$ can be made positive --- otherwise
$c_{323}=c_{223}=c_{213}=0$ --- we have
$\delta P_6 < 0$.
If $c_{323}=0$, then we have $a_1=a_2=0$ and $c_1=c_2=0$ to make
the variation of 
the RDMs effectively diagonal,  but we can
still make $b_1$ and $b_2$ positive.  Then we can have $\delta P_6<0$.
Similarly, if $|c_{331}|^2=|c_{232}|^2$, we must have
$c_1=c_2=0$, but we can still have either $a_1\neq 0$, $a_2\neq 0$
or $b_1\neq 0$, $b_2\neq 0$ (or both).  Then we can have 
$\delta P_6<0$.

Thus, if $c_{331}\neq 0$ and $c_{213}\neq 0$ at the same time, we
can always lower the value of $P_6$ if it is negative.  
Hence, if $c_{331}$ and
$c_{213}$ are both nonzero, the function $P_6$ does not have a 
negative minimum.
\end{proof}

Finally we deal with the cases with two pairs of degenerate
eigenvalues to complete the proof of inequality (\ref{ineq6}).
\begin{lemma}
Suppose that the minimum of the function $P_6$ is negative.
Then it does not occur
at a state with
$\lambda_2^{(1)}=\lambda_3^{(1)}$ and 
$\lambda_1^{(2)} = \lambda_2^{(2)}$ and without any other 
degeneracy.
\end{lemma}

\begin{proof}
If $(c_{231},c_{331}) = (0,0)$, then we can use
a unitary transformation on
the first qutrit to make $c_{233}$ vanish.  Then $P_6 \geq 0$
by Lemma~\ref{lemma16}.
Thus, we can assume that $(c_{231},c_{331})\neq (0,0)$.
If $(c_{233},c_{333})=0$, then 
we have $P_6 \geq 0$ by Lemma~\ref{lemma213}.
Thus, we can also assume $(c_{233},c_{333})\neq 0$.  By
Proposition~\ref{prop5} we conclude that
the vectors $(c_{231},c_{331})$ and $(c_{233},c_{333})$ are
orthogonal to each other.  By the same proposition
$(c_{232},c_{332})$ must be orthogonal to both of these vectors.
This implies that $(c_{232},c_{332})=(0,0)$.  Then by a unitary
transformation on the first qutrit we
can have $c_{233}=0$, and we conclude that $P_6\geq 0$ by 
Lemma~\ref{lemmanice}.
\end{proof}

\begin{lemma}
Suppose that the minimum of the function $P_6$ is negative.
Then it does not occur at 
a state with
$\lambda_2^{(1)}=\lambda_3^{(1)}$ and 
$\lambda_1^{(3)} = \lambda_2^{(3)}$ and without any other 
degeneracy.
\end{lemma}

\begin{proof}
Suppose that $(c_{231},c_{331}) = (0,0)$.
Then we can have $c_{233}=0$ while maintaining
$c_{231}=c_{331}=0$ by a unitary transformation on the first qutrit.
Then $P_6 \geq 0$ by Lemma~\ref{lemma16}.
Next, suppose $c_{213}=c_{313}=c_{223}=c_{323}=0$.  
First we use a unitary transformation on 
the first qutrit to let $c_{233}=0$.  Then a unitary transformation
on the third qutrit can be used to have $c_{231}=0$.  
Since $c_{213}=c_{223}=0$, by Proposition~\ref{prop6} we have
$$
|c_{331}|^2+|c_{333}|^2 \leq \lambda_3^{(3)} =
\lambda_3^{(3)}-|c_{213}|^2-|c_{223}|^2\,.
$$
We use this together with $|c_{211}|^2
+|c_{221}|^2\leq \lambda_2^{(2)}$
to have
\begin{eqnarray}
P_6 
& = & \lambda_1^{(1)}-\lambda_2^{(1)}+\lambda_2^{(2)}-\lambda_3^{(2)}
+ 2\lambda_2^{(3)}+\lambda_3^{(3)} \nonumber \\
& \geq & \lambda_1^{(1)}-\lambda_2^{(1)}
- \lambda_3^{(2)}+2\lambda_2^{(3)}\nonumber \\
&& + |c_{211}|^2+|c_{221}|^2 + 
|c_{331}|^2+|c_{333}|^2 + |c_{213}|^2+|c_{223}|^2\,.\nonumber
\end{eqnarray}
By substituting the expression of $\lambda_I^{(a)}$ in terms of
$|c_{ijk}|^2$ we find
$$
P_6  \geq  -2(|c_{231}|^2+|c_{233}|^2)=0\,.
$$
Thus, it is enough to show that if
$(c_{231},c_{331},c_{232},c_{332}) \neq (0,0,0,0)$, then
$(c_{223},c_{323}) = (c_{213},c_{313}) = (0,0)$.

If the matrix
\begin{equation}
\left( \begin{array}{cc} c_{231} & c_{331} \\ c_{232} & c_{332}
\end{array} \right)
\label{matmat}
\end{equation}
has rank less than two, then we can set $c_{231}=c_{331} = 0$ by
a unitary transformation on the third
qutrit.  Then a unitary transformation on
the first qutrit can be used to let $c_{233}=0$.  Then $P_6 \geq 0$.
Therefore we can assume that the matrix (\ref{matmat}) is regular.

First we use Proposition~\ref{prop9} after a qutrit permutation
$(123)\to (132)$ with $[ijk]=[213]$
and $[i'j'k']=[132]$ to show that $(c_{223},c_{323})=(0,0)$.
By considering the variation as given in Proposition
\ref{prop9} we find
\begin{eqnarray}
\delta P_6 & = & \delta \lambda_1^{(1)}-\delta \lambda_2^{(1)}
+ \delta \lambda_2^{(2)}-\delta\lambda_3^{(2)}
+ \delta\lambda_2^{(3)}-\delta\lambda_1^{(3)}\nonumber \\
& = & |a_2-b_1| -{\rm max}\,(a_1+a_2,b_1+b_2) 
-3\,{\rm min}\,(a_1+a_2,b_1+b_2)\,.\nonumber 
\end{eqnarray}
This can be made negative unless $a_1+a_2=0$ or $b_1+b_2=0$.  Since
the matrix in (\ref{matmat}) is regular, we can have $a_i \neq 0$ if
$b_i\neq 0$ and vice versa for both $i=1$ and $2$.  Thus, unless
$a_1=a_2=b_1=b_2=0$, we can have both $a_1+a_2$ and $b_1+b_2$ nonzero.
Hence $\delta P_6$ can be made negative.  Thus, unless
$(c_{223},c_{323})= (0,0)$, $P_6$ cannot have a minimum.
Next we change $k'$ from $2$ to $1$ to show that 
$(c_{213},c_{313})=(0,0)$.  The only difference here is that
we have $\delta \lambda_2^{(3)} = 0$ instead of
$\lambda_2^{(3)}=-a_1-a_2-b_1-b_2$.  Thus we find
$$
\delta P_6 
= |a_2-b_1| 
-2\,{\rm min}\,(a_1+a_2,b_1+b_2)\,.
$$
Again, this can be made negative unless $a_1+a_2=0$ or $b_1+b_2=0$.
Hence by the same argument as above we conclude that the function
$P_6$ cannot have a minimum unless $(c_{213},c_{313}) = (0,0)$.
\end{proof}

\begin{lemma}
Suppose that the minimum of the function $P_6$ is negative.
Then it does not occur at 
a state with
$\lambda_1^{(2)}=\lambda_2^{(2)}$ and 
$\lambda_1^{(3)} = \lambda_2^{(3)}$ and without any other 
degeneracy.
\end{lemma}

\begin{proof}
Assume that $c_{233}\neq 0$.  Then by lemmas in Sec.~\ref{perturb}
we find that the PNCs other than $c_{233}$ 
are $(c_{231},c_{232})$, $(c_{211},c_{212},c_{221},c_{222})$ and
$(c_{331},c_{332})$.
By a unitary transformation 
we can let $c_{212}=c_{221}=0$.  By 
orthogonality relations $c_{233}\overline{c_{23i}}=0$, $i=1,2$,
we have $c_{231}=c_{232}=0$.  Hence the PNCs are
$c_{233},c_{211},c_{222},c_{331}$ and $c_{332}$.
If $c_{332}=0$, then 
$|c_{331}|^2\geq |c_{222}|^2+|c_{233}|^2+|c_{211}|^2$ but
$|c_{233}|^2\geq |c_{331}|^2$ and $|c_{222}|^2\geq |c_{331}|^2$.
These imply that $c_{331}=c_{233}=c_{211}=c_{222}=0$,
which is a contradiction.  Hence $c_{332}\neq 0$ and by
an orthogonality relation we have $c_{331}=0$.
Then
$$
P_6 = -|c_{233}|^2-|c_{211}|^2+2|c_{222}|^2+ |c_{332}|^2 \geq 0
$$
because $|c_{332}|^2 \geq |c_{233}|^2+|c_{211}|^2+|c_{222}|^2$.

Next suppose that $c_{233}=0$. 
Suppose further that $(c_{231},c_{232})=(0,0)$.  Then, we can use
a unitary transformation on the third qutrit
to have $c_{331}=0$.  Then $P_6\geq 0$ by Lemma~\ref{lemma16}.
Hence we assume that $(c_{231},c_{232})\neq (0,0)$.
Let us use a unitary transformation to let $c_{232}=0$ for convenience.
Note that $L(2,3,1)=-2$ and $L(2,3,2)=0$.  Since $L(1,3,3)=1$,
we must have $c_{133}=0$ by Lemma~\ref{lemma3}.  
For $(c_{113},c_{123})$ to be nonzero, we must have
$L(1,1,3) < L(2,3,2)$, which is false, by Lemma~\ref{lemma5}.
Hence $(c_{113},c_{123})=(0,0)$.  Similarly, for 
$(c_{313},c_{323}) \neq (0,0)$ we must have
$L(3,1,3) < L(2,3,2)$, which is false.  Hence
$(c_{313},c_{323})=(0,0)$.
Thus, the PNCs of
the form $c_{ij3}$ are $c_{213}$, $c_{223}$ and $c_{333}$. (We have 
already shown that $c_{233}=0$.)

By applying Proposition~\ref{prop5} to the case with degeneracy in
the third qutrit,  we find that
$(c_{131},c_{132})$ and $(c_{331},c_{332})$ are mutually orthogonal and
are both orthogonal to $(c_{231},c_{232})$.  Since we have let
$c_{232}=0$, we obtain $c_{131}=c_{331}=0$ and either $c_{332}$ or
$c_{132}$ must vanish.  However, if $c_{332}=0$, then we have
$P_6 \geq 0$ by Lemma~\ref{lemmanice}.  Hence we can assume that
$c_{332}\neq 0$ and $c_{132}=0$.
Next, we use Proposition~\ref{prop9} after a qutrit permutation
$(123)\to (321)$ with
$[ijk]=[111]$ and $k'=2$.  Then we find
$$
\delta P_6  =  3(a_1+a_2+b_1+b_2) +2|a_2-b_1|
+ {\rm max}\,(a_1+a_2,b_1+b_2)\,. 
$$
By making all $a_i$ and $b_i$ nonpositive, we obtain 
$$
\delta P_6 \leq 3a_1+a_2+b_1 + 3b_2\,.
$$
The right-hand side can be made negative unless
$a_1=a_2=b_1=b_2=0$.
Hence $c_{111}=c_{112}=c_{121}=c_{122}=0$.

Next we apply Proposition~\ref{prop9} with 
$[ijk]=[113]$. The only difference is that
we have $\delta \lambda_1^{(1)} =0$ instead of
$\delta\lambda_1^{(1)}=a_1+a_2+b_1+b_2$.  Hence
$$
\delta P_6 =  2(a_1+a_2+b_1+b_2) + 2|a_2-b_1| 
+ {\rm max}\,(a_1+a_2,b_1+b_2)\,. 
$$
If $a_1$ or $b_2$ is nonzero, then $\delta P_6$ can be made
negative by making the nonzero one large and negative.
If $a_2$ and $b_1$ are both nonzero, then we can let $a_2=b_1 < 0$
$a_1=b_2=0$ to make $\delta P_6 < 0$.  If $\delta P_6 \geq 0$ and
if $c_{231}\neq 0$ and $c_{232}=0$, then the PNCs 
among $c_{3jk}$, $j,k=1,2$, 
is $c_{322}$ {\em after using a unitary transformation in the second
qutrit}.  If we undo this transformation, we have $c_{312}\neq 0$
as well.

Next let $[ijk]=[112]$ in Proposition~\ref{prop9}.  
Then $\delta \lambda_1^{(1)}=\delta \lambda_2^{(1)} =0$
and all other variations of the eigenvalues are the same.  Hence
$$
\delta P_6  =  a_1+a_2+b_1+b_2 + 2|a_2-b_1| 
+ {\rm max}\,(a_1+a_2,b_1+b_2)\,.
$$
If $a_1$ or $b_2$ is nonzero, then
by letting the nonzero one be negative and
letting $a_2=b_1=0$ we have
$$
\delta P_6 \leq a_1 + b_2 < 0\,.
$$
If $a_1=b_2=0$, then
$$
\delta P_6 = a_2+b_1 + 2|a_2-b_1| + {\rm max}\,(a_2,b_1)\,.
$$
If both $a_2$ and $b_1$ are nonzero, then by letting $a_2=b_1$ and
letting them be negative, we find
$$
\delta P_6 < a_2+b_1 < 0\,.
$$
By the same argument as in the $k=3$ case, we have only
$c_{212}$ and $c_{222}$ among $c_{2jk}$, $j,k=1,2$,
possibly nonzero if $\delta P_6 \geq 0$.  
Here we can use
a unitary transformation 
on the second qutrit to let $c_{212}=0$.

Thus, the following coefficients are the PNCs:
$$
c_{332},c_{322},c_{312},c_{222},c_{213},c_{223},c_{333},c_{231}.
$$
Recall that $c_{332}\neq 0$.  Hence, by 
orthogonality relations we have $c_{322}=c_{312}=0$.  We may assume
that $c_{213}\neq 0$ because otherwise
$\lambda_1^{(2)}=\lambda_2^{(2)}=0$ and  $P_6\geq 0$ by 
Lemma~\ref{lemma2}.  Then by the orthogonality relation 
$c_{213}\overline{c_{223}} = 0$, we have $c_{223}=0$.  From
$c_{332}\overline{c_{333}}=0$ we have $c_{333}=0$. Then
$c_{233}=c_{333}=0$, and by Lemma~\ref{lemma213} we have $P_6\geq 0$.
\end{proof}

Now we turn our attention to inequality (\ref{ineq7}), i.e.
$P_7 \geq 0$.

\begin{lemma}
The function $P_7$ is nonnegative if $\lambda_1^{(1)}=\lambda_2^{(1)}$,
$\lambda_1^{(2)}=\lambda_2^{(2)}$ or $\lambda_2^{(2)}=\lambda_3^{(2)}$.
\end{lemma}

\begin{proof}
If $\lambda_1^{(1)} =\lambda_2^{(1)}$, then $P_7\geq 0$ is
equivalent to
$$
2\lambda_1^{(2)}+\lambda_2^{(2)}+2\lambda_3^{(3)}+\lambda_2^{(3)}
\geq 2\lambda_1^{(1)}+\lambda_3^{(1)}\,.
$$
Since $\lambda_3^{(3)}\geq 1/3\geq \lambda_1^{(1)}$ and 
$\lambda_1^{(2)}\geq 0$, it is enough to show that
$$
\lambda_1^{(2)}+\lambda_2^{(2)}+\lambda_3^{(3)}+\lambda_2^{(3)}
\geq \lambda_1^{(1)}+\lambda_3^{(1)}\,.
$$
This follows from inequality (\ref{ineq3}) since 
$\lambda_2^{(1)}\geq \lambda_1^{(1)}$.
If $\lambda_1^{(2)}=\lambda_2^{(2)}$, then $P_7 \geq 0$ is equivalent
to
$$
2\lambda_2^{(2)}+\lambda_1^{(2)}+2\lambda_3^{(3)}+\lambda_2^{(3)}
\geq 2\lambda_2^{(1)}+\lambda_3^{(1)}\,,
$$
which follows from inequality (\ref{ineq6}).  Finally, if 
$\lambda_2^{(2)}=\lambda_3^{(2)}$, then 
$\lambda_1^{(2)} = 1-2\lambda_2^{(2)}$.
Hence 
\begin{eqnarray}
P_7 & = & 2-3\lambda_2^{(2)}+2\lambda_3^{(3)}+\lambda_2^{(3)}
-2\lambda_2^{(1)}-\lambda_3^{(1)}\nonumber \\
& = &  
2-3\lambda_2^{(2)} + \lambda_3^{(3)}-\lambda_1^{(3)}
- \lambda_2^{(1)} + \lambda_1^{(1)}\,. \nonumber
\end{eqnarray}
Since $\lambda_2^{(2)}$ and $\lambda_2^{(1)}$ are both less than
or equal to $1/2$, we have $P_7 \geq 0$.
\end{proof}

\noindent
Thus, the only degeneracies we need to consider are
$\lambda_2^{(1)}=\lambda_3^{(1)}$, $\lambda_1^{(3)}=\lambda_2^{(3)}$
and $\lambda_2^{(3)}=\lambda_3^{(3)}$.
\begin{lemma}
If $c_{231}=c_{221}=c_{331}=0$, then $P_7\geq 0$.
\label{lemma21}
\end{lemma}

\begin{proof}
Writing $P_7$ as
$$
P_7 = \lambda_1^{(1)}-\lambda_2^{(1)}
+ \lambda_1^{(2)}-\lambda_3^{(2)}
+ 2\lambda_3^{(3)}+\lambda_2^{(3)}
$$
and noting that $|c_{232}|^2+|c_{233}|^2 \leq \lambda_3^{(3)}$, which 
follows from Proposition~\ref{prop6} and the remark following it, 
we have
$$
P_7 
=  \lambda_2^{(3)}+\lambda_3^{(3)} + \lambda_1^{(1)}-\lambda_2^{(1)}
+ \lambda_1^{(2)}-\lambda_3^{(2)}+ |c_{232}|^2+|c_{233}|^2\,.
$$
Then, by expressing the eigenvalues in terms of $|c_{ijk}|^2$ we find
that the negative terms are $-2|c_{231}|^2-|c_{221}|^2-|c_{331}|^2$.
Hence,
if $c_{231}=c_{221}=c_{331}=0$, then $P_7 \geq 0$.
\end{proof}

\begin{lemma}
If $\lambda_2^{(1)}=\lambda_3^{(1)}$ and
$\lambda_1^{(3)}=\lambda_2^{(3)}=\lambda_3^{(3)}$, then
$P_7 \geq 0$.
\end{lemma}

\begin{proof}
Consider the vectors $(c_{231},c_{232},c_{233})$ and
$(c_{331},c_{332},c_{333})$.
We can use a unitary
transformation on the third qutrit to make $c_{231}$ and $c_{331}$
both vanish.  Then
we use a unitary transformation 
on the first qutrit to let
$c_{221}=0$.  Note that the condition $(c_{231},c_{331})=(0,0)$
is preserved.  Hence,
$c_{231}=c_{331}=c_{221}=0$ and $P_7\geq 0$ by the previous lemma.
\end{proof}

\begin{lemma}
Suppose that the minimum of the function $P_7$ is negative.
Then it does not occur at a state satisfying any of the
following conditions:
\begin{enumerate}
\item $\lambda_2^{(1)} = \lambda_3^{(1)}$ with no other 
degeneracy;
\item $\lambda_1^{(3)} =\lambda_2^{(3)}$ with no other
degeneracy;
\item $\lambda_2^{(3)} = \lambda_3^{(3)}$ with no other
degeneracy.
\end{enumerate}
\end{lemma}

\begin{proof}
Case 1.  If $(c_{231},c_{331})=(0,0)$, then a
unitary transformation on the first qutrit can be used
to have $c_{221}=0$.  Then we have $P_7\geq 0$ by Lemma~\ref{lemma21}.
Next let $(c_{231},c_{331})\neq (0,0)$.  By Lemma~\ref{lemma8}
we have $\lambda_1^{(1)}=0$ or $\lambda_1^{(2)}=\lambda_3^{(3)}=0$.
The latter is impossible.  Hence $\lambda_1^{(1)}=0$.
Then $\lambda_2^{(1)}=\lambda_3^{(1)}=1/2$.
Use a
unitary transformation to let $c_{331}=0$.
Then, since all $(c_{2jk},c_{3jk})$ must be orthogonal to
$(c_{231},c_{331})$ for 
$\left[jk\right]\neq \left[31\right]$ by Proposition 
\ref{prop5}, we have $c_{2jk}=0$ except for $c_{231}$.
This implies that $|c_{231}|^2=\lambda_2^{(1)}=1/2$.  Hence
$\lambda_1^{(3)}\geq 1/2$.  This is impossible.  

Case 2. Assume that $(c_{231},c_{232})\neq (0,0)$.  If
$c_{ij3}\neq 0$ for some $[ij]$, then we must have
$L(i,j,3) \leq L(2,3,2)=-1$.  This is impossible because $r_3=2$.
If $c_{ij3}=0$ for all 
$\left[ij\right]$, then $\lambda_3^{(3)}=0$, which is
impossible.  Hence $c_{231}=c_{232}=0$.
If $c_{ij3}\neq 0$ and if $(c_{221},c_{222})$ or $(c_{331},c_{332})$
is not $(0,0)$,  then we must have
$L(i,j,3) \leq L(2,2,2) =L(3,3,2)=0$.  
Since $r_3=2$, we must have
$p_i=-2$, $q_j=0$, i.e. $\left[ij\right]=\left[23\right]$.  Thus, if
$(c_{221},c_{222})$ or $(c_{331},c_{332})$ is not $(0,0)$, then
the only nonzero coefficient of the form $c_{ij3}$ is $c_{233}$.
If the vectors $(c_{221},c_{222})$ and $(c_{331},c_{332})$ are
linearly dependent, then a unitary transformation can be used
to have $c_{221}=c_{331}=0$.  Then by
Lemma~\ref{lemma21} we have $P_7\geq 0$.
If they are linearly independent,
by Proposition~\ref{prop5} applied to the case with
degeneracy in the third qutrit we conclude that
all the other vectors $(c_{ij1},c_{ij2})$ must vanish
since $\left[ij\right]=\left[22\right]$ and 
$\left[33\right]$ are the only pairs with 
$R^{(3)}(i,j)\equiv p_i+q_j = -1$.
Then the PNCs are
$$
c_{221},c_{222},c_{331},c_{332},c_{233}.
$$
We can let $c_{221}=0$ by a unitary transformation on the
third qutrit.
Then by an orthogonality relation we find
$c_{332}=0$. [Note that the vectors $(c_{221},c_{222})$ and
$(c_{331},c_{332})$ are linearly independent.]
Hence,
$c_{222}$, $c_{233}$ and $c_{331}$ are the PNCs.
Then $|c_{331}|^2=\lambda_1^{(3)}=\lambda_3^{(1)}=1/3$ and
$\lambda_2^{(1)}=|c_{222}|^2+|c_{233}|^2=2/3$.  This is impossible.
Hence, we must have $c_{231}=c_{221}=c_{331}=0$ and $P_7\geq 0$
by Lemma~\ref{lemma21}.

Case 3. 
Assume that $c_{231}\neq 0$.  If $(c_{ij2},c_{ij3})\neq (0,0)$,
then $L(i,j,2)\leq L(2,3,1)=-2$.  This is impossible.  Since
$c_{ij3}=0$ for all $\left[ij\right]$ would imply that $\lambda_3^{(3)}=0$,
we conclude that $c_{231}=0$. If $c_{221}$ or $c_{331}$ is nonzero
and if $(c_{ij2},c_{ij3})\neq (0,0)$, then we must have
$L(i,j,2) \leq L(2,2,1)=L(3,3,1)=-1$.  This is satisfied only
by $(c_{232},c_{233})$.  Then, by an orthogonality relation we have
$c_{232}=0$ or $c_{233}=0$.  Hence, either $\lambda_2^{(3)}=0$, which
yields $P_7\geq 0$ by Lemma~\ref{lemma2}, or $\lambda_3^{(3)}=0$, 
which is impossible. Thus, $c_{231}=c_{221}=c_{331}=0$
and $P_7\geq 0$ by Lemma~\ref{lemma21}.
\end{proof}

\begin{lemma}
Suppose that the minimum of the function $P_7$ is negative.  
Then it does not
occur at a state with
$\lambda_1^{(3)}=\lambda_2^{(3)}=\lambda_3^{(3)}$ and without any
other degeneracy.
\end{lemma}

\begin{proof}
First we establish the following result, which is similar to
Proposition~\ref{prop5}: if $(c_{ij1},c_{ij2},c_{ij3})$ and
$(c_{i'j'1},c_{i'j'2},c_{i'j'3})$ are nonzero, then
either $R^{(3)}(i,j)=R^{(3)}(i'j')$ or these vectors are orthogonal
to each other.  This can be shown as follows.  Consider the variation
$$
\delta c_{ijK} = \alpha c_{i'j'K}\,,\ \ 
\delta c_{i'j'K} = -\overline{\alpha}c_{ijK}\ \ {\rm for\ all}\ K
$$
and $\delta c_{IJK}=0$ for all other coefficients.
We note first that the variation of the third RDM vanishes,
The variation of $P_7$ is found to be
$$
\delta P_7 = 2[R^{(3)}(i,j)-R^{(3)}(i',j')]
\sum_{k=1}^3 {\rm Re}\,(\alpha c_{i'j'k}\overline{c_{ijk}})\,.
$$
Hence, if $\delta P_7 \geq 0$ for all $\alpha$, we must have
either $R^{(3)}(i,j)=R^{(3)}(i',j')$ or
$\sum_{k=1}^3 c_{i'j'k}\overline{c_{ijk}} = 0$.
We note that $R^{(3)}(2,3)=-2$, $R^{(3)}(2,2)=R^{(3)}(3,3)=-1$,
and $R^{(3)}(i,j)\geq 0$ for all other $\left[ij\right]$.

If $c_{23k}=0$ for all $k$, then we can have $c_{221}=c_{331}=0$
by a unitary transformation on the third qutrit and have $P_7\geq 0$
by Lemma~\ref{lemma21}.  Assume that
$(c_{233},c_{232},c_{231})$ is nonzero.  Let
$c_{232}=c_{231}=0$ by a unitary transformation on the third qutrit.  
Then, if $c_{33k}=0$ for all $k$, we can
use a third-qutrit unitary transformation between $|1\>$ and $|2\>$
to have $c_{221}=0$ while
keeping $c_{232}=c_{231}=c_{331}=0$.  Then we have $P_7 \geq 0$
by Lemma~\ref{lemma21}.  So we assume that $c_{33k}\neq 0$ for some
$k$.  By orthogonality of $(c_{231},c_{232},c_{233})$ and
$(c_{331},c_{332},c_{333})$ and by the condition
$c_{231}=c_{232}=0$ we have $c_{333}=0$.  We choose
$c_{332}\neq 0$ and $c_{331}=0$.  If $c_{221}=0$ for the vector
$(c_{221},c_{222},c_{223})$, then 
$P_7\geq 0$ by Lemma~\ref{lemma21}.  So let $c_{221}\neq 0$.   Then,
all the other vectors $(c_{ij1},c_{ij2},c_{ij3})$ are orthogonal to
three linearly independent vectors.  
Hence they must all vanish. Then, 
the PNCs are
$c_{233}$, $c_{332}$ and $c_{221}$
since $c_{222}=0$ due to the relation
$c_{222}\overline{c_{221}} = 0$.
Then, $|c_{332}|^2\geq |c_{233}|^2+|c_{221}|^2$ and
$|c_{233}|^2\geq |c_{332}|^2$ 
imply that $|c_{332}|^2=|c_{233}|^2$, $c_{221}=0$.
Hence $\lambda_1^{(3)}=0$, which contradicts the degeneracy assumption.
\end{proof}

\noindent
The following lemma completes the proof of inequality (\ref{ineq7}).
\begin{lemma}
Suppose that the minimum of $P_7$ is negative.  Then
it does not occur in either of the following
two cases:
\begin{enumerate}
\item $\lambda_2^{(1)}=\lambda_3^{(1)}$ and
$\lambda_2^{(3)}=\lambda_3^{(3)}$ with no
other degeneracy;
\item $\lambda_2^{(1)}=\lambda_3^{(1)}$ and
$\lambda_1^{(3)}=\lambda_2^{(3)}$ with no
other degeneracy.
\end{enumerate}
\end{lemma}

\begin{proof}
Case 1.  If $(c_{231},c_{331})=(0,0)$, then
we can use a unitary transformation on the
first qutrit to have $c_{221}=0$, and use
Lemma~\ref{lemma21} to conclude that $P_7\geq 0$.
Suppose $(c_{231},c_{331})\neq (0,0)$.  
If $(c_{1j2},c_{1j3})\neq (0,0)$, then we must have
$L(1,j,2)\leq L(3,3,1)=-1$, which is impossible.  Hence 
$(c_{1j2},c_{1j3})=(0,0)$ for all $j$. Next let
$(c_{2j2},c_{3j2},c_{2j3},c_{3j3})\neq (0,0,0,0)$.  Then by 
considering the variation given in Proposition~\ref{prop9} after
a qutrit permutation
$(123)\to (132)$ with $[ijk]\rightarrow [22j]$ and $j'=1$, we find
\begin{eqnarray}
\delta P_7 & = & \delta \lambda_1^{(1)}-\delta \lambda_2^{(1)}
+ \delta \lambda_1^{(2)}-\delta\lambda_3^{(2)}
+\delta \lambda_3^{(3)}-\delta\lambda_1^{(3)}\nonumber \\
& = & |a_2-b_1|+{\rm max}\,(a_1+a_2,b_1+b_2)
+ (4-j)(a_1+a_2+b_1+b_2)\,.
\label{identic}
\end{eqnarray}
If $j=1$ or $2$, then by choosing $a_i$ and $b_i$ to be nonpositive,
we have
$$
\delta P_7 \leq |a_2-b_1| + 2(a_1+a_2+b_1+b_2) < 0
$$
if some of $a_i$ and $b_i$ is nonzero.  For $j=3$, if $a_i$
and $b_i$ are nonpositive, then
$$
\delta P_7 = |a_2-b_1| + 2\,{\rm max}\,(a_1+a_2,b_1+b_2)
+ {\rm min}\,(a_1+a_2,b_1+b_2)\,.
$$
If $a_1\neq 0$ or $b_2\neq 0$ or if both $a_2$ and $b_1$ are nonzero,
then we can have $\delta P_7 < 0$.  If $a_2$ is the only nonzero
coefficient, then we find
$\delta P_7 = |a_2| + a_2 = 0$, and similarly for $b_1$.  The cases
$a_2\neq 0$ and $b_1\neq 0$ are related by unitary transformations.
By choosing the latter, we find that
$c_{233}=c_{332}=c_{333}=0$.  Since we have already concluded that
there are no coefficients of the form $c_{ij3}$ other than
$c_{233}$ and $c_{333}$, we have
$\lambda_3^{(3)}=0$, which is a contradiction.

Case 2.  It is enough to show that $(c_{231},c_{331})=(0,0)$
for the same reason as in case 1.  Assume that
\begin{equation}
\left( \begin{array}{cc}
c_{231} & c_{331} \\ c_{232} & c_{332} \end{array}
\right)
\neq
\left( \begin{array}{cc}
0 & 0 \\ 0 & 0 \end{array}\right)\,.
\label{mattt}
\end{equation}
If this matrix has rank one, then we can use a unitary transformation
on the third qutrit 
to have $c_{231}=c_{331}=0$.  Then, we can have $c_{221}=0$ using
a unitary transformation on the first
qutrit and
conclude that $P_7 \geq 0$ by Lemma~\ref{lemma21}.  Therefore we
may assume that the matrix (\ref{mattt}) has rank two.
Suppose further that $c_{1j3} \neq 0$.  Then we must have
$L(1,j,3)\leq L(3,3,2) = 0$, which is impossible.  Hence
$c_{1j3}=0$ for all $j$.  Next, suppose that
$(c_{2j3},c_{3j3})\neq (0,0)$.  Then by letting $(abc)=(132)$ and
$[ijk]\rightarrow [213]$ and $j'=3$, $k'\to j$ in 
Proposition~\ref{prop9},  we find
$$
\delta P_7 = |a_2-b_1|-{\rm min}\,(a_1+a_2,b_1+b_2)
- (4-j)(a_1+a_2+b_1+b_2)\,.
$$
This formula will be identical with (\ref{identic}) if we let
$a_i \to -a_i$ and $b_i \to -b_i$.  Hence $\delta P_7$ can be made
negative if $(c_{213},c_{313})$ or $(c_{223},c_{323})$ is nonzero.  
For $j=3$, then we can have $\delta P_7 < 0$
unless $a_2$ or $b_1$ is the only possibly nonzero one among
$a_i$ and $b_i$.  Since the matrix (\ref{mattt}) has rank two,
we must have both $c_{231}$ and $c_{332}$ nonzero after diagonalizing
it.  Hence, if $a_i$
can be made nonzero, then so can $b_i$, and vice versa.  Hence
we can conclude that unless $a_i$ and $b_i$, $i=1,2$, 
are all zero, we can
have $\delta P_7 < 0$.  Hence we must have $c_{ij3}=0$ for all 
$\left[ij\right]$,
i.e. $\lambda_3^{(3)} = 0$, which is a contradiction.
\end{proof}

\section{Construction of some states}
\label{constr1}

We have shown that the inequalities listed in Theorem~\ref{main}
are necessary for a state to exist with a given E-point.
Our next task is to show
that these inequalities guarantee that there is a quantum 
state having these eigenvalues.
The following elementary fact will be useful.
\begin{lemma}
Suppose that the angles $\theta$ and $\varphi$ are constrained by
$$
u\sin 2\theta - v\sin 2\varphi = 0\,, 
$$
where $u$ and $v$ are real constants,
and that there are no other constraints on them.  Then, 
$\theta + \varphi$ can take any value and the range of
the function
$$
f \equiv u \cos 2 \theta + v\cos 2\varphi
$$
is given by $||u| - |v|| \leq |f| \leq |u| + |v|$. 
\label{obvious}
\end{lemma}

\begin{proof}
The constraint equation can be written as
$$
(u-v)\sin(\theta+\varphi)\cos(\theta-\varphi)
+ (u+v)\cos(\theta+\varphi)\sin(\theta-\varphi)=0\,.
$$
For any value of $\theta+\varphi$, we can find a value of
$\theta-\varphi$ so that this equation is satisfied.  Hence,
the combination $\theta+\varphi$ is unconstrained.  Note that
\begin{eqnarray}
f^2 & = & (u\cos 2\theta + v\cos 2\varphi)^2 
+ (u\sin 2\theta - v\sin 2\varphi)^2 \nonumber \\
& = & u^2 + v^2 + 2uv\cos 2(\theta + \varphi)\,.\nonumber
\end{eqnarray}
This shows that $|f|$ ranges between $(u+v)^2$ and $(u-v)^2$.  Hence, the
range of $f$ is given by $||u|-|v||\leq |f|\leq |u|+|v|$.
(Since $f\to -f$ for $\theta \to \theta + \pi/2$ and
$\varphi + \pi/2$, the range is invariant under $f\to -f$.)
\end{proof}

\noindent
First we construct
a convex subset of $S$.
\begin{proposition}
There is a state with any E-point in the convex
set $S_0$ bounded by hyperplanes with the following corner points:
\begin{eqnarray}
&& \left[B,B,B\right],\ \left[B,A,A\right],\ \left[A,B,B\right],\
\left[O,B,B\right],\ \left[A,A,A\right],\nonumber \\
&& 
\left[O,A,A\right],\ \left[O,O,O\right],\ 
\left[A,B,\sixth\sixth\twthirds\right],\
\left[A,\quarter\quarter\half,0\quarter\thquarters\right] \nonumber
\end{eqnarray}
and those obtained by qutrit permutations from them.
\label{prop10}
\end{proposition}

\begin{proof}
First consider the following PNCs:
$c_{333},c_{322},c_{232},c_{223},c_{311},c_{131},c_{113}$.
Suppose that we add $c_{122}$, $c_{212}$ and $c_{221}$ to 
this list.  Then the only
off-diagonal elements in the RDMs
that become nonzero are the ones connecting $|1\>$ and $|3\>$ in
each qutrit.  Thus, each RDM can be
diagonalized by a simple unitary transformation.  Motivated by
this observation, we consider the state with
the PNCs given as follows.  First let
\begin{eqnarray}
&& \left( \begin{array}{c} b_{333} \\ b_{133} \end{array}\right)
= \left( \begin{array}{c} a\cos\theta_1 \\  a\sin\theta_1
\end{array}\right)\,,\ \ 
\left( \begin{array}{c} b_{331} \\ b_{131} \end{array}\right)
= \left( \begin{array}{c} -g\sin\theta_1 \\ g\cos\theta_1
\end{array}\right),\nonumber \\
&& \left( \begin{array}{c} b_{313} \\ b_{113} \end{array}\right)
= \left( \begin{array}{c} -h\sin\theta_1 \\ h\cos\theta_1
\end{array}\right)\,,\ \ 
\left( \begin{array}{c} b_{311} \\ b_{111} \end{array}\right)
= \left( \begin{array}{c} f\cos\theta_1 \\ f\sin\theta_1
\end{array}\right)\,, \nonumber
\end{eqnarray}
where $a$, $h$, $g$ and $f$ are real constants.  Next we define
$$
\left(\begin{array}{c} d_{i3k} \\ d_{i1k} \end{array}\right)
= \left( \begin{array}{cc} \cos\varphi_1 & -\sin\varphi_1 \\
\sin\varphi_1 & \cos\varphi_1 \end{array}\right)
\left( \begin{array}{c} b_{i3k} \\ b_{i1k} \end{array}\right)\,,
$$
where $\left[ik\right] = \left[11\right]$, $\left[13\right]$, 
$\left[31\right]$ or $\left[33\right]$.  Finally we let
$$
\left(\begin{array}{c} c_{ij3} \\ c_{ij1} \end{array}\right)
= \left( \begin{array}{cc} \cos\chi_1 & -\sin\chi_1 \\
\sin\chi_1 & \cos\chi_1 \end{array}\right)
\left( \begin{array}{c} d_{ij3} \\ d_{ij1} \end{array}\right)
$$
with $\left[ij\right]=\left[11\right]$, $\left[13\right]$, 
$\left[31\right]$ or $\left[33\right]$.
We also let
\begin{eqnarray}
&& c_{322} = b\cos\theta_2\,,\ \ c_{122}=b\sin\theta_2\,,\ \
c_{232}=c\cos\varphi_2\,,\ \ c_{212}=c\sin\varphi_2\,,\nonumber \\
&& c_{223} = d\cos\chi_2\,,\ \ c_{221} = d\sin\chi_2\,, \nonumber
\end{eqnarray}
where $b$, $c$ and $d$ are real constants.
Then we find $a^2 + b^2 + c^2 + d^2 + f^2 + g^2 + h^2 = 1$ and
\begin{eqnarray}
\lambda_2^{(1)} & = & c^2 + d^2\,, \nonumber \\
\lambda_2^{(2)} & = & b^2 + d^2\,, \nonumber \\
\lambda_2^{(3)} & = & b^2 + c^2\,, \nonumber
\end{eqnarray}
and
\begin{eqnarray}
\lambda_3^{(1)}-\lambda_1^{(1)} & = & 
b^2 \cos 2\theta_2 +(a^2+f^2-g^2-h^2)\cos 2\theta_1\,, \nonumber \\
\lambda_3^{(2)}-\lambda_1^{(2)} & = & 
c^2\cos 2\varphi_2 + (a^2+g^2-f^2-h^2)\cos 2\varphi_1\,,\nonumber \\
\lambda_3^{(3)}-\lambda_1^{(3)} & = & 
d^2\cos 2\chi_2 + (a^2+h^2-f^2-g^2)\cos 2\chi_1\,. \nonumber
\end{eqnarray}
The orthogonality relations are
\begin{eqnarray}
b^2\sin 2\theta_2 + (a^2+f^2-g^2-h^2)\sin 2\theta_1 & = & 0\,,
\nonumber \\
c^2 \sin 2\varphi_2 + (a^2+g^2-f^2-h^2)\sin 2\varphi_1 & = & 0\,,
\nonumber \\
d^2 \sin 2\chi_2 + (a^2+h^2-f^2-g^2)\sin 2\chi_1 & = & 0\,. \nonumber
\end{eqnarray}
We require that
\begin{eqnarray}
a^2 + f^2 - g^2 - h^2 & \geq & 0\,, \nonumber \\
a^2 + g^2 - f^2 - h^2 & \geq & 0\,,\nonumber \\
a^2 + h^2 - f^2 - g^2 & \geq & 0\,.\nonumber
\end{eqnarray}
Then, by Lemma~\ref{obvious} we conclude that
$\lambda_3^{(a)} - \lambda_1^{(a)}$ have the ranges given by the
following inequalities:
\begin{eqnarray}
&& |a^2+f^2-g^2-h^2 - b^2|\leq \lambda_3^{(1)}-\lambda_1^{(1)}
\leq a^2 + f^2 - g^2 - h^2 + b^2\,, \label{lam311} \\  
&& |a^2+ g^2 - f^2 -h^2 -c^2| \leq \lambda_3^{(2)}-\lambda_1^{(2)}
\leq a^2 + g^2 -f^2 -h^2 + c^2\,,\label{lam312}\\ 
&& |a^2+h^2 -f^2-g^2-d^2| \leq \lambda_3^{(3)}-\lambda_1^{(3)}
\leq a^2 + h^2 - f^2-g^2 + d^2\,. \label{lam313}
\end{eqnarray}
The equations for $\lambda_2^{(a)}$ and the inequalities for
$\lambda_3^{(a)}-\lambda_1^{(a)}$ can be satisfied for all corner
points listed here.
We give $a^2$, $b^2$ etc. that are nonzero 
for each corner point:
\begin{eqnarray}
\left[B,B,B\right] & : &
a^2=1/4,\ b^2=c^2=d^2=1/6,\ f^2=g^2=h^2=1/12,\nonumber \\
\left[B,A,A\right] & : &
a^2=b^2=1/3,\ c^2=d^2=1/6,\nonumber \\
\left[A,B,B\right] & : &
a^2=1/3,\ b^2=1/12,\ c^2=d^2=1/4,\  f^2=1/12, \nonumber \\
\left[O,B,B\right] & : &
a^2=b^2=f^2=1/3,\nonumber \\
\left[A,A,A\right] & : &
a^2=b^2=c^2=d^2=1/4,\nonumber \\
\left[O,A,A\right] & : &
a^2=b^2=1/2,\nonumber \\
\left[O,O,O\right] & : &
a^2=1,\nonumber\\
\left[A,B,\sixth\sixth\twthirds\right] & : &
a^2=d^2=1/3,\ c^2=f^2=1/6,\nonumber \\
\left[A,0\quarter\thquarters,\quarter\quarter\half\right] & : &
a^2=1/2,\ c^2=d^2=1/4. \nonumber
\end{eqnarray}
The E-points obtained by qutrit permutations can be realized
because this construction is invariant under qutrit permutations.
If two sets of eigenvalues $\lambda_i^{(a)I}$ and
$\lambda_i^{(a)II}$ are given by $a^2 = a_I^2$, $b^2=b_I^2$, etc. and
$a^2=a_{II}^2$, $b^2=b_{II}^2$, 
etc., respectively, then any set of eigenvalues of the form 
$\lambda_i^{(a)} = \alpha \lambda_i^{(a)I} 
+ (1-\alpha)\lambda_i^{(a)II}$ with
$0\leq \alpha \leq 1$ is given by
$a^2= \alpha a_I^2 + (1-\alpha)a_{II}^2$,
$b^2 = \alpha b_I^2 + (1-\alpha)b_{II}^2$, etc. with inequalities
(\ref{lam311})--(\ref{lam313}) being satisfied.  Hence the
set of states constructed here forms a convex set in the space of
E-points.
\end{proof}

It will be useful later 
to construct states whose E-points are on
the boundary of inequality (\ref{ineq1}).  This inequality with
$(abc)=(123)$ reads
$$
P_1 \equiv 
\lambda_1^{(2)}+\lambda_2^{(2)}+\lambda_1^{(3)}+\lambda_2^{(3)}
- \lambda_1^{(1)}-\lambda_2^{(1)}\geq 0\,. 
$$
{}From the proof of Theorem~\ref{general} we find that
to have $P_1=0$ we need $c_{ijk} = 0$ for all $[ijk]$ except
those with $i=3$ or $j=3$.
We also find that unless
$\lambda_2^{(1)}=\lambda_3^{(1)}$, we must have
$c_{332}=c_{331}=0$.  Thus, a generic state satisfying 
the equality $P_1=0$ has only the following PNCs:
\begin{equation}
c_{131},c_{231}, c_{132},c_{232},c_{113},c_{213},c_{123},c_{223},
c_{333}.
\label{cs}
\end{equation}
Notice that for all these coefficients $c_{ijk}$ we have
$L(i,j,k)=0$.
The orthogonality relations are as follows:
\begin{eqnarray}
c_{131}\overline{c_{231}} + c_{132}\overline{c_{232}}
+ c_{113}\overline{c_{213}} + c_{123}\overline{c_{223}} & = & 0\,,
\label{const1} \\
c_{113}\overline{c_{123}} + c_{213}\overline{c_{223}} & = & 0\,, \\
c_{131}\overline{c_{132}}+c_{231}\overline{c_{232}} & = & 0\,.
\end{eqnarray}
The last two constraints are satisfied by letting
\begin{eqnarray}
&& c_{113}= a\cos\theta,\ \ c_{213} = a\sin\theta,\ \
c_{123} = -b\sin\theta,\ \ c_{223} = b\cos\theta, \nonumber \\
&& c_{131} = d\cos\varphi,\ \ c_{231} = -d\sin\varphi,\ \
c_{132} = f\sin\varphi,\ \ c_{232} = f\cos\varphi, \nonumber
\end{eqnarray}
where $a$, $b$, $d$, $f$, $theta$ and $\varphi$ are real numbers.
Let us write the eigenvalues $\lambda_i^{(a)}$ as
$\Lambda_i^{(a)}$ for later purposes.  Then
$$
\Lambda_1^{(2)}=a^2,\ \
\Lambda_2^{(2)}=b^2,\ \
\Lambda_1^{(3)}=d^2,\ \
\Lambda_2^{(3)}=f^2
$$
and
\begin{eqnarray}
\Lambda_2^{(1)}+\Lambda_1^{(1)} & = & 
\Lambda_2^{(2)}+\Lambda_1^{(2)} + \Lambda_2^{(3)}+\Lambda_1^{(3)}\,,
\label{border} \\
\Lambda_2^{(1)}-\Lambda_1^{(1)}
& = & (\Lambda_2^{(2)}-\Lambda_1^{(2)})\cos 2\theta
+ (\Lambda_2^{(3)}-\Lambda_1^{(3)})\cos 2\varphi\,. 
\end{eqnarray}
The constraint (\ref{const1}) reads
$$
(\Lambda_2^{(2)}-\Lambda_1^{(2)})\sin 2\theta
= (\Lambda_2^{(3)}-\Lambda_1^{(3)})\sin 2\varphi\,. 
$$
Hence, by Lemma~\ref{obvious} the range
of $\Lambda_2^{(1)}-\Lambda_1^{(1)}$ is given by
\begin{equation}
|(\Lambda_2^{(2)}-\Lambda_1^{(2)})-(\Lambda_2^{(3)}-\Lambda_1^{(3)})|
\leq \Lambda_2^{(1)}-\Lambda_1^{(1)}
\leq \Lambda_2^{(2)}-\Lambda_1^{(2)}+\Lambda_2^{(3)}-\Lambda_1^{(3)}
\,.
\label{between}
\end{equation}
In summary, there is a state with any E-point satisfying
(\ref{border}) and (\ref{between}), and
$\Lambda_2^{(a)} \geq \Lambda_1^{(a)}$ for all $a$.  
Notice that it is not
necessary for the eigenvalues to satisfy
$\Lambda_3^{(a)} \geq \Lambda_2^{(a)}$, or even
$\Lambda_3^{(a)}\geq \Lambda_1^{(a)}$.
It is clear that the states constructed here form a convex set
in the space of E-points.  It can readily be seen that the
corner points of this boundary hyperplane listed in
the proof of Proposition~\ref{prop111} all satisfy 
(\ref{between}) and, of course, (\ref{border}).  Hence
all E-points on the boundary hyperplane (\ref{border})
have corresponding states.
This construction of states can be slightly generalized
to prove the following lemma, which will be useful later.
\begin{lemma}
There is a state with any E-point in the simplex with
the following corner points:
$$
\left[O,O,O\right]\,\ 
\left[B,O,B\right]\,\ 
\left[B,B,O\right]\,\ 
\left[A,O,A\right]\,\ 
\left[A,A,O\right]\,\
\left[B,0\third\twthirds,0\third\twthirds\right],\ 
\left[B,A,A\right].
$$
\label{lemma27}
\end{lemma}

\begin{proof}
Let us add $c_{322}$ to the list of PNCs~(\ref{cs}).  
Then we find that no new orthogonality
relation will be introduced.  Let us define $\Lambda_i^{(1)}$ 
to be the sum over $J$ and $K$ of $|c_{iJK}|^2$, where $c_{iJK}$
appear in the list~(\ref{cs}), and similarly for
$\Lambda_j^{(2)}$ and $\Lambda_k^{(3)}$.  Then the eigenvalues
$\lambda_i^{(a)}$ can be expressed as follows:
\begin{eqnarray}
&& \lambda_1^{(1)} = \Lambda_1^{(1)},\ 
\lambda_2^{(1)} = \Lambda_2^{(1)},\ 
\lambda_3^{(1)} = \Lambda_3^{(1)} + |c_{322}|^2\,,
\nonumber \\
&& \lambda_1^{(2)} = \Lambda_1^{(2)},\ 
\lambda_2^{(2)}=\Lambda_2^{(2)}+|c_{322}|^2\,,\nonumber \\
&& \lambda_1^{(3)} =\Lambda_1^{(3)},\
\lambda_2^{(3)} = \Lambda_2^{(3)}+|c_{322}|^2\,. \label{exprsd}
\end{eqnarray}
The preceding construction shows that
if $\lambda_i^{(a)}$ are given in this manner and if $\Lambda_i^{(a)}$
satisfy (\ref{border}) and (\ref{between}), then there is
a state with the eigenvalues $\lambda_i^{(a)}$.  Furthermore, if
we require that $\Lambda_2^{(a)} \geq \Lambda_1^{(a)}$ for all $a$,
then the set of E-points
satisfying all these conditions is convex.

All points except $[B,A,A]$ satisfy
the conditions given here with $c_{322}=0$ since they
are on the boundary~(\ref{border}) with
$\lambda_i^{(a)} = \Lambda_i^{(a)}$ for all $i$ and $a$.  Thus,
all we need to do is find $\Lambda_i^{(a)}$ [which satisfy
(\ref{border}) and (\ref{between})] and $c_{322}$
such that the eigenvalues $[B,A,A]$ are expressed
as in~(\ref{exprsd}).  We can do so with
$|c_{322}|^2=1/6$, $\Lambda_1^{(1)}=\Lambda_2^{(1)}=1/3$,
$\Lambda_3^{(1)}=1/6$, $\Lambda_1^{(2)}=\Lambda_1^{(3)}=0$,
$\Lambda_2^{(2)}=\Lambda_2^{(3)}=1/3$ (and with
$\Lambda_3^{(2)}=\Lambda_3^{(3)}=1/2$ as a result).
\end{proof}

We will need the states given by the following PNCs:
\begin{eqnarray}
&& c_{132} =  a\cos\theta\,,\ \ 
c_{232} = -a\sin\theta,\ \nonumber \\
&& c_{112} =  r\sin\alpha\sin\theta\,,\ \
c_{212} = r\sin\alpha \cos\theta,\nonumber \\
&& c_{122}= r \cos\alpha \sin\theta\,,\ \
c_{222}  = r\cos\alpha\cos\theta,\nonumber \\
&& c_{131} = f\sin\theta\,,\ \
c_{231}  = f\cos\theta,\nonumber \\
&& c_{113} = p\sin\alpha \cos\theta - q \cos\alpha \cos\varphi\,,
\nonumber \\
&& c_{213} = - p\sin\alpha \sin\theta - q\cos\alpha \sin\varphi\,,
\nonumber \\
&& c_{123} = p \cos\alpha\cos\theta + q \sin\alpha \cos\varphi\,,
\nonumber \\
&& c_{223} = - p\cos\alpha \sin\theta + q \sin\alpha \sin\varphi\,,\ \
c_{333} = g, \nonumber
\end{eqnarray}
where $a$, $f$, $g$, $p$, $q$, $r$, $\theta$, $\varphi$ and 
$\alpha$ are real numbers.
(Note that the PNCs will be
those for the boundary $P_1=0$ if $r=0$ although the parametrization
is slightly different.)
The only nontrivial orthogonality relations are
\begin{eqnarray}
(r^2 + f^2 - a^2 - p^2)\sin 2\theta & = & -q^2\sin 2\varphi\,,
\label{ssconst} \\
(r^2 + p^2 - q^2)\sin 2\alpha & = & 2pq \cos(\theta+\varphi)
\cos 2\alpha\,.  \label{alphaeq}
\end{eqnarray}
The eigenvalues are given by
\begin{eqnarray}
\lambda_1^{(1)} & = & (a^2+p^2)\cos^2\theta
+ (r^2 + f^2)\sin^2\theta + q^2\cos^2\varphi\,, \nonumber \\
\lambda_2^{(1)} & = & (a^2+p^2)\sin^2\theta
+ (r^2 + f^2)\cos^2\theta + q^2\sin^2\varphi\,, \nonumber \\
\lambda_1^{(2)} & = & (r^2 + p^2)\sin^2\alpha + q^2\cos^2\alpha
- pq\sin 2\alpha\,\cos(\theta+\varphi)\,, \nonumber \\
\lambda_2^{(2)} & = & (r^2+p^2)\cos^2\alpha + q^2 \sin^2\alpha
+ pq\sin 2\alpha\,\cos(\theta+\varphi)\,, \nonumber \\
\lambda_1^{(3)} & = & f^2\,, \nonumber \\
\lambda_2^{(3)} & = & a^2 + r^2\,.\nonumber
\end{eqnarray}
Let us introduce the variable $s^2 \equiv p^2 + q^2$.  Then
we have
\begin{eqnarray}
\lambda_1^{(1)} + \lambda_2^{(1)}
& = & a^2 + r^2 + f^2 + s^2\,, \label{solve1}\\
\lambda_1^{(2} + \lambda_2^{(2)}
& = & r^2 + s^2, \label{solve2}\\
\lambda_2^{(3)} & = & a^2 + r^2\,, \label{solve3}\\
\lambda_1^{(3)} & = & f^2\,. 
\end{eqnarray}
Notice that these equations can be solved for $a^2$, $r^2$, $f^2$
and $s^2$.
Then note that Eq.~(\ref{alphaeq}) can be satisfied by letting
\begin{eqnarray}
\sin 2\alpha & = & \frac{2pq\cos(\theta + \varphi)}
{\sqrt{(r^2+p^2-q^2)^2 + 4p^2 q^2 \cos^2(\theta+\varphi)}}\,, 
\nonumber \\
\cos 2\alpha & = & \frac{r^2 + p^2 - q^2}
{\sqrt{(r^2+p^2-q^2)^2 + 4p^2 q^2\cos^2(\theta+\varphi)}}
\nonumber
\end{eqnarray}
unless $r^2+p^2-q^2=pq\cos(\theta+\varphi)=0$.
The eigenvalues $\lambda_1^{(2)}$ and
$\lambda_2^{(2)}$ can be given as 
\begin{eqnarray}
\lambda_1^{(2)} & = & \frac{1}{2}
\left[ r^2 + p^2 + q^2 - 
\sqrt{(r^2+p^2 - q^2)^2 + 4p^2q^2\cos^2(\theta +
\varphi)}\right]\,, \nonumber \\
\lambda_2^{(2)} & = & \frac{1}{2}
\left[ r^2 + p^2 + q^2 + 
\sqrt{(r^2+p^2 - q^2)^2 + 4p^2q^2\cos^2(\theta +
\varphi)}\right]\,. \nonumber
\end{eqnarray}
These equations are valid
even if $r^2+p^2-q^2 = pq\cos(\theta +\varphi)=0$.

Suppose that $a^2$, $r^2$, $f^2$ and $s^2$ are fixed at some values
satisfying
\begin{eqnarray}
s^2 - r^2 & \geq & 0\,,\nonumber \\
a^2-f^2-2r^2 & \geq  & 0\,. \nonumber
\end{eqnarray}
We first show that $\lambda_1^{(2)}$ can
take any value between $0$ and $(r^2+s^2)/2$.
By the first part of Lemma~\ref{obvious} we find that
the combination $\theta+\varphi$ can vary freely.
Hence $\lambda_1^{(2)}$
can take any value as long as
$$
\frac{1}{2}\left[ r^2+s^2 -\sqrt{(r^2+p^2-q^2)^2 + 4p^2q^2}\right]
\leq \lambda_1^{(2)} \leq
\frac{1}{2}\left[ r^2 + s^2 - |r^2+p^2-q^2|\right]\,.
$$
By varying $q^2$ from $0$ to $(r^2+s^2)/2$, which is possible by
the assumption that $r^2 \leq s^2$, we can see that
$\lambda_1^{(2)}$ varies from $0$ to $(r^2+s^2)/2$.

However, let us impose the condition $r^2 \leq \lambda_1^{(2)}$
or, equivalently, $\lambda_2^{(2)} \leq s^2$.  We will show, 
under this condition, that
$\lambda_2^{(1)}-\lambda_1^{(1)}$ can take any value
between
$|a^2-f^2-2r^2-\lambda_2^{(2)}+\lambda_1^{(2)}|$
and
$|a^2 -f^2 -2r^2 + \lambda_2^{(2)}-\lambda_1^{(2)}|$.
Note first that
\begin{equation}
\lambda_1^{(2)}\lambda_2^{(2)} = q^2[r^2+p^2\sin^2(\theta+\varphi)]\,.
\label{sin2}
\end{equation}
This equation to have a solution for $\theta+\varphi$ if and only if
$$
q^2 r^2 \leq \lambda_1^{(2)}\lambda_2^{(2)} \leq
q^2(r^2+p^2) = q^2(\lambda_1^{(2)}+\lambda_2^{(2)}-q^2)\,.
$$
These inequalties are satisfied if
$\lambda_1^{(2)}\leq q^2 \leq \lambda_2^{(2)}$.
Assuming that $\lambda_1^{(2)} > r^2$, we have
$q^2 > r^2 \geq 0$ and 
$p^2 = s^2 - q^2 \geq s^2 - \lambda_2^{(2)} = \lambda_1^{(2)}-r^2 > 0$.
Then from (\ref{sin2}) we find 
$\sin^2(\theta+\varphi)$ as a function of $q^2$ with
$\lambda_1^{(2)}$ fixed as
$$
\sin^2(\theta+\varphi) 
= \frac{\lambda_1^{(2)}\lambda_2^{(2)}-q^2r^2}{p^2q^2}\,.
$$
By substituting this in 
\begin{eqnarray}
|\lambda_2^{(1)} - \lambda_1^{(1)}|^2
& = & \left[ (r^2+f^2-a^2-p^2)\cos 2\theta - q^2 \cos 2\varphi 
\right]^2 \nonumber \\
&& + \left[ (r^2 + f^2 - a^2 - p^2)\sin 2\theta
+ q^2 \sin 2\varphi\right]^2\nonumber \\
& = & (r^2 + f^2 - a^2 - p^2 - q^2)^2 \nonumber \\
&& + 4(r^2 + f^2 -a^2 - p^2)q^2\sin^2(\theta + \varphi)\,, \nonumber
\end{eqnarray}
we obtain
\begin{eqnarray}
|\lambda_2^{(1)} - \lambda_1^{(1)}|^2
& = & (r^2 + f^2 - a^2 - p^2 - q^2)^2 \nonumber \\
&& + 4(r^2 + f^2 -a^2 - p^2)
\frac{\lambda_1^{(2)}\lambda_2^{(2)}-q^2r^2}{p^2}\,. \nonumber
\end{eqnarray}
If we substitute $q^2=\lambda_1^{(2)}$ and $q^2 =\lambda_2^{(2)}$,
then the right-hand side becomes
$|a^2 - f^2-2r^2+\lambda_2^{(2)}-\lambda_1^{(2)}|^2$
and $|a^2-f^2-2r^2-\lambda_2^{(2)}+\lambda_1^{(2)}|^2$, respectively.
Hence, by continuity, $|\lambda_2^{(1)}-\lambda_1^{(1)}|^2$
takes any value between these values as $q^2$ ranges from
$\lambda_1^{(2)}$ to $\lambda_2^{(2)}$.  If $\lambda_1^{(2)}=r^2$,
then Eq.~(\ref{sin2}) can be solved by letting
$q^2 = s^2 = \lambda_2^{(2)}$.  Then we have $p^2=0$, and
$\sin^2(\theta+\varphi)$ can take any value in $[0,1]$.  Hence,
$\lambda_2^{(2)}-\lambda_1^{(2)}$ can take any value between
$|a^2-f^2-2r^2-\lambda_2^{(2)}+\lambda_1^{(2)}|$ and
$|a^2-f^2-2r^2+\lambda_2^{(2)}+\lambda_1^{(2)}|$. This range 
contains the desired range.
The result obtained here can be summarized as follows.
\begin{lemma}
Suppose that
\begin{eqnarray}
r^2 & \equiv & \lambda_1^{(2)} + \lambda_2^{(2)}
+ \lambda_1^{(3)}+ \lambda_2^{(3)} - \lambda_1^{(1)} 
- \lambda_2^{(1)} \geq 0\,,\nonumber \\
a^2 & \equiv & \lambda_1^{(1)} + \lambda_2^{(1)}
- \lambda_1^{(2)} - \lambda_2^{(2)} - \lambda_1^{(3)} \geq 0\,,
\nonumber \\
s^2 & \equiv & \lambda_1^{(1)} + \lambda_2^{(1)} - \lambda_1^{(3)}
- \lambda_2^{(3)}\geq 0\,. \nonumber
\end{eqnarray}
Let $f^2 \equiv \lambda_1^{(3)}$.
If $r^2 \leq \lambda_1^{(2)}\leq\lambda_2^{(2)}\leq s^2$,
$a^2 - f^2-2r^2 \geq 0$ and
$$
|a^2 -f^2 -2r^2 -\lambda_2^{(2)}+\lambda_1^{(2)}|
\leq \lambda_2^{(1)}-\lambda_1^{(1)} \leq 
a^2-f^2 -2r^2 +\lambda_2^{(2)}-\lambda_1^{(2)}\,,
$$
then there is a state with these eigenvalues.
The set of E-points satisfying these conditions is
convex.
\label{lemma28}
\end{lemma}

\begin{proof}
The first three conditions guarantee that Eqs.~(\ref{solve1}),
(\ref{solve2}) and (\ref{solve3}) can be solved for
$a^2$, $r^2$ and $s^2$.  Convexity is obvious from the form of
the conditions imposed.
\end{proof}

\noindent
The following fact, emphasized by Bravyi~\cite{bravyi}, is useful
in constructing states.
\begin{lemma}
Consider the set of states such that $c_{ijk}\neq 0$ only if
the triple $[ijk]$ is in a set $E$.  Suppose that
if $[ijk]$ and $[i'j'k']$ are in $E$ and if
$[ijk]\neq [i'j'k']$, then 
at most one of the equations $i=i'$, $j=j'$ and $k=k'$ holds. (This
implies that there are no nontrivial orthogonality relations.)
Then the set of E-points for these states is convex.
\label{useful}
\end{lemma}

\begin{proof}
Let $\lambda_l^{(a)I}$ and $\lambda_l^{(a)II}$ be the eigenvalues
of the states with the specified nonzero coefficients. Let
$A_{ijk} \equiv |c_{ijk}|^2$ and write
$A_{ijk} = A_{ijk}^I$ and $A_{ijk}^{II}$ for these states.  Then, for
any $\lambda_l^{(a)}$ given by
$$
\lambda_l^{(a)} = \alpha \lambda_l^{(a)I} 
+ (1-\alpha)\lambda_l^{(a)II}
$$
with $0\leq \alpha \leq 1$
for all $a$ and $l$,
a corresponding state is obtained by letting
$$
A_{ijk} = \alpha A_{ijk}^{I} + (1-\alpha)A_{ijk}^{II}
$$
for all $[ijk]$.
\end{proof} 

\section{Construction of the set $S$}
\label{constr2}

We have constructed states with some E-points in the convex set
$S$ in Proposition~\ref{prop10} and Lemma~\ref{lemma27}.  We will
start constructing states with their E-points
in outlying subregions of $S$.  First
we divide the region $S$ into two: one region $S_1$ that satisfies
\begin{equation}
P_8 \equiv 
2\lambda_2^{(2)}+\lambda_3^{(2)}+2\lambda_1^{(3)}+\lambda_2^{(3)}
- 2\lambda_2^{(1)} - \lambda_3^{(1)} \leq 0 \label{P8}
\end{equation}
and the other which satisfies $P_8 \geq 0$.  The only corner point
of $S$ with $P_8 < 0$ is 
$[A,\quarter\quarter\half,0\quarter\thquarters]$ which
has $P_8 = -1/4$.
The corner points satisfying $P_8=0$ are
\begin{eqnarray}
&& \left[O,O,O\right],\ 
\left[B,B,O\right],\
\left[A,A,O\right],\
\left[A,O,A\right],\nonumber \\ 
&& \left[A,B,A\right],\ 
\left[0\third\twthirds,B,0\third\twthirds\right],\ 
\left[A,\sixth\sixth\twthirds,\sixth\sixth\twthirds\right],\ 
\left[A,B,\sixth\sixth\twthirds\right]. 
\label{P8eq0}
\end{eqnarray}
Using this observation, 
we show that the set $S_1$ consists of three simplices.
\begin{lemma}
The subset of $S$ satisfying $P_8 \leq 0$, i.e the convex set $S_1$,
consists of the three
simplices $C_1$, $C_2$ and $C_3$ with the following five E-points 
being corner points of all three simplices:
$$
\left[O,O,O\right],\ 
\left[B,B,O\right],\ 
\left[A,A,O\right],\ 
\left[A,B,\sixth\sixth\twthirds\right],\ 
\left[A,\quarter\quarter\half,0\quarter\thquarters\right],
$$
and with the additional corner points for each simplex given by
\begin{eqnarray}
C_1 & : & \left[A,B,A\right],\
\left[0\third\twthirds,B,0\third\twthirds\right];
\nonumber \\
C_2 & : & \left[A,O,A\right],\
\left[A,B,A\right]; \nonumber \\
C_3 & : & \left[A,O,A\right],\ 
\left[A,\sixth\sixth\twthirds,\sixth\sixth\twthirds\right].
\nonumber 
\end{eqnarray}
\end{lemma}

\begin{proof}
Since the E-point $[A,\quarter\quarter\half,0\quarter\thquarters]$
is the only E-point satisfying $P_8 < 0$, the boundary hyperplanes
of $S_1$ are $P_8=0$ and those of $S$ that
contain $[A,\quarter\quarter\half,0\quarter\thquarters]$.
The latter are $\lambda_1^{(1)}=0$, $\lambda_2^{(1)}=\lambda_3^{(1)}$,
$\lambda_1^{(2)}=\lambda_2^{(2)}$, $\lambda_1^{(3)}=0$,
(\ref{ineq2}) with $(abc)=(231)$, (\ref{ineq3}) with $(abc)=(132)$,
(\ref{ineq6}) with $(abc)=(132)$ and (\ref{ineq7}) with
$(abc)=(132)$.  It is enough to show that
$C_1\cup C_2 \cup C_3$ is bounded by these hyperplanes.

Let us define the following functions:
\begin{eqnarray}
P_9 & \equiv & 2\lambda_1^{(1)}+\lambda_2^{(1)}
+ 2\lambda_1^{(3)} +
\lambda_2^{(3)}-2\lambda_1^{(2)}-\lambda_2^{(2)},\nonumber \\
P_{10} & \equiv & \lambda_2^{(1)} - \lambda_2^{(2)}-\lambda_2^{(3)}\,.
\end{eqnarray}
Then the boundary hyperplanes 
of the simplex $C_1$
can be given as follows, where the E-point listed with
each boundary hyperplane is the one which is not
on the hyperplane:
\begin{eqnarray}
&& \left[O,O,O\right]\,:\,(\ref{ineq7}), (abc)=(132);\ \  
\left[B,B,O\right]\,:\,\lambda_1^{(1)}=0;\nonumber \\
&& \left[A,A,O\right]\,:\, \lambda_1^{(2)}=\lambda_2^{(2)};\ \ 
\left[A,B,\sixth\sixth\twthirds\right]\,:\,
\lambda_1^{(3)}=0;\nonumber \\
&& \left[A,\quarter\quarter\half,0\quarter\thquarters\right]\,:\,
P_8=0;\ \ 
\left[A,B,A\right]\,:\, (\ref{ineq3}), (abc)=(132);\nonumber \\
&& \left[0\third\twthirds,B,0\third\twthirds\right]
\,:\, P_9=0.\nonumber
\label{C1list}
\end{eqnarray}
Those for $C_2$ are
\begin{eqnarray}
&& \left[O,O,O\right]\,:\,\lambda_2^{(1)}=\lambda_3^{(1)};\ \
\left[B,B,O\right]\,:\,\lambda_1^{(1)}=0;\nonumber \\
&& \left[A,A,O\right]\,:\,\lambda_1^{(2)}=\lambda_2^{(2)};\ \ 
\left[A,B,\sixth\sixth\twthirds\right]
\,:\,\lambda_1^{(3)}=0;\nonumber \\
&& \left[A,\quarter\quarter\half,0\quarter\thquarters\right]\,:\,
P_8=0;\ \
\left[A,O,A\right]\,:\, P_9=0;\nonumber \\
&& \left[A,B,A\right]\,:\, P_{10}=0.
\end{eqnarray}
Those for $C_3$ are
\begin{eqnarray}
&& \left[O,O,O\right]\,:\,\lambda_2^{(1)}=\lambda_3^{(1)};\ \ 
\left[B,B,O\right]\,:\,\lambda_1^{(1)}=0;\nonumber \\
&& \left[A,A,O\right]\,:\,\lambda_1^{(2)}=\lambda_2^{(2)};\ \ 
\left[A,B,\sixth\sixth\twthirds\right]
\,:\,(\ref{ineq2}), (abc)=(231);\nonumber \\
&& \left[A,\quarter\quarter\half,0\quarter\thquarters\right]\,:\,
P_8=0;\ \
\left[A,O,A\right]\,:\, (\ref{ineq6}), (abc)=(132);\nonumber \\
&& \left[A,\sixth\sixth\twthirds,\sixth\sixth\twthirds\right]\,:\, 
P_{10}=0.
\end{eqnarray}
The E-point $[0\third\twthirds,B,0\third\twthirds]$ is
the only corner point of $C_1$, $C_2$ or $C_3$
with $P_9 < 0$, and all other corner points
of $C_1$ satisfy $P_9=0$.  Hence all E-points in $C_1$
satisfy $P_9 \leq 0$.
The E-point
$[A,\sixth\sixth\twthirds,\sixth\sixth\twthirds]$ is the
only E-point of $C_1$, $C_2$ or $C_3$ 
with $P_{10} > 0$, and all other corner points
of $C_3$ satisfy $P_{10} =0$.  Hence all E-points in $C_3$
satisfy $P_{10}\geq 0$. All corner points of
$C_2$ satisfy $P_9=P_{10}=0$ except
$[A,O,A]$ with $P_9=1$ and $P_{10}=0$
and $[A,B,A]$ with $P_9=0$ and $P_{10}=-1/3$.
Hence all E-points in $C_2$ satisfy
$P_9 \geq 0$ and $P_{10} \leq 0$.
Thus, the intersections $C_1\cap C_2$ and
$C_2\cap C_3$ are hyperplanes $P_9 =0$ and
$P_{10}=0$, respectively.  The corner points
of these boundary hyperplanes are
\begin{eqnarray}
&& P_9 : \left[A,B,A\right],
\left[A,\quarter\quarter\half,0\quarter\thquarters\right], 
\left[A,B,\sixth\sixth\twthirds\right],
\left[A,A,O\right],
\left[B,B,O\right],
\left[O,O,O\right];\nonumber \\
&& P_{10}:
\left[A,O,A\right], 
\left[A,\quarter\quarter\half,0\quarter\thquarters\right],
\left[A,B,\sixth\sixth\twthirds\right],
\left[A,A,O\right],
\left[B,B,O\right], 
\left[O,O,O\right]. \nonumber
\end{eqnarray}
Thus, the simplices $C_1$, $C_2$ and $C_3$ are
glued together at these boundaries, and all other
boundaries are boundaries of the convex set $S_1$.
Hence $S_1=C_1\cup C_2\cup C_3$.
\end{proof}
Now we construct states with their E-points in 
the simplices $C_1$, $C_2$ and $C_3$.
\begin{lemma}
There is a state with any E-point in $S_1=C_1\cup C_2\cup C_3$.
\label{lemma30}
\end{lemma}

\begin{proof}
Since $C_2 \subset S_0$,
there is a state with any E-point
in $C_2$ due to Proposition~\ref{prop10}.
Next we construct a state with any E-point in
the simplex $C_1$.  From the list (\ref{C1list})
we see that the boundary obtained by omitting $[A,B,A]$ is
\begin{equation}
\lambda_2^{(2)}+\lambda_3^{(2)}+\lambda_2^{(3)}+\lambda_1^{(3)}
- \lambda_2^{(1)}- \lambda_3^{(1)} = 0. \label{C1eq}
\end{equation}
This suggests that the states may be constructed by using
Lemma~\ref{lemma28} after letting
$(\lambda_1^{(1)},\lambda_2^{(1)})\to
(\lambda_2^{(1)},\lambda_3^{(1)})$
and
$(\lambda_1^{(2)},\lambda_2^{(2)})
\to(\lambda_2^{(2)},\lambda_3^{(2)})$. 
This is allowed since the only inequalities among $\lambda_i^{(a)}$
that we need for this lemma to hold are
$\lambda_2^{(a)}\geq \lambda_1^{(a)}$ for $a=1,2$.
For an E-point satisfying equality (\ref{C1eq}), we have, using the
notation of this lemma,
\begin{eqnarray}
r^2 & = & \lambda_2^{(2)} + \lambda_3^{(2)}
+ \lambda_1^{(3)}+ \lambda_2^{(3)} - \lambda_2^{(1)} 
- \lambda_3^{(1)} = 0\,, \nonumber \\
a^2 & = & \lambda_2^{(1)} + \lambda_3^{(1)}
- \lambda_2^{(2)} - \lambda_3^{(2)} - \lambda_1^{(3)}
= \lambda_2^{(3)} \geq 0\,,\nonumber \\
s^2 - \lambda_3^{(2)} 
& = & \lambda_2^{(1)} + \lambda_3^{(1)} - \lambda_1^{(3)}
- \lambda_2^{(3)} -\lambda_3^{(2)}= \lambda_2^{(2)} \geq 0
\nonumber
\end{eqnarray}
and $f^2 = \lambda_1^{(3)}$.
Hence we have 
$r^2 \leq \lambda_2^{(2)} \leq \lambda_3^{(2)}\leq s^2$.
We also have  
$$
a^2-f^2 -2r^2 = \lambda_2^{(3)}-\lambda_1^{(3)} \geq 0\,.
$$
The possible range of $\lambda_3^{(1)}-\lambda_2^{(1)}$ is
$$
|\lambda_2^{(3)}-\lambda_1^{(3)}-\lambda_3^{(2)}+\lambda_2^{(2)}|
\leq \lambda_3^{(1)}-\lambda_2^{(1)} \leq
\lambda_2^{(3)}-\lambda_1^{(3)}+\lambda_3^{(2)}-\lambda_2^{(2)}\,.
$$
These are equivalent to the following three inequalities if
Eq.~(\ref{C1eq}) is satisfied:
inequality (\ref{ineq5}) with
$(abc)=(231)$, inequality (\ref{ineq6}) with
$(abc)=(132)$ and inequality (\ref{ineq7}) with 
$(abc)=(132)$.  Hence these are all satisfied by the corner points
satisfying (\ref{C1eq}).  For $[A,B,A]$, we find
$r^2 = 1/6$, $a^2=1/3$, $s^2-\lambda_3^{(2)}=1/6$, and
$a^2 - f^2 -2r^2= 0$.  
Thus, $|2r^2+f^2-a^2 \pm \lambda_3^{(2)}-\lambda_2^{(2)}| = 0$, 
and the condition
$$
|a^2-f^2-2r^2 - \lambda_3^{(2)}+\lambda_2^{(2)}|
\leq \lambda_3^{(1)}-\lambda_2^{(1)}
\leq a^2 -f^2 -2r^2 + \lambda_3^{(2)} -\lambda_2^{(2)}
$$ 
is satisfied by $[A,B,A]$.
Therefore, by Lemma~\ref{lemma28} all E-points in $C_1$
have corresponding states.

The construction of states with their E-points in $C_3$ is similar.
We note that the boundary of $C_3$ obtained
by omitting $[A,B,\sixth\sixth\twthirds]$ is
\begin{equation}
\lambda_1^{(1)}+\lambda_3^{(1)}+\lambda_1^{(3)}
+\lambda_2^{(3)} - \lambda_1^{(2)}-\lambda_3^{(2)}=0\,. \label{thiseq}
\end{equation}
This suggests that we can apply Lemma~\ref{lemma28}
with $\lambda_2^{(1)}\leftrightarrow \lambda_3^{(1)}$,
$\lambda_2^{(3)}\leftrightarrow \lambda_3^{(3)}$ and then with
$(1)\to (2)\to (3) \to (1)$.
For an E-point satisfying (\ref{thiseq}) we have, using the notation
of this lemma,
\begin{eqnarray}
r^2 & = & \lambda_1^{(1)} + \lambda_3^{(1)}
+ \lambda_1^{(3)}+ \lambda_2^{(3)} - \lambda_1^{(2)} 
- \lambda_3^{(2)} = 0\,, \nonumber \\
a^2 & = & \lambda_1^{(2)} + \lambda_3^{(2)}
- \lambda_1^{(3)} - \lambda_2^{(3)} - \lambda_1^{(1)} 
= \lambda_3^{(1)}\,, \nonumber \\
s^2 -\lambda_2^{(3)} & = & 
\lambda_1^{(2)} + \lambda_3^{(2)} - \lambda_1^{(1)}
- \lambda_3^{(1)} - \lambda_2^{(3)} = \lambda_1^{(3)}
\nonumber 
\end{eqnarray}
and $f^2=\lambda_1^{(1)}$.
So these are all nonnegative on 
the E-points satisfying (\ref{thiseq}).
For $[A,B,\sixth\sixth\twthirds]$ we find
$r^2 = 1/6$, $f^2=0$, $a^2=1/3$, $s^2-\lambda_2^{(3)}=0$.
For the E-points satisfying (\ref{thiseq}) we have
$a^2-f^2 - 2r^2=\lambda_3^{(1)}-\lambda_1^{(1)}\geq 0$.  For 
$[A,B,\sixth\sixth\twthirds]$ we have $a^2-f^2 -2r^2= 0$.  Then, for
$C_3$ to be realized by the states constructed here it is sufficient
to have
\begin{equation}
|2r^2+f^2-a^2 +\lambda_2^{(3)}-\lambda_1^{(3)}|
\leq \lambda_3^{(2)}-\lambda_1^{(2)}\leq
-2r^2-f^2+a^2 + \lambda_2^{(3)}-\lambda_1^{(3)}  \label{sufsuf}
\end{equation}
for all corner points.
This is satisfied by $[A,B,\sixth\sixth\twthirds]$ since
$\lambda_2^{(3)}-\lambda_1^{(3)}=\lambda_3^{(2)}-\lambda_1^{(2)}=0$.
For the E-points satisfying (\ref{thiseq}), Equation~(\ref{sufsuf})
is equivalent to inequality (\ref{ineq4}) with $(abc)=(123)$,
inequality (\ref{ineq6}) with $(abc)=(132)$ and
$$
2\lambda_3^{(1)}+\lambda_1^{(1)}+2\lambda_1^{(3)}
+\lambda_2^{(3)} \geq  2\lambda_1^{(2)}+\lambda_3^{(2)}\,.
$$
This inequality 
can readily be verified by using inequality (\ref{ineq3}) with
$(abc)=(231)$ and $\lambda_3^{(1)}\geq 1/3\geq \lambda_1^{(2)}$.
\end{proof}

Thus, we have constructed a state with any E-point in $S$
with an additional condition $P_8\leq 0$. Since the intersection of
$S$ and the hyperplane $P_8=0$ is the 5-dimensional convex set
with corner points~(\ref{P8eq0}),
the convex subset of $S$ with the condition $P_8 \geq 0$ are
those of $S$ excluding $[A,\quarter\quarter\half,0\quarter\thquarters]$.
By symmetry we can
construct a state with any E-point in $S$ satisfying
\begin{equation}
2\lambda_2^{(c)}+\lambda_3^{(c)}+2\lambda_1^{(b)}+\lambda_2^{(b)}
- 2\lambda_2^{(a)} - \lambda_3^{(a)} \leq 0
\end{equation}
for any $(abc)$.  Thus, our next task is to construct a state with
any E-point in $S$ with additional conditions
\begin{equation}
2\lambda_1^{(b)}+\lambda_2^{(b)}+2\lambda_2^{(c)}+\lambda_3^{(c)}
-2\lambda_2^{(a)}-\lambda_3^{(a)} \geq 0
\label{ineq8}
\end{equation}
for all $(abc)$.  This set of E-points is the convex set with
all the corner points inherited from $S$ except for
$[A,\quarter\quarter\half,0\quarter\thquarters]$ and those obtained by
qutrit permutations from it.  Let us denote this set by $S_2$.
{}From the construction of $S_1$ it is clear that inequalities
(\ref{ineq2}), (\ref{ineq3}), (\ref{ineq6}) and (\ref{ineq7}) are
redundant once we impose (\ref{ineq8}).  We give a more
straightforward proof of this fact.
\begin{proposition}
Inequalities (\ref{ineq2}), (\ref{ineq3}), (\ref{ineq6}) and
(\ref{ineq7}) follow from
inequalities (\ref{ineq1}), (\ref{ineq4}), (\ref{ineq5}) and
(\ref{ineq8}).
\end{proposition}

\begin{proof}
Inequalities (\ref{ineq6}) and (\ref{ineq7}) follow immediately from
inequality (\ref{ineq8}) by noting that 
$\lambda_2^{(b)}\geq \lambda_1^{(b)}$ and
$\lambda_3^{(c)}\geq \lambda_2^{(c)}$, respectively.
Inequality (\ref{ineq5}) is equivalent to
$$
2\lambda_3^{(a)} + \lambda_2^{(a)} \leq
2\lambda_2^{(b)}+\lambda_1^{(b)}+ 2\lambda_3^{(c)}+\lambda_2^{(c)}\,.
$$
By adding this and inequality (\ref{ineq8}) and dividing by three 
we obtain inequality (\ref{ineq3}).  Inequality (\ref{ineq1}) is
equivalent to
$$
2\lambda_3^{(a)}+\lambda_1^{(a)} \leq
2\lambda_2^{(b)}+\lambda_1^{(b)} + 2\lambda_3^{(c)}+\lambda_1^{(c)}\,.
$$
On the other hand, inequality (\ref{ineq8}) is equivalent to
$$
2\lambda_1^{(a)} + \lambda_3^{(a)}
\leq 2\lambda_1^{(b)}+\lambda_2^{(b)} +
2\lambda_1^{(c)}+\lambda_3^{(c)}\,.
$$
By adding these inequalities together and dividing by three we obtain
inequality (\ref{ineq2}).
\end{proof}

We observe that among the corner points of $S_2$ (in fact of $S$) 
the E-point
$[B,0\third\twthirds,0\third\twthirds]$ is the only E-point 
which violates the inequality 
$$
\tilde{P}_9 \equiv 2\lambda_1^{(2)}+\lambda_2^{(2)}
+2\lambda_1^{(3)}+\lambda_2^{(3)} - 2\lambda_1^{(1)}-\lambda_2^{(1)}
\geq 0\,.
$$
It has
$\tilde{P}_{9}=-1/3$.  The E-points which satisfy $\tilde{P}_{9}=0$ are
\begin{eqnarray}
&& \left[O,O,O\right],\ \left[A,A,O\right],\ 
\left[A,O,A\right],\ \left[B,O,B\right],\nonumber \\
&& \left[B,B,O\right],\ \left[B,A,A\right],\ 
\left[B,A,\sixth\sixth\twthirds\right],\
\left[B,\sixth\sixth\twthirds,A\right]. \nonumber 
\end{eqnarray}
These E-points are the corner points of the convex subset of $S_2$
with the additional condition $\tilde{P}_{9} \leq 0$ as we will 
show below.
\begin{lemma}
Let $S_3$ be the convex subset of $S_2$ obtained by imposing an 
additional
inequality $\tilde{P}_{9}\leq 0$.  Then 
$S_3 = D_1 \cup D_2 \cup D_3$ where $D_1$, $D_2$
and $D_3$ are simplices.  The following E-points are the corner points
of all three simplices:
$$
\left[B,0\third\twthirds,0\third\twthirds\right],\ 
\left[O,O,O\right],\ \left[B,B,O\right],\
\left[B,O,B\right],\ \left[B,A,A\right].
$$
The additional corner points of each simplex are as follows:
\begin{eqnarray}
D_1 & : & \left[A,A,O\right],\ 
\left[B,A,\sixth\sixth\twthirds\right]; \nonumber \\
D_2 & : & \left[A,A,O\right],\ \left[A,O,A\right];\nonumber \\
D_3 & : & \left[A,O,A\right],\ 
\left[B,\sixth\sixth\twthirds,A\right].
\end{eqnarray}
\end{lemma}

\begin{proof}
Since the E-point $[B,0\third\twthirds,0\third\twthirds]$
is the only E-point satisfying $\tilde{P}_9 < 0$, 
the boundary hyperplanes
of $S_3$ are $\tilde{P}_9=0$ and the boundary hyperplanes of $S_2$ that
contain $[B,\third\twthirds,0\third\twthirds]$. 
The latter are $\lambda_1^{(1)}=\lambda_2^{(1)}$, 
$\lambda_2^{(1)}=\lambda_3^{(1)}$,
$\lambda_1^{(2)}=0$, $\lambda_1^{(3)}=0$,  (\ref{ineq8}) with
$(abc)=(231)$, (\ref{ineq8}) with $(abc)=(321)$,
(\ref{ineq1}) with $(abc)=(123)$, (\ref{ineq5}) with $(abc)=(123)$
and
(\ref{ineq5}) with $(abc)=(132)$.  It is enough to show that
$D_1\cup D_2 \cup D_3$ is bounded by these hyperplanes.  We define
\begin{eqnarray}
Q_1 & \equiv & \lambda_2^{(1)}-\lambda_1^{(1)}
-\lambda_2^{(2)}+\lambda_1^{(2)}+\lambda_2^{(3)}-\lambda_1^{(3)}\,,
\nonumber \\
Q_2 & \equiv & \lambda_2^{(1)}-\lambda_1^{(1)}
+\lambda_2^{(2)}-\lambda_1^{(2)}-\lambda_2^{(3)}+\lambda_1^{(3)}\,.
\nonumber
\end{eqnarray}
First we note that all E-points in $D_1$ satisfy $Q_1 \leq 0$ and
$Q_2 \geq 0$, all E-points in $D_2$ satisfy $Q_1 \geq 0$ and
$Q_2 \geq 0$, and all E-points in $D_3$ satisfy $Q_1 \geq 0$ and
$Q_2 \leq 0$.
Let us
list the boundaries of each simplex together 
with the E-point omitted to obtain each boundary 5-simplex:
\begin{eqnarray}
D_1 & : & \left[B,0\third\twthirds,0\third\twthirds\right]\,:\,
\tilde{P}_9=0;\ \left[O,O,O\right]\,:\,\lambda_2^{(1)}=\lambda_3^{(1)};
\nonumber \\
&& \left[B,O,B\right]\,:\,(\ref{ineq8}), (abc)=(321);\ 
\left[B,B,O\right]\,:\,\lambda_1^{(2)}=0;\nonumber \\
&& \left[B,A,A\right]\,:\,(\ref{ineq5}), (abc)=(132);\ 
\left[A,A,O\right]\,:\,\lambda_1^{(1)}=\lambda_1^{(2)};\nonumber \\
&& \left[B,A,\sixth\sixth\twthirds\right]\,:\, Q_1=0; \nonumber
\end{eqnarray}
\begin{eqnarray}
D_2 & : & 
\left[B,0\third\twthirds,0\third\twthirds\right]\,:\,
\tilde{P}_9=0;\ \left[O,O,O\right]\,:\,\lambda_2^{(1)}=\lambda_3^{(1)};
\nonumber \\
&& \left[B,O,B\right]\,:\,\lambda_1^{(3)}=0;\ 
\left[B,B,O\right]\,:\,\lambda_1^{(2)}=0;\nonumber \\
&& \left[B,A,A\right]\,:\,(\ref{ineq1}), (abc)=(123);\ 
\left[A,A,O\right]\,:\,Q_2=0;\nonumber \\
&& \left[A,O,A\right]\,:\, Q_1=0; \nonumber
\end{eqnarray}
\begin{eqnarray}
D_3 & :& 
\left[B,0\third\twthirds,0\third\twthirds\right]\,:\,
\tilde{P}_9=0;\ \left[O,O,O\right]\,:\,\lambda_2^{(1)}=\lambda_3^{(1)};
\nonumber \\
&& \left[B,O,B\right]\,:\,\lambda_1^{(3)}=0;\ 
\left[B,B,O\right]\,:\,(\ref{ineq8}), (abc)=(231);\nonumber \\
&& \left[B,A,A\right]\,:\,(\ref{ineq5}), (abc)=(123);\ 
\left[A,O,A\right]\,:\,\lambda_1^{(1)}=\lambda_1^{(2)};\nonumber \\
&& \left[B,\sixth\sixth\twthirds,A\right]\,:\, Q_2=0; \nonumber
\end{eqnarray}
Thus, the simplices 
$D_1$ and $D_2$ are glued together at the common boundary
$Q_1=0$ with corner points
$[B,0\third\twthirds,0\third\twthirds]$,
$[O,O,O]$, $[B,O,B]$, $[B,B,O]$, $[B,A,A]$ and $[A,A,O]$, and
$D_2$ and $D_3$ at 
$Q_2=0$ with corner points
$[B,0\third\twthirds,0\third\twthirds]$,
$[O,O,O]$, $[B,O,B]$, $[B,B,O]$, $[B,A,A]$ and $[A,O,A]$.  The other
boundaries are all boundary hyperplanes of $S_3$.  Hence,
$S_3 = D_1 \cup D_2 \cup D_3$.
\end{proof}

\noindent
Next we construct states which have E-points in
this region.
\begin{lemma}
There is a state with any E-point in $S_3$.
\end{lemma}

\begin{proof}
Note that states with their E-points in $D_2$ have been
constructed by Lemma~\ref{lemma27}.
In order to construct states with their E-points in $D_1$ we consider
the following PNCs:
$c_{333},c_{113},c_{223},c_{121},c_{231},c_{322},c_{132}$.
Since there are no nontrivial orthogonality relations, the set of
E-points obtained from these coefficients is
convex by Lemma~\ref{useful}.  
Hence it suffices to construct the corner points of $D_1$
using this set. This can be done as follows (with
$A_{ijk}\equiv |c_{ijk}|^2$) :
\begin{eqnarray}
&& \left[B,0\third\twthirds,0\third\twthirds\right]\,:\,
A_{132}=A_{223}=A_{333}=1/3;\nonumber \\
&& \left[O,O,O\right]\,:\, A_{333}=1;\nonumber \\
&& \left[B,O,B\right]\,:\, A_{333}=A_{231}=A_{132}=1/3;\nonumber \\
&& \left[B,B,O\right]\,:\, A_{333}=A_{113}=A_{223}=1/3;\nonumber \\
&& \left[B,A,A\right]\,:\, A_{132}=A_{223}=1/3,\ A_{322}=A_{333}=1/6;
\nonumber \\
&& \left[A,A,O\right]\,:\, A_{333}=A_{223}=1/2;\nonumber \\
&& \left[B,A,\sixth\sixth\twthirds\right]\,:\,
A_{121}=A_{132}=1/6,\ A_{333}=A_{223}=1/3.\nonumber
\end{eqnarray}
The simplex $D_3$ is obtained from $D_1$ by the qutrit permutation
$(2)\leftrightarrow (3)$.
\end{proof}

\noindent
Thus, we have constructed states with their E-points in the
convex subset of $S_2$ satisfying $\tilde{P}_9\leq 0$.  By symmetry we
can construct states with their E-points in $S_2$ if they satisfy
$$
P_9^{(abc)} \equiv
2\lambda_1^{(b)}+\lambda_2^{(b)}+2\lambda_1^{(c)}
+ \lambda_2^{(c)}- 2\lambda_1^{(a)}-\lambda_2^{(a)} \leq 0
$$
for some $(abc)$.  Let us denote the convex subset of $S_2$ satisfying
$P_9^{(abc)} \geq 0$ for all $(abc)$ by $S_4$.  
Our task is now reduced to
constructing states with
their E-points in $S_4$.  The corner points of $S_4$ are those of
$S$ excluding $[A,\quarter\quarter\half,0\quarter\thquarters]$
and $[B,0\third\twthirds,0\third\twthirds]$ and those obtained by
qutrit permutations from them.
The following lemma, which allows us to reduce the convex set $S_4$
further by removing $[A,A,A]$ from the corner points, can be proved
by using Proposition~\ref{prop10}.
\begin{lemma}
The convex subset of $S_4$ satisfying
\begin{equation}
Q_3 \equiv 
1 + \lambda_1^{(1)}-\lambda_2^{(1)}+\lambda_1^{(2)}-\lambda_2^{(2)}
+\lambda_1^{(3)}-\lambda_2^{(3)} \leq 0  \label{Q3}
\end{equation}
is a simplex with the following corner points:
$$
\left[A,A,A\right],\ \left[O,A,A\right],\ \left[A,O,A\right],\ 
\left[A,A,O\right],\ \left[B,A,A\right],\ \left[A,B,A\right],\
\left[A,A,B\right].
$$
There is a state with any E-point in this simplex.
\end{lemma}

\begin{proof}
It is straightforward to check that $Q_3 > 0$ for any corner point
of $S_4$ not listed here.  
We have $Q_3 = -1/2$ for $[A,A,A]$ and $Q_3=0$ for
the other E-points in the list.
The simplex with the corner points given here is bounded by
the hyperplane $Q_3=0$ and the boundary hyperplanes of $S_4$ that
contain $[A,A,A]$, which are
$\lambda_1^{(a)}=0$ and $\lambda_2^{(a)}=\lambda_3^{(a)}$,
$a=1,2,3$.  Hence, this simplex is the subset of $S_4$ with the
condition~(\ref{Q3}).
All E-points in this 
simplex can be realized by quantum states by Proposition
\ref{prop10} since it is a subset of $S_0$.
\end{proof}

\noindent
Thus, our task is further reduced to constructing the convex set
$S_5$ whose corner points are those of $S_4$ excluding $[A,A,A]$.
The following proposition almost accomplishes this task.
\begin{proposition}
There is a state with any E-point in
the convex set $S_6^{(1)}$ bounded by hyperplanes
with the same corner points as $S_5$ excluding $[O,A,A]$.
\label{prop12}
\end{proposition}

\begin{proof}
Consider the following set of PNCs:
$$
\{ c_{333},c_{113},
c_{131},c_{223},c_{232},c_{211},c_{321},c_{312},c_{122}\}\,.
$$
Note that this set is invariant under the qutrit permutation
$(2)\leftrightarrow (3)$.  Note also that there are no nontrivial
orthogonality relations to be satisfied.  Hence, the set of 
the E-points
realized by the states considered here is convex by
Lemma~\ref{useful}.  Therefore all we need to do is find
the values of $A_{ijk}=|c_{ijk}|^2$ for each corner point of $S_5$ 
except for $[O,A,A]$.  This can be done as follows:
\begin{eqnarray}
&& \left[O,O,O\right]\,:\,A_{333}=1;\nonumber \\
&& \left[B,B,B\right]\,:\,A_{333}=A_{113}=A_{133}=\cdots=A_{122}=1/9;
\nonumber \\
&& \left[B,A,B\right]\,:\,A_{333}=A_{223}=A_{131}=A_{232}=A_{321}
=A_{122}=1/6;\nonumber \\
&&
\left[A,B,B\right]\,:\,A_{333}=A_{223}=A_{211}=A_{232}=A_{321}=A_{312}
=1/6;\nonumber \\
&& \left[B,B,O\right]\,:\,A_{333}=A_{113}=A_{223}=1/3;\nonumber \\
&& \left[O,B,B\right]\,:\,A_{333}=A_{312}=A_{321}=1/3;\nonumber \\
&& \left[B,A,A\right]\,:\,A_{333}=A_{122}=1/3,\ A_{232}=A_{223}=1/6;
\nonumber \\
&& \left[A,B,A\right]\,:\,A_{333}=A_{232}=1/6,\ A_{312}=A_{223}=1/3;
\nonumber \\
&& \left[A,A,O\right]\,:\,A_{333}=A_{223}=1/2;\nonumber \\
&& \left[A,\sixth\sixth\twthirds,\sixth\sixth\twthirds\right]\,:\,
A_{211}=A_{232}=A_{223}=1/6,\ A_{333}=1/2;\nonumber \\
&& \left[\sixth\sixth\twthirds,A,\sixth\sixth\twthirds\right]\,:\,
A_{223}=A_{122}=A_{321}=1/6,\ A_{333}=1/2;\nonumber \\
&& \left[A,B,\sixth\sixth\twthirds\right]\,:\,
A_{211}=A_{312}=1/6,\ A_{333}=A_{223}=1/3;\nonumber \\
&& \left[\sixth\sixth\twthirds,B,A\right]\,:\,
A_{333}=A_{312}=1/3,\ A_{223}=A_{122}=1/6;\nonumber \\
&& \left[B,\sixth\sixth\twthirds,A\right]\,:\,
A_{122}=A_{113}=1/6,\ A_{333}=A_{232}=1/3. \nonumber
\end{eqnarray}
The E-points which are obtained from these by the qutrit
permutation $(2)\leftrightarrow(3)$ can also be obtained by letting
$A_{ijk} \to A_{ikj}$ in this list.
\end{proof}

\noindent
There are two convex sets $S_6^{(2)}$ and $S_6^{(3)}$ where
all corner points of $S_5$
are realized except $[A,O,A]$ and except $[A,A,O]$,
respectively, which are constructed similarly.  
Therefore all simplices whose corner points
are also corner points of $S_5$ 
can be realized except
those with all three E-points
$[O,A,A]$, $[A,O,A]$ and $[A,A,O]$ as corner points.
This observation can be used to finish the construction of
states with their E-points in $S_5$.
\begin{proposition}
There is a state with any E-point in $S_5$.
\end{proposition}

\begin{proof}
Any E-point $X$ in the set $S_5$ can be represented as
\begin{equation}
X = \sum_{s} \alpha_s Y_s,  \label{XEQ}
\end{equation}
where $Y_s$ are the corner points of $S_5$ and where the
$\alpha_s$ are nonnegative and satisfy $\sum_{s}\alpha_s = 1$.
If the coefficients $\alpha_s$ for the corner points
$[A,\sixth\sixth\twthirds,\sixth\sixth\twthirds]$,
$[\sixth\sixth\twthirds,A,\sixth\sixth\twthirds]$ and
$[\sixth\sixth\twthirds,\sixth\sixth\twthirds,A]$
are zero, then the corresponding E-point is realized by
a state by Proposition \ref{prop10}.  Also, if the coefficient
$\alpha_s$ for $[A,A,O]$, $[A,O,A]$ or $[O,A,A]$ vanishes,
then, the corresponding E-point is realized by 
Proposition \ref{prop12} and the remark following it.  

Suppose that
the coefficient for $[A,\sixth\sixth\twthirds,\sixth\sixth\twthirds]$
is $\alpha_1^{(1)}$ and that for $[O,A,A]$ is $\alpha_2^{(1)}$ for an 
E-point $X$ in (\ref{XEQ}).  Note that
$$
2\left[A,\sixth\sixth\twthirds,\sixth\sixth\twthirds\right]
+ \left[O,A,A\right]
= \left[A,A,O\right] + \left[A,O,A\right] + \left[O,B,B\right].
$$
Suppose that $\alpha_1^{(1)} \geq 2\alpha_2^{(1)}$.
Then
\begin{eqnarray}
 \alpha_1^{(1)}
\left[A,\sixth\sixth\twthirds,\sixth\sixth\twthirds\right]
+ \alpha_2^{(1)}\left[O,A,A\right] 
& = &
(\alpha_1^{(1)} -2\alpha_2^{(1)})
\left[A,\sixth\sixth\twthirds,\sixth\sixth
\twthirds\right]\nonumber \\
&& + \alpha_2^{(1)}\left\{
\left[A,A,O\right] + \left[A,O,A\right] + \left[O,B,B\right]\right\}
\,. \nonumber
\end{eqnarray}
Thus, the E-point $X$ 
can be reexpressed in such a way that the coefficient
for $\left[O,A,A\right]$ vanishes.
On the other hand, if $\alpha_1^{(1)} \leq 2\alpha_2^{(1)}$, 
then we can write
\begin{eqnarray}
\alpha_1^{(1)}
\left[A,\sixth\sixth\twthirds,\sixth\sixth\twthirds\right]
+ \alpha_2^{(1)}\left[O,A,A\right] & = & 
\left(\alpha_2^{(1)} -\half\alpha_1^{(1)}\right)
\left[O,A,A\right]
\nonumber \\
&& + \half \alpha_1^{(1)}\left\{
\left[A,A,O\right] + \left[A,O,A\right] + \left[O,B,B\right]\right\}.
\nonumber 
\end{eqnarray}
Hence, the E-point $X$
can be reexpressed in such a way that the coefficient
for $\left[A,\sixth\sixth\twthirds,\sixth\sixth\twthirds\right]$ 
vanishes.  By symmetry
the same conclusions can be made for the case where
the coefficients for both 
$[\sixth\sixth\twthirds,A,\sixth\sixth\twthirds]$ and
$[A,O,A]$ are nonzero or where those for both
$[\sixth\sixth\twthirds,\sixth\sixth\twthirds,A]$ and
$[A,A,O]$ are nonzero.

Let the coefficients for 
$[\sixth\sixth\twthirds,A,\sixth\sixth\twthirds]$ and
$[\sixth\sixth\twthirds,\sixth\sixth\twthirds,A]$ be
$\alpha_1^{(2)}$ and $\alpha_1^{(3)}$, respectively,
and those for
$[A,O,A]$ and $[A,A,O]$ be
$\alpha_2^{(2)}$ and $\alpha_2^{(3)}$, respectively, 
for the E-point $X$.  If
$\alpha_1^{(a)} \geq 2\alpha_2^{(a)}$ for some $a$, then the E-point 
$X$ can be reexpressed in such a way that one of the coefficients 
$\alpha_2^{(a)}$ vanishes.  Then by Proposition \ref{prop12} we can
conclude that E-point $X$ has a corresponding state.  If 
$\alpha_1^{(a)} < 2\alpha_2^{(a)}$ for all $a$, then by successively
applying the argument above to each qutrit, we can make 
$\alpha_1^{(a)}$ vanish for all $a$. (Note that the coefficients
$\alpha_2^{(a)}$ never decrease in this process.)  Then by Proposition
\ref{prop10} the E-point $X$ has a corresponding state.
\end{proof}

\section{Concluding remark}
\label{conclude}

It is surprising that there is a very simple condition on the
RDMs for the $n$-qubit pure quantum
state~\cite{hss,bravyi} for all $n$.  The analogous 
condition for the qutrit system 
has turned out to be more complicated as shown in this paper.
Nevertheless, the result, Theorem~\ref{main}, has some structure,
and it is probable that there is a proof simpler than the one
given here. 
It is natural to investigate the corresponding problem for the
$n$-qudit pure quantum system for all $n$.  There are two features in
our result which may be true in the general case.  One is that the
set of the allowed E-points is convex.  The other is that
the boundaries of this set are hyperplanes in the space of 
E-points.  We conclude this paper by a proposition
which suggests that the boundaries are indeed hyperplanes
in the $n$-qudit case.

An $n$-qudit quantum state is given by
\begin{equation}
|\Psi\> \equiv \sum_{1\leq i_1,i_2,\ldots,i_n\leq d}
c_{i_1i_2\cdots i_n}|i_1\>\otimes|i_2\>\otimes \cdots 
\otimes |i_n\> \label{state}
\end{equation}
with  $\sum_{i_1,i_2,\ldots,i_n}|c_{i_1i_2\cdots i_n}|^2 = 1$,
where $|1\>$, $|2\>$,...,$|d\>$ form an orthonormal basis.
As before we assume that the RDMs
are diagonal.  Let $\lambda_i^{(a)}$ be the $i$th smallest 
eigenvalue of the RDM for the
$a$th qudit.  We regard the set of these eigenvalues as a point
in the $n(d-1)$-dimensional Euclidean space parametrized by
$\lambda_i^{(a)}$, $1\leq i\leq d-1$, $1 \leq a \leq n$ and
call this set an E-point as before. It is
convenient to define the lattice E-point $L_{i_1 i_2\ldots i_n}$
as the E-point given by $\lambda_{i_a}^{(a)}=1$ and,
as a result, $\lambda_i^{(a)}=0$ if $i\neq i_a$.  Note that these
E-points do not correspond to a physical state unless
$i_1=i_2=\cdots=i_n=d$ because they do not satisfy the conditions
$\lambda_i^{(a)} \leq \lambda_j^{(a)}$ for $i < j$. For a given
set $N$ of $n$-tuples we define
$V_{N}$ to be the convex set whose elements $\lambda$ are the E-points
given as
$$
\lambda = \sum_{[i_1i_2\cdots i_n]\in N}A_{i_1i_2\cdots i_n} 
L_{i_1i_2\cdots i_n}\,,
$$
where  $0\leq A_{i_1i_2\cdots i_n}\leq 1$ and
$\sum_{[i_1i_2\cdots i_n]\in N}A_{i_1i_2\cdots i_n} = 1$.
The following
lemma is useful in proving our proposition.
\begin{lemma}
Consider a state of the
form (\ref{state}) with the eigenvalues $\Lambda_i^{(a)}$, 
$i=1,...,d$, $a=1,...,n$, such that $c_{i_1 i_2 \cdots i_n} \neq 0$
only if $[i_1 i_2\ldots i_n] \in N$.
Then the corresponding E-point 
$\Lambda\equiv (\Lambda_i^{(a)})$ is in $V_N$.
\label{penult}
\end{lemma}

\begin{proof}
An E-point $\lambda \equiv (\lambda_i^{(a)})$ in $V_N$ is given by
$$
\lambda_i^{(a)} = \sum_{(j_1,j_2,\ldots,j_n)\in N}
A_{j_1j_2\ldots j_n}\delta_{j_a i}\,,
$$
where $0\leq A_{j_1 j_2\ldots j_n} \leq 1$ and
$\sum_{[i_1 i_2\ldots i_n]\in N} A_{i_1 i_2\ldots i_n} = 1$.
These are the formulae for $\Lambda_i^{(a)}$ for the state
(\ref{state}) if we make the identification 
$A_{i_1 i_2\cdots i_n} = |c_{i_1 i_2\cdots i_n}|^2$.
\end{proof}

\begin{proposition}
Let $U$ be an open set in the $d(n-1)$-dimensional space of
E-points.  Suppose that the eigenvalues are not
degenerate in $U$.
Suppose further that an E-point $\lambda\equiv (\lambda_i^{(a)})$ 
has a corresponding state if and 
only if $f(\lambda)\geq 0$, where $f$ is a function
with a continuous and nonzero gradient and
that the hypersurface $f(\lambda)=0$ is connected.
Then this hypersurface is part of the hyperplane
$V_N$ for some set of $n$-tuples $N$.
\end{proposition}

\begin{proof}
Suppose that a state with $\lambda_i^{(a)} = \Lambda_i^{(a)}$ 
is in $U$ and on the boundary so that
$f(\Lambda)=0$, where $\Lambda \equiv (\Lambda_i^{(a)})$,
and that both $c_{i_1 i_2 \cdots i_n}$ and
$c_{i'_1 i'_2 \cdots i'_n}$ are nonzero for this state.
We consider the variation
$$
\delta c_{i_1 i_2\cdots i_n} = \alpha c_{i'_1 i'_2 \cdots i'_n}\,,\ \
\delta c_{i'_1 i'_2 \cdots i'_n}
= -\overline{\alpha} c_{i_1 i_2\cdots i_n}
$$
with all other $\delta c_{I_1 I_2\cdots I_n}$ vanishing.
Then the variation of the function $f$ is given by
\begin{eqnarray}
\delta f & = & \sum_{a=1}^n\sum_{i=1}^{d-1}
\left. \frac{\partial f}{\partial \lambda_i^{(a)}}
\right|_{\lambda=\Lambda}
\delta\lambda_i^{(a)}\nonumber \\
& = & \sum_{a=1}^n \left. \left[
\frac{\partial f}{\partial \lambda_{i_a}^{(a)}}
- \frac{\partial f}{\partial \lambda_{i'_a}^{(a)}}\right]
\right|_{\lambda=\Lambda}
\cdot 2{\rm Re}\,(\alpha c_{i'_1 i'_2\ldots i'_n}
\overline{c_{i_1i_2\ldots i_n}})\,,\nonumber
\end{eqnarray}
where we have defined 
$\partial f/\partial\lambda_d^{(a)} = 0$ for all $a$, because there
are no degenerate eigenvalues.
For $\delta f$ to be nonnegative for all $\alpha$ we must have
\begin{equation}
\sum_{a=1}^{n} \left.\frac{\partial f}{\partial \lambda_{i_a}^{(a)}}
\right|_{\lambda=\Lambda}
= \sum_{a=1}^{n}\left.
\frac{\partial f}{\partial \lambda_{i'_a}^{(a)}}
\right|_{\lambda=\Lambda}\,. \label{bon}
\end{equation}
Define a linear function $F$ by
$$
F(\lambda) = \sum_{a=1}^n \sum_{i=1}^{d-1} 
\left. \frac{\partial f}{\partial\lambda_i^{(a)}}
\right|_{\lambda=\Lambda}\lambda_i^{(a)}\,.
$$
Then Eq.~(\ref{bon}) implies that
$F(L_{i_1i_2\cdots i_n})$ must be the same for all
$[i_1 i_2 \ldots i_n] \in \tilde{N}$, 
where $\tilde{N}$ is the set of $n$-tuples
such that $c_{i_1 i_2\cdots i_n}\neq 0$ for the given state.
Let this value be $v$.
Then the equation $F(\lambda)=v$ determines a hyperplane
containing all lattice E-points $L_{i_1 i_2 \cdots i_n}$
such that
$[i_1 i_2\ldots i_n]\in \tilde{N}$.  This implies that the convex set
$V_{\tilde{N}}$ is contained in a hyperplane.  Thus,
$V_{\tilde{N}} \subset V_N$ with $\tilde{N} \subset N$, where
$V_N$ is a hyperplane. Hence, by Lemma~\ref{penult} we find
$\Lambda \in V_{\tilde{N}} \subset V_N$.
Thus, all E-points on $f(\lambda)=0$ are on a hyperplane $V_N$ for 
some $N$.   Since there are only a finite number of such hyperplanes,
and since the connected hypersurface 
$f(\lambda)=0$ has a tangent hyperplane
everywhere, this hypersurface must coincide with a single hyperplane
$V_N$.
\end{proof}

\acknowledgments

\noindent
The author thanks Tony Sudbery and Jason Szulc for useful discussions
at an early stage of this work.

\

\end{document}